\documentclass[journal]{IEEEtran} 

\usepackage{url}
\usepackage{epsfig,algorithm,algorithmic,amsmath,amssymb,color,subfigure,graphicx}
\usepackage{epstopdf}
\usepackage{enumerate}
\usepackage[normalem]{ulem}

\usepackage{graphicx,accents}
\usepackage{amsfonts,amssymb,latexsym,cite,ifthen,multirow,array}

\author{
Namrata Vaswani and Jinchun Zhan
\thanks{
N. Vaswani and J. Zhan are with the Dept. of Electrical and Computer Engineering, Iowa State University, Ames, IA, USA. Email: namrata@iastate.edu.
This work was supported by NSF grant CCF-0917015 and CCF-1117125.
}
}

 \DeclareMathOperator{\sgn}{sgn}

\title{Recursive Recovery of Sparse Signal Sequences from Compressive Measurements: A Review} 

\begin{document}
\maketitle
\setlength{\arraycolsep}{0.03cm}
\newcommand{\xhat}{\hat{x}}
\newcommand{\xpred}{\hat{x}_{t|t-1}}
\newcommand{\Ppred}{P_{t|t-1}}
\newcommand{\ty}{\tilde{y}_t}
\newcommand{\tty}{\tilde{y}_{t,\text{res}}}
\newcommand{\tw}{\tilde{w}_t}
\newcommand{\ttw}{\tilde{w}_{t,f}}
\newcommand{\betahat}{\hat{\beta}}

\newcommand{\ypast}{y_{1:t-1}}
\newcommand{\sone}{S_{*}}
\newcommand{\sinf}{{S_{**}}}
\newcommand{\smax}{S_{\max}}
\newcommand{\smin}{S_{\min}}
\newcommand{\samax}{S_{a,\max}}
\newcommand{\Nhat}{{\hat{N}}}

\newcommand{\Dnum}{D_{num}}
\newcommand{\pss}{p^{**,i}}
\newcommand{\fr}{f_{r}^i}

\newcommand{\A}{{\cal A}}
\newcommand{\Z}{{\cal Z}}
\newcommand{\B}{{\cal B}}
\newcommand{\R}{{\cal R}}
\newcommand{\reg}{{\cal G}}
\newcommand{\const}{\mbox{const}}

\newcommand{\trace}{\mbox{tr}}

\newcommand{\hsim}{{\hspace{0.0cm} \sim  \hspace{0.0cm}}}
\newcommand{\he}{{\hspace{0.0cm} =  \hspace{0.0cm}}}

\newcommand{\vect}[2]{\left[\begin{array}{cccccc}
     #1 \\
     #2
   \end{array}
  \right]
  }

\newcommand{\matr}[2]{ \left[\begin{array}{cc}
     #1 \\
     #2
   \end{array}
  \right]
  }
\newcommand{\vc}[2]{\left[\begin{array}{c}
     #1 \\
     #2
   \end{array}
  \right]
  }

\newcommand{\gdot}{\dot{g}}
\newcommand{\Cdot}{\dot{C}}
\newcommand{\re}{\mathbb{R}}
\newcommand{\n}{{\cal N}}  
\newcommand{\N}{{\overrightarrow{\bf N}}}  
\newcommand{\chat}{\tilde{C}_t}
\newcommand{\chati}{\chat^i}

\newcommand{\cmin}{C^*_{min}}
\newcommand{\twi}{\tilde{w}_t^{(i)}}
\newcommand{\twj}{\tilde{w}_t^{(j)}}
\newcommand{\wi}{{w}_t^{(i)}}
\newcommand{\twio}{\tilde{w}_{t-1}^{(i)}}

\newcommand{\tWi}{\tilde{W}_n^{(m)}}
\newcommand{\tWj}{\tilde{W}_n^{(k)}}
\newcommand{\Wi}{{W}_n^{(m)}}
\newcommand{\tWio}{\tilde{W}_{n-1}^{(m)}}

\newcommand{\ds}{\displaystyle}

\newcommand{\SAR}{S$\!$A$\!$R }
\newcommand{\MAR}{MAR}
\newcommand{\MMRF}{MMRF}
\newcommand{\AR}{A$\!$R }
\newcommand{\GMRF}{G$\!$M$\!$R$\!$F }
\newcommand{\DTM}{D$\!$T$\!$M }
\newcommand{\MSE}{M$\!$S$\!$E }
\newcommand{\RCS}{R$\!$C$\!$S }
\newcommand{\uomega}{\underline{\omega}}
\newcommand{\y}{v}
\newcommand{\x}{w}
\newcommand{\lu}{\mu}
\newcommand{\g}{g}
\newcommand{\s}{{\bf s}}
\newcommand{\bft}{{\bf t}}
\newcommand{\refmap}{{\cal R}}
\newcommand{\totrefl}{{\cal E}}
\newcommand{\beq}{\begin{equation}}
\newcommand{\eeq}{\end{equation}}
\newcommand{\bdm}{\begin{displaymath}}
\newcommand{\edm}{\end{displaymath}}
\newcommand{\hatz}{\hat{z}}
\newcommand{\hatu}{\hat{u}}
\newcommand{\tilz}{\tilde{z}}
\newcommand{\tilu}{\tilde{u}}
\newcommand{\hhatz}{\hat{\hat{z}}}
\newcommand{\hhatu}{\hat{\hat{u}}}
\newcommand{\tilc}{\tilde{C}}
\newcommand{\hatc}{\hat{C}}
\newcommand{\tim}{n}

\newcommand{\ssp}{\renewcommand{\baselinestretch}{1.0}}
\newcommand{\defd}{\mbox{$\stackrel{\mbox{$\triangle$}}{=}$}}
\newcommand{\goes}{\rightarrow}
\newcommand{\tends}{\rightarrow}
\newcommand{\defn}{:=} 
\newcommand{\se}{&=&}
\newcommand{\sdefn}{& \defn  &}
\newcommand{\sle}{& \le &}
\newcommand{\sge}{& \ge &}
\newcommand{\plusminus}{\stackrel{+}{-}}
\newcommand{\Ey}{E_{Y_{1:t}}}
\newcommand{\ey}{E_{Y_{1:t}}}

\newcommand{\equivto}{\mbox{~~~which is equivalent to~~~}}
\newcommand{\nonzero}{i:\pi^n(x^{(i)})>0}
\newcommand{\nonzeroc}{i:c(x^{(i)})>0}

\newcommand{\supn}{\sup_{\phi:||\phi||_\infty \le 1}}

\newcommand{\eps}{\epsilon}
\newcommand{\udq}{\underline{D_Q}}
\newcommand{\td}{\tilde{D}}
\newcommand{\epsinv}{\epsilon_{inv}}
\newcommand{\al}{\mathcal{A}}

\newcommand{\bfx} {\bf X}
\newcommand{\bfy} {\bf Y}
\newcommand{\bfz} {\bf Z}
\newcommand{\ddas}{\mbox{${d_1}^2({\bf X})$}}
\newcommand{\ddbs}{\mbox{${d_2}^2({\bfx})$}}
\newcommand{\dda}{\mbox{$d_1(\bfx)$}}
\newcommand{\ddb}{\mbox{$d_2(\bfx)$}}
\newcommand{\xinc}{{\bfx} \in \mbox{$C_1$}}
\newcommand{\eqa}{\stackrel{(a)}{=}}
\newcommand{\eqb}{\stackrel{(b)}{=}}
\newcommand{\eqe}{\stackrel{(e)}{=}}
\newcommand{\leqc}{\stackrel{(c)}{\le}}
\newcommand{\leqd}{\stackrel{(d)}{\le}}

\newcommand{\leqa}{\stackrel{(a)}{\le}}
\newcommand{\leqb}{\stackrel{(b)}{\le}}
\newcommand{\leqe}{\stackrel{(e)}{\le}}
\newcommand{\leqf}{\stackrel{(f)}{\le}}
\newcommand{\leqg}{\stackrel{(g)}{\le}}
\newcommand{\leqh}{\stackrel{(h)}{\le}}
\newcommand{\leqi}{\stackrel{(i)}{\le}}
\newcommand{\leqj}{\stackrel{(j)}{\le}}

\newcommand{\w}{{W^{LDA}}}
\newcommand{\halpha}{\hat{\alpha}}
\newcommand{\hsigma}{\hat{\sigma}}
\newcommand{\slmax}{\sqrt{\lambda_{max}}}
\newcommand{\slmin}{\sqrt{\lambda_{min}}}
\newcommand{\lmax}{\lambda_{max}}
\newcommand{\lmin}{\lambda_{min}}

\newcommand{\da} {\frac{\alpha}{\sigma}}
\newcommand{\chka} {\frac{\check{\alpha}}{\check{\sigma}}}
\newcommand{\sumo}{\sum _{\underline{\omega} \in \Omega}}
\newcommand{\distance}{d\{(\hatz _x, \hatz _y),(\tilz _x, \tilz _y)\}}
\newcommand{\col}{{\rm col}}
\newcommand{\rcs}{\sigma_0}
\newcommand{\CalR}{{\cal R}}
\newcommand{\df}{{\delta p}}
\newcommand{\dq}{{\delta q}}
\newcommand{\dZ}{{\delta Z}}
\newcommand{\pprime}{{\prime\prime}}

\newcommand{\vn}{N}

\newcommand{\bv}{\begin{vugraph}}
\newcommand{\ev}{\end{vugraph}}
\newcommand{\bi}{\begin{itemize}}
\newcommand{\ei}{\end{itemize}}
\newcommand{\ben}{\begin{enumerate}}
\newcommand{\een}{\end{enumerate}}
\newcommand{\be}{\protect\[}
\newcommand{\ee}{\protect\]}
\newcommand{\bean}{\begin{eqnarray*} }
\newcommand{\eean}{\end{eqnarray*} }
\newcommand{\bea}{\begin{eqnarray} }
\newcommand{\eea}{\end{eqnarray} }
\newcommand{\nn}{\nonumber}
\newcommand{\ba}{\begin{array} }
\newcommand{\ea}{\end{array} }
\newcommand{\ep}{\mbox{\boldmath $\epsilon$}}
\newcommand{\epp}{\mbox{\boldmath $\epsilon '$}}
\newcommand{\Lep}{\mbox{\LARGE $\epsilon_2$}}
\newcommand{\und}{\underline}
\newcommand{\pdif}[2]{\frac{\partial #1}{\partial #2}}
\newcommand{\odif}[2]{\frac{d #1}{d #2}}
\newcommand{\dt}[1]{\pdif{#1}{t}}
\newcommand{\urho}{\underline{\rho}}

\newcommand{\spc}{{\cal S}}
\newcommand{\tspc}{{\cal TS}}

\newcommand{\uv}{\underline{v}}
\newcommand{\us}{\underline{s}}
\newcommand{\uc}{\underline{c}}
\newcommand{\utheta}{\underline{\theta}^*}
\newcommand{\ualpha}{\underline{\alpha^*}}

\newcommand{\uxy}{\underline{x}^*}
\newcommand{\uxyj}{[x^{*}_j,y^{*}_j]}
\newcommand{\arcl}[1]{arclen(#1)}
\newcommand{\one}{{\mathbf{1}}}

\newcommand{\uxyjt}{\uxy_{j,t}}
\newcommand{\E}{\mathbb{E}}

\newcommand{\rhomat}{\left[\begin{array}{c}
                        \rho_3 \ \rho_4 \\
                        \rho_5 \ \rho_6
                        \end{array}
                   \right]}
\newcommand{\deltat}{\tau} 
\newcommand{\deltatt}{\Delta t_1}
\newcommand{\ceil}[1]{\ulcorner #1 \urcorner}

\newcommand{\xxi}{x^{(i)}}
\newcommand{\txi}{\tilde{x}^{(i)}}
\newcommand{\txj}{\tilde{x}^{(j)}}

\newcommand{\mi}[1]{{#1}^{(m,i)}}

\newcommand{\cred}{} 
\newtheorem{theorem}{Theorem}[section]
\newtheorem{lem}[theorem]{Lemma}
\newtheorem{sigmodel}[theorem]{Model}
\newtheorem{corollary}[theorem]{Corollary}
\newtheorem{definition}[theorem]{Definition}
\newtheorem{remark}[theorem]{Remark}
\newtheorem{example}[theorem]{Example}
\newtheorem{ass}[theorem]{Assumption}
\newtheorem{proposition}[theorem]{Proposition}
\newtheorem{fact}[theorem]{Fact}
\newcommand{\bd}{\begin{definition}}
\newcommand{\ed}{\end{definition}}
\newcommand{\poly}{\text{poly}}
\newcommand{\kron}{\otimes}

\newcommand{\rest}{{\T_\text{rest}}}
\newcommand{\zetahat}{\hat{\zeta}}
\newcommand{\sm}{\text{small}}
\newcommand{\tDelta}{{\tilde{\Delta}_u}}
\newcommand{\tDeltae}{{\tilde{\Delta}_e}}
\newcommand{\tT}{{\tilde{\T}}}
\newcommand{\add}{{\text{add}}} 

\newcommand{\thr}{{\text{thr}}}
\newcommand{\delthr}{{\text{del-thr}}}
\newcommand{\delbound}{{b}}
\newcommand{\err}{{\text{err}}}
\newcommand{\Q}{{\cal Q}}

\newcommand{\Bernoulli}{{\text{Bernoulli}}}
\newcommand{\del}{{\text{del}}}
\newcommand{\dett}{{\text{add}}}   
\newcommand{\CSres}{{\text{CSres}}}
\newcommand{\diff}{{\text{diff}}}
\newcommand{\st}{k}
\newcommand{\sd}{u}
\newcommand{\sde}{e}
\newcommand{\sn}{s}
\newcommand{\Sp}{\check{S}}
\newcommand{\sno}{m} 
\newcommand{\mno}{n} 

\renewcommand{\N}{{\mathcal{N}}}
\newcommand{\T}{{\mathcal{T}}}
\renewcommand{\Nhat}{{{\hat{\mathcal{N}}}}}
\newcommand{\tjset}{\mathrm{addtimes}_j}
%
 \newcounter{threshcounter}
 \renewcommand{\thethreshcounter}{D\arabic{threshcounter}}
 \newcommand{\threshcnt}[1]{\refstepcounter{threshcounter} \label{#1} \thetag{\thethreshcounter}}
 \newcounter{algcounter}
 \renewcommand{\thealgcounter}{A\arabic{algcounter}}
 \newcommand{\algcnt}[1]{\refstepcounter{algcounter} \label{#1} \thetag{\thealgcounter}}
 \newcounter{emcounter}
 \renewcommand{\theemcounter}{E\arabic{emcounter}}
 \newcommand{\emcnt}[1]{\refstepcounter{emcounter} \label{#1} \thetag{\theemcounter}}
\newcommand{\supp}{\text{supp}}

\begin{abstract}
In this article, we review the literature on design and analysis of recursive algorithms for reconstructing a time sequence of sparse signals from compressive measurements. The signals are assumed to be sparse in some transform domain or in some dictionary. Their sparsity patterns can change with time, although, in many practical applications, the changes are gradual. An important class of applications where this problem occurs is dynamic projection imaging, e.g., dynamic magnetic resonance imaging (MRI) for real-time medical applications such as interventional radiology, or dynamic computed tomography.
\end{abstract}

\section{Introduction} \label{intro}
In this paper, we review the literature on the design and analysis of {recursive} algorithms for causally reconstructing a time sequence of sparse signals from a limited number of linear measurements (compressive measurements). The signals are assumed to be sparse, or approximately sparse, in some transform domain referred to as the sparsity basis. Their sparsity pattern (support set of the sparsity basis coefficients' vector) can change with time. The signals could also be sparse in a known dictionary and everything described in this article will apply. The term ``recursive algorithms", refers to algorithms that only use past signals' estimates and the current measurements' vector to get the current signal's estimate \footnote{In some other works, ``recursive estimation" is also used to refer to recursively estimating a single signal as more of its measurements come in. This should not be confused with our definition.}.

The problem of recovering a sparse signal from a small number of its linear measurements has been studied for a long time. In the signal processing literature, the works of Mallat and Zhang \cite{mallat} (matching pursuit), Chen and Donoho (basis pursuit) \cite{bpdn}, Feng and Bresler \cite{feng_bresler,feng_bresler_2} (spectrum blind recovery of multi-band signals), Gorodnistky and Rao \cite{gorod_rao,bdrao_focuss} (a reweighted minimum 2-norm algorithm for sparse recovery) and Wipf and Rao \cite{sbl} (sparse Bayesian learning for sparse recovery) were among the first works on this topic. The papers by Candes, Romberg, Tao and by Donoho \cite{donoho,decodinglp,candes} introduced the compressive sensing (CS) problem. The idea of CS is to compressively sense signals that are sparse in some known domain and then use sparse recovery techniques to recover them. The most important contribution of \cite{donoho,decodinglp,candes} was that these works provided practically meaningful conditions for exact sparse recovery using basis pursuit. In the last decade since these papers appeared, this problem has received a lot of attention. Often the terms ``sparse recovery" and ``CS" are used interchangeably. We also do this in this article. The CS problem occurs in a large class of applications. Examples include magnetic resonance imaging (MRI), computed tomography (CT), and various other projection imaging applications, where measurements are acquired one linear projection at a time. The ability to reconstruct from fewer measurements is useful for these applications since it means that less time is needed for completing an MR or a CT scan.

Consider the dynamic CS problem, i.e., the problem of recovering a time sequence of sparse signals. Most of the initial solutions for this problem consist of batch algorithms. These can be split into two categories depending on what assumption they use on the time sequence. The first category is batch algorithms that solve the multiple measurements' vectors (MMV) problem. These use the assumption that the support set of the sparse signals {\em does not} change with time \cite{wipf2007empirical,tropp2006algorithms2,chen2006theoretical,eldar2009compressed,davies_eldar}. The second category is batch algorithms that treat the entire time sequence as a single sparse spatiotemporal signal by assuming Fourier sparsity along the time axis \cite{singlepixelvideo,sparsedynamicmri,jung_etal}. However, in many situations neither of these assumptions is valid.
Moreover, even when these are valid assumptions, batch algorithms are offline, slower, and their memory requirement increases linearly with the sequence length. In this work, we focus on recursive algorithms for solving the dynamic CS problem. Their computational and storage complexity is much lower than that of the batch techniques and is, in fact, comparable to that of simple-CS solutions. At the same time, the number of measurements required by these algorithms for exact or accurate recovery is significantly smaller than what simple-CS solutions need, as long as an accurate estimate of the first sparse signal is available. Hereafter  ``simple-CS" refers to dynamic CS solutions that recover each sparse signal in the sequence independently without  using any information from past or future frames.%

\begin{figure}
\centering
\includegraphics [width=9cm,height=3cm]{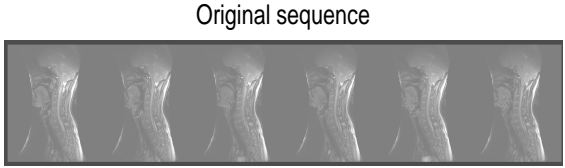} 
\vspace{-0.4in}
\caption{\small{
We show a dynamic MRI sequence of the vocal tract (larynx) that was acquired when the person was speaking a vowel. 
}}
\label{examples}
\end{figure}

\begin{figure}
\centerline{
\subfigure[{slow support changes (adds)}]{
\includegraphics [width=4cm,height=3cm]{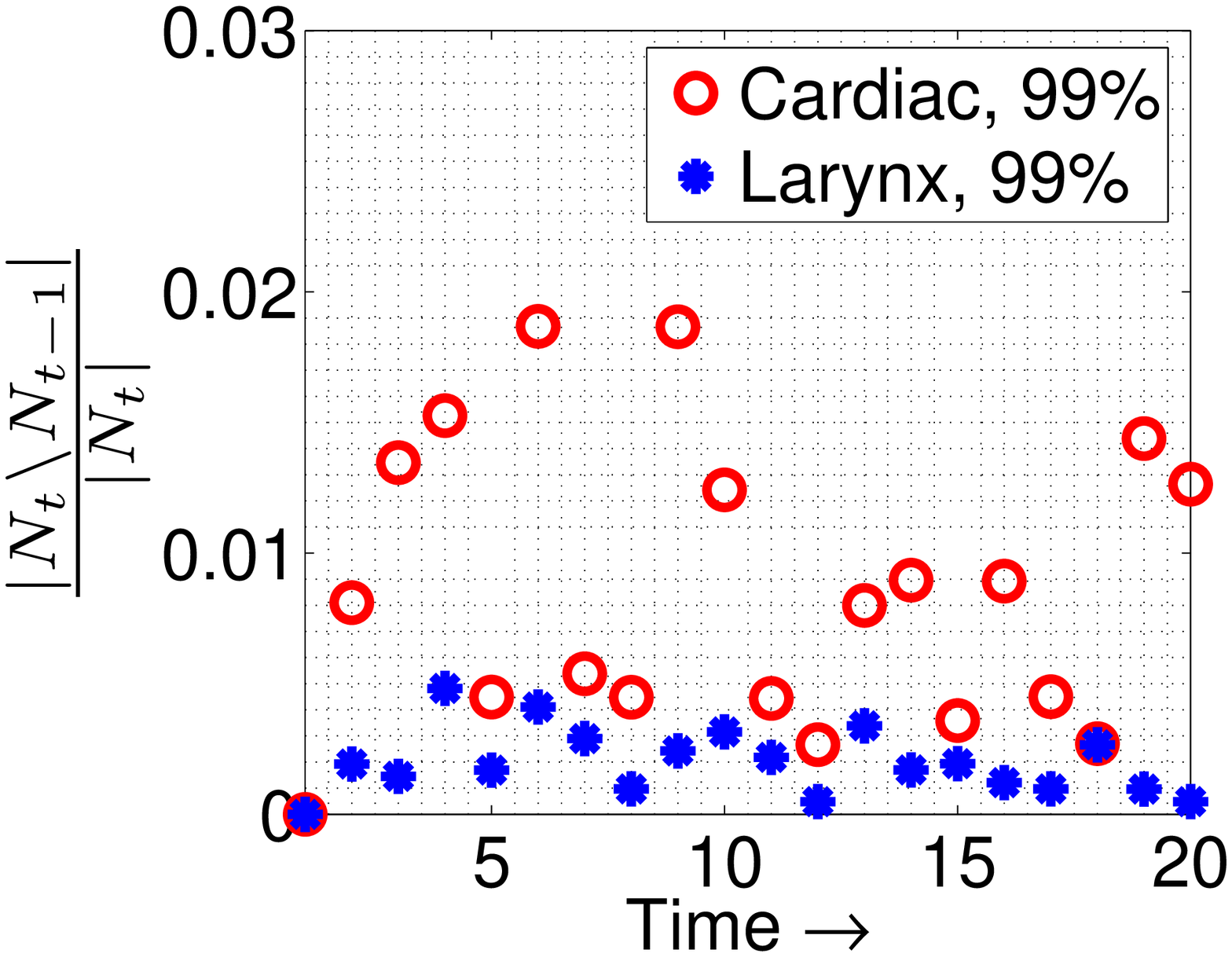}
}
\subfigure[{slow support changes (removals)}]{
\includegraphics[width=4cm,height=3cm]{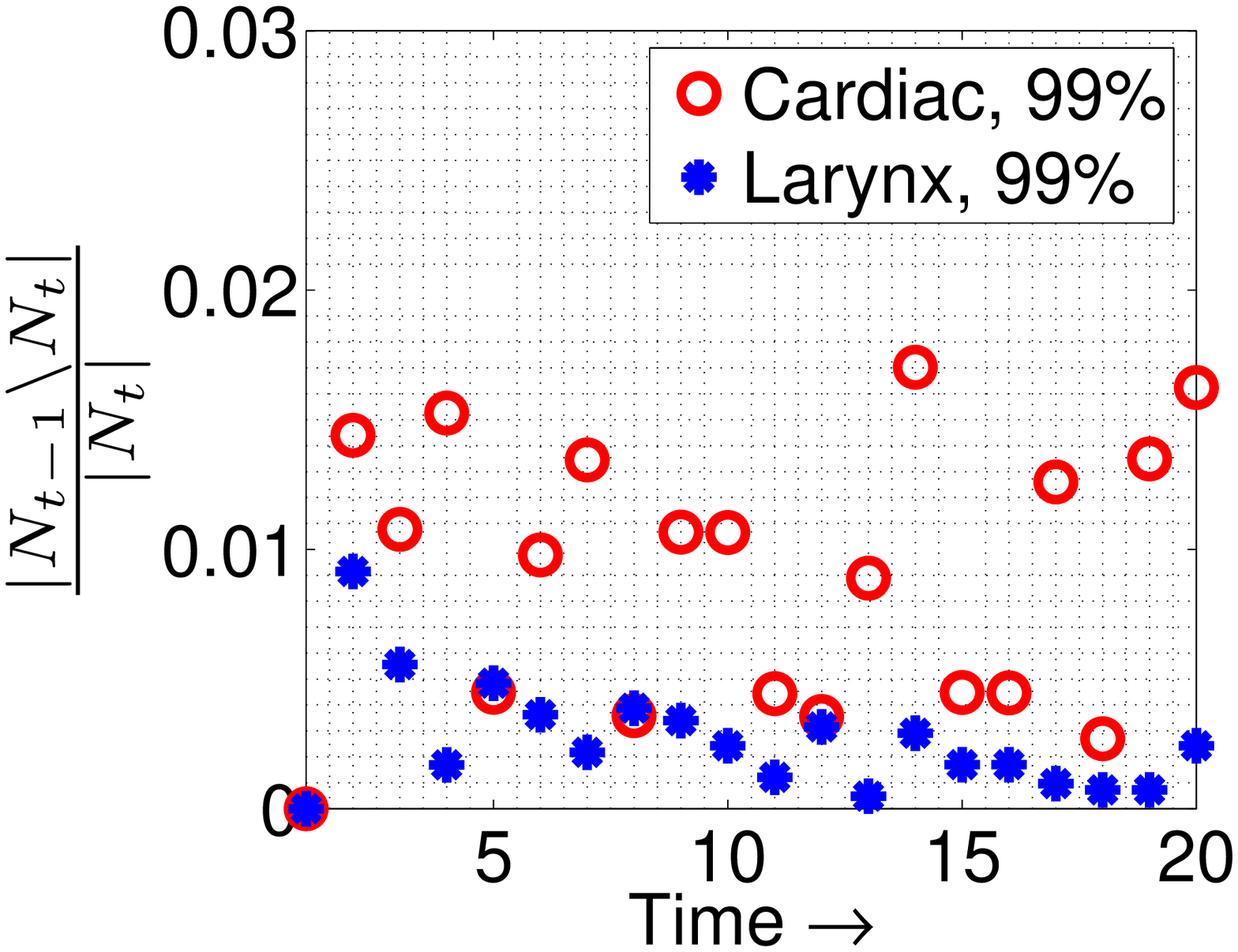}
}
}
\vspace{-0.05in}
\caption{\small{
In these figures, $\N_t$ refers to the 99\%-energy support of the 2D discrete wavelet transform (two-level Daubechies-4 2D DWT) of the larynx sequence shown in Fig. \ref{examples} and of a cardiac sequence. The 99\%-energy support size, $|\N_t|$, varied between 6-7\% of the image size in both cases. 
We plot the number of additions (top) and the number of removals (bottom) as a fraction of the support size. {\em Notice that all support change sizes are less than 2\% of the support size.}
}}
\label{suppchange}
\end{figure}

The above problem occurs in multiple applications. In fact, it occurs in virtually every application where CS is useful and there exists a sequence of sparse signals. For a comprehensive list of CS applications, see \cite{igorcarron,rice}. One common class of applications is dynamic MRI or dynamic CT. We show an example of a vocal tract (larynx) MR image sequence in Fig. \ref{examples}. Notice that the images are piecewise smooth and hence wavelet sparse. As shown in Fig. \ref{suppchange}, their sparsity pattern in the wavelet transform domain changes with time, but the changes are slow. We discuss this and other applications in Sec. \ref{apps}.



\subsection{Paper Organization}
The rest of this article is organized as follows. We summarize the notation and provide a short overview of some of the approaches for solving the static sparse recovery or CS problem in Section \ref{bgnd}. Next, in Section \ref{probdef}, we define the recursive dynamic CS problem, discuss its applications and explain why new approaches are needed to solve it.
We split the discussion of the proposed solutions into three sections. In Section \ref{supp_knowledge}, we discuss algorithms that only exploit slow support change. Under this assumption and if the first signal can be recovered accurately, our problem can be reformulated as one of sparse recovery with partial support knowledge. We describe solutions to this reformulated problem and their guarantees (which is also of independent interest). In Section \ref{supp_sig_knowledge}, we discuss algorithms that also exploit slow signal value change, by again reformulating a static problem first. In Section \ref{recrecon}, we first briefly explain how the solutions from the previous two sections apply to the recursive dynamic CS problem. Next, we describe solutions that were designed in the recursive dynamic CS context. Algorithm pseudo-code and the key ideas of how to set parameters automatically are also given. Error stability over time results are also discussed here. In Section \ref{track_adapt_filt}, we explain tracking-based and adaptive-filtering-based solutions and their pros and cons compared with previously described solutions.
Numerical experiments comparing the various approaches both on simulated data and on dynamic MRI sequences are discussed in Section \ref{sims}. In Section \ref{future}, we describe work on related problems and how it can be used in conjunction with recursive dynamic CS. Open questions for future work are also summarized. We conclude in Section \ref{conclude}.


\section{Notation and Background}
\label{bgnd}

\subsection{Notation}
\label{notation}
We define $[1,m]:=[1,2,\dots m]$.
For a set $\T$, we use $\T^c$ to denote the complement of $\T$ w.r.t. $[1,m]$, i.e., $\T^c := \{i \in [1,m]: i \notin \T \}$. The notation $|\T|$ denotes the size (cardinality) of the set $\T$. The set operation $\setminus$ denotes set set difference, i.e., for two sets $\T_1,\T_2$, $\T_1 \setminus \T_2:= \T_1 \cap \T_2^c$.
We use $\emptyset$ to denote the empty set.
For a vector, $v$, and a set, $\T$, $v_\T$ denotes the $|\T|$ length sub-vector containing the elements of $v$ corresponding to the indices in the set $\T$. Also, $\|v\|_k$ denotes the $\ell_k$ norm of a vector $v$. If just $\|v\|$ is used, it refers to $\|v\|_2$. When $k=0$, $\|v\|_0$ counts the number of nonzero elements in the vector $v$. This is referred to the $\ell_0$ norm even though it does not satisfy the properties of a norm. A vector $v$ is sparse if it has many zero entries. We use $\text{supp}(v):=\{i: v_i \neq 0\}$ to denote the support set of a sparse vector $v$, i.e., the set of indices of its nonzero entries.
%
%
For a matrix $M$, $\|M\|_k$ denotes its induced $k$-norm, while just $\|M\|$ refers to $\|M\|_2$. $M'$ denotes the transpose of $M$ and $M^\dag$ denotes its Moore-Penrose pseudo-inverse. For a tall matrix, $M$, $M^\dag : = (M'M)^{-1}M'$.
For a matrix $A$, $A_\T$ denotes the sub-matrix obtained by extracting the columns of $A$ corresponding to the indices in $\T$.
We use $I$ to denote the identity matrix.

\subsection{Sparse Recovery or Compressive Sensing (CS)} \label{cs_review}
The goal of sparse recovery or CS is to reconstruct an $m$-length sparse signal, $x$, with support $\N$, from an $n$-length measurement vector, $y:=Ax$ (noise-free case), or from $y:=Ax+w$ with $\|w\|_2 \le \eps$ (noisy case), when $A$ has more columns than rows, i.e., when $n<m$. Consider the noise-free case. Let $s = |\N|$. It is easy to show that this problem is solved if we can find the sparsest vector satisfying $y=A\beta$, i.e., if we can solve
\bea
\min_\beta {\| \beta \|_0} \  \text{subject to} \  y = A \beta,
\label{ell0min}
\eea
and if any set of $2s$ columns of $A$ are linearly independent \cite{decodinglp}. However doing this is impractical since it requires a combinatorial search. The complexity of solving (\ref{ell0min}) is exponential in the support size. 
In the last two decades, many practical (polynomial complexity) approaches have been developed. Most of the classical approaches can be split as follows (a) convex relaxation approaches, (b) greedy algorithms, (c) iterative thresholding methods, and (d) sparse Bayesian learning (SBL) based methods. 
Besides these, there are many more solution approaches that have appeared more recently. 

The convex relaxation approaches are also referred to as $\ell_1$ minimization programs since these replace the $\ell_0$ norm in (\ref{ell0min}) by the $\ell_1$ norm which is the closest norm to $\ell_0$ that is convex. Thus, in the noise-free case, one solves
\bea
\min_\beta {\| \beta \|_1}  \  \text{subject to} \  y = A \beta
\label{bp}
\eea
The above program is referred to as {\em basis pursuit (BP)} \cite{bpdn}. Clearly, it is a convex program. In fact it is easy to show that it is actually a linear program\footnote{Let $x^+$ denote a sparse vector whose $i$-th index is zero wherever $x_i \le 0$ and is equal to $x_i$ otherwise. Similarly let $x^-$ be a vector whose $i$-th index is zero wherever $x_i \ge 0$ and is equal to $-x_i$ otherwise. Then, clearly, $x = x^+ - x^-$ and $\|x\|_1 = \sum_{i=1}^m (x^+)_i + \sum_{i=1}^m (x^-)_i$. Thus \eqref{bp} is equivalent to the linear program $\min_{\beta^+,\beta^-} {\sum_{i=1}^m (\beta^+)_i + \sum_{i=1}^m (\beta^-)_i}  \  \text{subject to} \  y = A (\beta^+ - \beta^-)$.}.
{\em Since this was the program analyzed in the papers that introduced the term Compressed Sensing (CS) \cite{decodinglp,candes,donoho}, some later works wrongly refer to this program itself as ``CS".} In the noisy case, the constraint is replaced by $\|y-A \beta\|_2 \le \eps$ where $\eps$ is the bound on the $\ell_2$ norm of the noise, i.e.,
\bea
\min_\beta {\| \beta \|_1}  \  \text{subject to} \  \|y-A \beta\|_2 \le \eps.
\label{bpnoisy}
\eea
This is referred to as {\em BP-noisy.}
In  problems where the noise bound is not known, one can solve an unconstrained version of this problem by including the data term as a soft constraint (the ``Lagrangian version"):
\bea
\min_\beta  \gamma \| \beta \|_1 + 0.5 \|y - A \beta\|_2^2.
\label{ubpdn}
\eea
The above is referred to as {\em BP denoising (BPDN)} \cite{bpdn}. BPDN  solves an unconstrained convex optimization problem which is faster to solve than BP-noisy which has constraints. Of course, both are equivalent in the sense that for given $\epsilon$, there exists a $\gamma(\epsilon)$ such that the solutions to both coincide. However it is hard to find such a mapping.

Another class of solutions for the CS problem consists of greedy algorithms. These get an estimate of the support of $x$ in a greedy fashion. This is done by finding one, or a set of, indices of the columns of $A$ that have the largest correlation with the measurement residual from the previous iteration. The first known greedy algorithms were Matching Pursuit \cite{mallat} and later orthogonal matching pursuit (OMP)  \cite{omp_pati,OMP_tropp}. MP and OMP add only one index at a time. Later algorithms such as subspace pursuit \cite{subspacepursuit} and CoSaMP \cite{cosamp} add multiple indices at a time, but also include a step to delete indices.
%
%
%
%
The greedy algorithms are designed assuming that the columns of $A$ have unit $\ell_2$ norm. However when this does not hold, it is easy to reformulate the problem in order to ensure this.

\begin{remark}[matrices $A$ without unit $\ell_2$-norm columns]
Any matrix can be converted into a matrix with unit $\ell_2$-norm columns by right multiplying it with a diagonal matrix $D$ that contains $\|A_i\|_2^{-1}$ as its entries. Here $A_i$ is the $i$-th column of $A$. 
Thus, for any matrix $A$, $A_{\text{normalized}} = A D$. 
Whenever normalized columns of $A$ are needed, one can rewrite $y=Ax$ as $y = A D D^{-1} x = A_{\text{normalized}} \tilde{x}$ and first recover $\tilde{x}$ from $y$ and then obtain $x = D \tilde{x}$.%
\label{A_norm}
\end{remark}

A third solution approach for CS is Iterative Hard Thresholding (IHT) \cite{iht_blum,iht_blum2}. This is an iterative algorithm that proceeds by hard thresholding the current ``estimate'' of $x$ to $s$ largest elements. Let $H_s(a)$ denote the hard thresholding operator which zeroes out all but the $s$ largest magnitude elements of the vector $a$.  Let $\xhat^k$ denote the estimate of $x$ at the $k^{th}$ iteration. It proceeds as follows.
\beq
\xhat^0 =0, \ \xhat^{i+1} = H_s(\xhat^i + A'(y - A \xhat^i) ).
\label{iht}
\eeq

Another commonly used approach to solving the sparse recovery problem is sparse Bayesian learning (SBL) \cite{tipping,sbl}. In SBL, one models the sparse vector $x$ as consisting of independent Gaussian components with zero mean and variances $\gamma_i$ for the $i$-th component. The observation noise is assumed to be independent identically distributed (i.i.d.) Gaussian with zero mean and variance $\sigma^2$. SBL consists of an expectation maximization (EM)-type algorithm to estimate the hyper-parameters $\{\sigma^2, \gamma_1, \gamma_2, \dots \gamma_m\}$ from the observation vector $y$ using evidence maximization (type-II maximum likelihood). Since the true $x$ is sparse, it can be argued that the estimates of a lot of the $\gamma_i$'s will be zero or nearly zero (and can be zeroed out). Once the hyper-parameters are estimated, SBL computes the maximum a posteriori (MAP) estimate of $x$. This has a simple closed form expression under the assumed joint Gaussian model.


\subsection{Restricted Isometry Property and Null Space Property} \label{rip_nsp}
In this section we describe some of the properties introduced in recent work that are either sufficient or necessary and sufficient to ensure exact sparse recovery. 
\bd
The {\em restricted isometry constant (RIC)}, $\delta_s(A)$, for a matrix $A$, is the smallest real number satisfying
\bea
(1- \delta_s) \|b\|_2^2 \le \|A b\|_2^2 \le (1 + \delta_s) \|b\|_2^2
\label{def_delta}
\eea
for all $s$-sparse vectors $b$  \cite{decodinglp}.
A matrix $A$ satisfies the {\em RIP of order $s$} if $\delta_s(A) < 1$.
\ed
It is easy to see that $(1 - \delta_s) \le \|{A_\T}'A_\T\| \le (1 + \delta_s)$, $\|({A_\T}'A_\T)^{-1}\| \le 1/(1- \delta_s)$ and $\|{A_\T}^\dag\| \le 1/\sqrt{(1- \delta_s)}$ for sets $\T$ with $|\T|\le s$.
\bd
The {\em restricted orthogonality constant (ROC)}, $\theta_{s,\tilde{s}}$, for a matrix $A$, is the smallest real number satisfying
\bea
| {b_1}'{A_{\T_1}}'A_{\T_2} b_2 | \le \theta_{s,\tilde{s}} \ \|b_1\|_2 \ \|b_2\|_2 
\label{def_theta}
\eea
for all disjoint sets $\T_1, T_2 \subseteq [1,m]$ with $|\T_1| \le s$, $|\T_2| \le \tilde{s}$, $s+\tilde{s} \le m$, and for all vectors $b_1$, $b_2$ of length $|\T_1|$, $|\T_2|$ \cite{decodinglp}.
\ed
It is not hard to show that $\|{A_{\T_1}}'A_{\T_2}\| \le \theta_{s,\tilde{s}}$ \cite{just_lscs} and that $\theta_{s,\tilde{s}} \le \delta_{s+\tilde{s}}$ \cite{decodinglp}.

The following result was proved in \cite{candes_rip}.
\begin{theorem}[Exact recovery and error bound for BP and BP-noisy]
Denote the solution of BP, (\ref{bp}), by $\xhat$. In the noise-free case, i.e., when $y:=Ax$, if $\delta_s(A) < \sqrt{2}-1$, then $\xhat = x$ (BP achieves exact recovery).
In the noisy case, i.e., when $y:=Ax+w$ with $\|w\|_2 \le \eps$, if $\delta_{2s}(A) < 0.207$, then
\[
\|x - \xhat\|_2 \le C_1(2s) \eps \le 7.50 \eps \text{ where }  C_1(k) := \frac{4 \sqrt{1+\delta_k}}{1 - 2 \delta_k}
\]
where $\xhat$ is the the solution of BP-noisy, (\ref{bpnoisy}).
\label{BPthm}
\end{theorem}

With high probability (whp), random Gaussian matrices and various other random matrix ensembles satisfy the RIP of order $s$ whenever the number of measurements $n$ is of the order of $s \log m$ and $m$ is large enough \cite{decodinglp}. 

The null space property (NSP) is another property used to prove results for exact sparse recovery \cite{nsp,nsp_cohen}. NSP ensures that every vector $v$ in the null space of $A$ is not too sparse. NSP is known to be a necessary and sufficient condition for exact recovery of $s$-sparse vectors \cite{nsp,nsp_cohen}.
\bd
A matrix $A$ satisfies the {\em null space property (NSP)} of order $s$ if, for any vector $v$ in the null space of $A$,
$$\|v_S\|_1 < 0.5 \|v\|_1, \ \text{for all sets $S$ with $|S| \le s$}$$
\ed

\section{The Problem, Applications and Motivation}  \label{probdef}

\subsection{Problem Definition} \label{probform}


Let $t$ denote the discrete time index. We would like to recover a sparse $n$-length vector sequence $\{x_t\}$ from undersampled and possibly noisy measurements $\{y_t\}$ satisfying
\bea
y_t := A_t x_t + w_t, \   \|w_t\|_2 \le \eps,
\label{obsmod}
\eea
where $A_t:= H_t \Phi$ is an $n_t \times m$ matrix with $n_t < m$ and $w_t$ is bounded noise. Here $H_t$ is the measurement matrix and $\Phi$ is an orthonormal matrix for the sparsity basis. Alternatively, it can be a dictionary matrix. In the above formulation, $z_t:= \Phi x_t$ is actually the signal (or image arranged as a 1D vector) whereas $x_t$ is its representation in the sparsity basis or dictionary $\Phi$. For example, in MRI of wavelet sparse images, $H_t$ is a partial Fourier matrix and $\Phi$ is the matrix corresponding to the inverse 2D discrete wavelet transform (DWT).

We use $\N_t$ to denote the support set of $x_t$, i.e., $$\N_t:= \text{supp}(x_t) =  \{i: (x_t)_i \neq 0\}.$$
When we say $x_t$ is sparse, it means that $|\N_t| \ll m$.
Let $\xhat_t$ denote an estimate of $x_t$. The goal is to recursively reconstruct $x_t$ from $y_0,y_1, \dots y_t$, i.e., use only $\xhat_0, \xhat_1, \dots, \xhat_{t-1}$ and $y_t$ for reconstructing $x_t$. {\cred It is assumed that the first signal, $x_0$, can be accurately recovered from $y_0$. The simplest way to ensure this is by using more measurements at $t=0$ and using a simple-CS solution to recover $x_0$. Other possible approaches, such as using other prior knowledge or using a batch CS approach for an initial short sequence, are described in Section \ref{initialize}.
}

In order to solve the above problem, one can leverage the practically valid assumption of slow support (sparsity pattern) change \cite{kfcsicip,just_lscs,isitmodcs}:
\bea
|\N_t \setminus \N_{t-1}| \approx |\N_{t-1} \setminus \N_t| \ll |\N_t|. \label{ssc}
\eea
Notice from Fig. \ref{suppchange} that this is valid for dynamic MRI sequences.
A second assumption that can also be exploited is that of slow signal value change:
\bea
\|(x_t - x_{t-1})_{\N_{t-1} \cup \N_t}\|_2  \ll \|(x_t)_{\N_{t-1} \cup \N_t}\|_2. \label{ssvc}
\label{ssvc}
\eea
This, of course, is a commonly used assumption in almost all past work on tracking algorithms as well as in work on adaptive filtering algorithms. Notice that we can also write the above assumption as $\|x_t - x_{t-1}\|_2  \ll \|x_t\|_2$. This is true too since for $i \notin {\N_{t-1} \cup \N_t}$,  $(x_t- x_{t-1})_i=0$.

Henceforth we refer to the above problem of trying to {\em design recursive algorithms for the dynamic CS problem while using slow support and/or slow signal value change} as the  {\em recursive dynamic CS} problem. Also, as noted in Sec. \ref{intro}, {\em ``simple-CS solutions"} refers to dynamic CS solutions that recover each sparse signal in the sequence independently without  using any information from past or future frames.

\subsection{Applications} \label{apps}
An important application where the above problem occurs is undersampled dynamic MRI for applications such as interventional radiology \cite{interventionalMR,kfcsmri}, MRI-guided surgery and functional MRI based tracking of brain activations in response to changing stimuli \cite{blindestlowcnr,fmrimodcs}.
Since MR data is acquired one Fourier projection at a time, {the ability to accurately reconstruct using fewer measurements directly translates into reduced scan times.
Shorter scan times along with online reconstruction can potentially enable real-time\footnote{None of the solutions that we describe in this article are currently able to run in ``real-time". The fastest method still needs about 5 seconds per frame of processing when implemented using MATLAB code on a standard desktop (see Table \ref{t_larynx}).} imaging of fast changing physiological phenomena, thus making many interventional MRI applications such as MRI-guided surgery feasible in the future \cite{interventionalMR}}. As explained in \cite{interventionalMR}, these are currently not feasible due to the extremely slow image acquisition speed of MR systems. 
To understand the MRI application, assume that all images are rearranged as 1D vectors. 
The MRI measurement vector at time $t$, $y_t$, satisfies $y_t = H_t z_t + w_t$ where $z_t$ is the $m_1 \times m_2$ image at time $t$ arranged as an $m=m_1m_2$ length vector and
$$H_t =I_{\mathcal{O}_t}{}' (F_{m_1} \kron F_{m_2}).$$
Here $\mathcal{O}_t$ is the set of indices of the observed discrete frequencies at time $t$, $F_m$ is the $m$-point discrete Fourier transform (DFT) matrix and $\kron$ denotes Kronecker product.
Observe that the notation $I_{\mathcal{O}_t}{}'M$ creates a sub-matrix consisting of the rows of $M$ with indices in the set $\mathcal{O}_t$.

Cross-sectional images of the brain, heart, larynx or other human organs are usually piecewise smooth, e.g., see Fig. \ref{examples}, and thus well modeled as being sparse in the wavelet domain.
Hence, in this case, $\Phi$ is the inverse 2D discrete wavelet transform (DWT) matrix. 
If $W_m$ is the inverse DWT matrix corresponding to a chosen 1D wavelet, e.g. Daubechies-4, then
$$\Phi = W_{m_1} \kron W_{m_2}.$$
Slow support change of the wavelet coefficients vector, $x_t:= \Phi^{-1} z_t$, is verified in Fig. \ref{suppchange}.
%
For large-sized images, the matrix $H_t$ or the matrix $\Phi$ as expressed above become too big to store in memory. Thus, in practice, one does not compute the DFT or DWT using matrix-vector multiplies but directly as explained in Sec. \ref{sims_mri}. This ensures that one never needs to store $H_t$ or $\Phi$ in memory, only $\mathcal{O}_t$ needs to be stored.

Another potential application of recursive dynamic CS solutions is single-pixel camera (SPC) based video imaging. The SPC is not a very practical tool for optical imaging since much faster cameras exist, but it is expected to be useful for imaging applications where the imaging sensors are very expensive, e.g., for short-wave-infrared (SWIR) imaging \cite{chen2015fpa, herman2015recent, sankaranarayanan2015video} or for compressive depth acquisition \cite{colacco2012compressive}. In fact many companies including start-ups such as InView and large research labs such as MERL and Bell Labs have built their own SPCs \cite{huang2013lensless}. In this application, $z_t$ is again the vectorized image of interest that can often be modeled as being wavelet sparse. The measurements are random-Gaussian or Rademacher, i.e., each element of $H_t$ is either an independent and identically distributed (i.i.d.) Gaussian with zero mean and unit variance or is i.i.d. $\pm 1$. This problem formulation is also used for CS-based (or dynamic CS based) online video compression/decompression. This can be useful in applications where the compression end needs to be implemented with minimal hardware (just matrix-vector multiplies) while significantly higher computational power is available for decompression, e.g., in a sensor network based sensing setup as in Cevher et al. \cite{reddy} (this involved a camera network and a CS-based background-subtraction solution).

A completely different application is online denoising of image sequences that are sparsifiable in a known sparsity basis or dictionary. In fact, denoising was the original problem for which basis pursuit denoising (BPDN) was introduced \cite{bpdn}. Let $\Phi$ denote the given dictionary or sparsity basis. For example, one could let it be the inverse Discrete Cosine Transform (DCT) matrix. Or, as suggested in \cite{elad}, one can let $\Phi = [I \ D]$ where $I$ is an identity matrix and $D$ is the inverse DCT matrix. The DCT basis is a good sparsifying basis for textures in an image while the canonical basis ($I$) is a good sparsifying basis for edges \cite{elad}. For this problem, $H_t = I$  and so $y_t = z_t + w_t = \Phi x_t + w_t$ and $x_t$ is the sparse vector at time $t$. Since image sequences are correlated over time, slow support change and slow signal value change are valid assumptions on $x_t$. A generalization of this problem is the dictionary learning or the sparsifying transform learning problem where the dictionary or sparsifying transform is also unknown \cite{ksvd,sparse_tfm_learn}. This is briefly discussed in Sec. \ref{future}.

Another application of recursive dynamic CS is speech/audio reconstruction from time-undersampled measurements. This was studied by Friedlander et al. \cite{friedlander}. In a traditional Analog-to-Digital converter, speech is uniformly sampled at 44.1kHz. With random undersampling and dynamic CS based reconstruction, one can sample speech at an average rate of $b*44.1$kHz where $b$ is the undersampling factor. In \cite{friedlander}, $b=0.25$ was used. Speech is broken down into segments of time duration $m \tau$ seconds where $\tau = 1/44100$ seconds. Then $z_t$ corresponds to the time samples of the $t$-th speech segment uniformly sampled using $\tau$ as the sampling interval. The measurement vector for the $t$-th segment satisfies $y_t:= I_{\mathcal{O}_t}{}'z_t$ where $\mathcal{O}_t$ contains $\lceil b*m \rceil$ uniformly randomly selected indices out of the set $\{1,2, \dots,m\}$. The DCT is known to form a good sparsifying basis for speech/audio \cite{friedlander}. Thus, for this application, $\Phi$ corresponds to the inverse DCT matrix. As explained in \cite{friedlander}, the support set corresponding to the largest DCT coefficients in adjacent blocks of speech does not change much from one block to the next and thus slow support change holds. 
Of course, once the $z_t$'s are recovered, they can be converted to speech using the usual analog interpolation filter.

Besides the above, the recursive dynamic CS problem has also been explored for various other applications such as dynamic optical coherence tomography (OCT) \cite{mocs_oct}; dynamic spectrum sensing in cognitive radios, e.g., \cite{liu2011collaborative}; and sparse channel estimation and data detection in orthogonal frequency division multiplexing (OFDM) systems \cite{oliver2008sparse}. 

Lastly, consider a solution approach to recursive robust principal components' analysis (RPCA).
In \cite{rpca}, RPCA was defined as a problem of separating a low-rank matrix $L$ and a sparse matrix $X$ from the data matrix $Y:=X+L$  \cite{rpca}\footnote{Here $L$ corresponds to the noise-free data matrix that is, by definition, low rank (its left singular vectors form the desired principal components); and $X$ models the outliers (since outliers occur infrequently they are well modeled as forming a sparse matrix).}. 
Recursive RPCA is then the problem of recovering a time sequence of sparse vectors, $x_t$, and vectors lying in a low-dimensional subspace, $\ell_t$, from $y_t := x_t + \ell_t$ in a recursive fashion, starting with an accurate knowledge of the subspace from which $\ell_0$ was generated. A solution approach for recursive RPCA, called Recursive Projected CS (ReProCS), was introduced in recent work \cite{rrpcp_allerton,rrpcp_perf,rrpcp_tsp,rrpcp_aistats}. As we explain, one of two key steps of ReProCS involves solving a recursive dynamic CS problem.
ReProCS assumes that the subspace from which $\ell_t$ is generated either remains fixed or changes slowly over time, and any set of basis vectors for this subspace are dense (non-sparse) vectors. At time $t$, suppose that an accurate estimate of the subspace from which $\ell_{t-1}$ is generated is available. Let $\hat{P}_{t-1}$ be a matrix containing its basis vectors. ReProCS then projects $y_t$ orthogonal to the range of $\hat{P}_{t-1}$ to get $\tilde{y}_t:=(I-\hat{P}_{t-1} \hat{P}_{t-1}{}') y_t$. Because of the slow subspace change assumption, doing this approximately nullifies $\ell_t$ and gives projected measurements of $x_t$. Notice that $\tilde{y}_t = A_t x_t + w_t$ where $A_t:=(I-\hat{P}_{t-1} \hat{P}_{t-1}{}')$ and $w_t:=A_t \ell_t$ is small ``noise" due to $\ell_t$ not being fully nullified.
The problem of recovering $x_t$ from $\tilde{y}_t$ is clearly a CS problem in small noise. If one considers the sequence of $x_t$'s, and if slow support or signal value change holds for them, then this becomes a recursive dynamic CS problem. One example situation where these assumptions hold is when $x_t$ is the foreground image sequence in a video analytics application\footnote{The background image sequence of a typical static camera video is well modeled as lying in a low-dimensional subspace (forms $\ell_t$) and the foreground image sequence is well-modeled as being sparse since it often consists of one or more moving objects (forms $x_t$) \cite{rpca}.}.
The denseness assumption on the subspace basis vectors ensures that RIP holds for the matrix $A_t$ \cite{rrpcp_perf}. This implies that $x_t$ can be accurately recovered from $\tilde{y}_t$. One can then recover $\hat{\ell}_t=y_t - \xhat_t$ and use this to update its subspace estimate, $\hat{P}_t$. A more recent recursive RPCA algorithm, GRASTA \cite{grass_undersampled}, also uses the above idea although both the projected CS and subspace update steps are solved differently.%

In all the applications described above except the last one, the signal of interest is only compressible (approximately sparse). Whenever we say ``slow support change", we are referring to the changes in the $b$\%-energy-support (the largest set containing at most $b$\% of the total signal energy). In the last application, the outlier sequence, e.g., the foreground image sequence for the video analytics application, is an exactly sparse image sequence. 


\subsection{Motivation: why are new techniques needed?} \label{motiv}
One question that comes to mind is why are new techniques needed to solve the recursive dynamic CS problem and why can we not use ideas from the adaptive filtering or the tracking literature applied to simple-CS solutions?
The reason is as follows. Adaptive filtering and tracking solutions rely on slow signal value change which is the only thing one can use for dense signal sequences. This can definitely be done for sparse, and approximately sparse, signal sequences as well, and does often result in good experimental results. We explain these ideas in Sec. \ref{track_adapt_filt}. However sparse signal sequences have more structure, e.g., slow support change, that can be exploited to (i) get better algorithms and (ii) prove stronger theoretical results. For example, neither tracking-based solutions nor adaptive-filtering-based solutions allow for exact recovery using fewer measurements than what simple-CS solutions need. For a detailed example refer to Sec. \ref{csres_sec}. Similarly, one cannot obtain stability over time results for these techniques under weaker assumptions on the measurement matrix than what simple-CS solutions need.

\section{Exploiting slow support change: Sparse recovery with partial support knowledge} \label{supp_knowledge}

%
When defining the recursive dynamic CS problem, we introduced two assumptions that are often valid in practice - slow support change and slow nonzero signal value change. In this section, we describe solutions that only exploit the slow support change assumption, i.e., (\ref{ssc}).
If the initial signal, $x_0$, can be recovered accurately, then, under this assumption, recursive dynamic CS can be reformulated as a problem of {\em sparse recovery using partial support knowledge} \cite{isitmodcs0, isitmodcs}. We can use the support estimate of $\xhat_{t-1}$, denoted $\Nhat_{t-1}$, as the ``partial support knowledge".
We give the reformulated problem below followed by the proposed solutions for it. Their dynamic counterparts are summarized later in Sec. \ref{recrecon}. 

If the support does not change slowly enough, but the change is still highly correlated and the correlation model is known, one can get an accurate support prediction by using the correlation model information applied to the previous support estimate. An algorithm based on this idea is described in \cite{rrpcp_isit}.

\subsection{Reformulated problem: Sparse recovery using partial support knowledge}\label{PKS}
The goal is to recover a sparse vector, $x$, with support $\N$, from noise-free measurements, $y:= Ax$, or from noisy measurements,  $y:= Ax + w$, when partial and possibly erroneous support knowledge, $\T$, is available \cite{isitmodcs0, isitmodcs}.

The true support $\N$ can be rewritten as
$$\N = \T \cup \Delta_u \setminus \Delta_e \ \ \text{where} \ \ \Delta_u: = \N \setminus \T, \ \Delta_e: = \T \setminus \N $$
Here $\Delta_u$ denotes the set of unknown (or missing) support entries while $\Delta_e$ denotes the set of extra entries in $\T$.
Let
$$s:=|\N|, k:=|\T|, u:=|\Delta_u|, e:=|\Delta_e|$$
It is easy to see that
$$s = k+u-e$$
We say the {\em support knowledge is accurate if} $u \ll s$ and $e \ll s$.

This problem is also of independent interest, since in many static sparse recovery applications, partial support knowledge is often available. For example, when using wavelet sparsity for an image with very little black background (most of its pixel are nonzero), most of its wavelet scaling coefficients will also be nonzero. Thus, the set of indices of the wavelet scaling coefficients could serve as accurate partial support knowledge.

\subsection{Least Squares CS-residual (LS-CS)}
The Least Squares CS-residual (LS-CS) algorithm  \cite{kfcspap,just_lscs} can be interpreted as the first solution for the above problem. It starts by computing an initial LS estimate of $x$ by assuming that its support set is equal to $\T$:
\[
\xhat_{\text{init}} = I_\T ({A_\T}'{A_\T})^{-1}{A_\T}' y_t.   \nn
\]
Using this, it computes 
\bea
\xhat \se \xhat_{\text{init}} + [\arg \min_b \|b\|_1 \ \text{s.t.} \ \|y - A \xhat_{\text{init}} - A b \|_2 \le \eps ].
\label{lscs}
\eea
This is followed by support estimation and computing a final LS estimate as described later in Sec. \ref{support_est}. 
The signal residual $\beta: = x -  \xhat_{\text{init}}$ satisfies
\bea
\beta \se I_\T ({A_\T}'{A_\T})^{-1} {A_\T}' ( A_{\Delta_u} x_{\Delta_u} + w) + I_{\Delta_u} x_{\Delta_u} \nn
\eea
If $A$ satisfies RIP of order at least $|\T|+|\Delta_u|$, $\|{A_\T}' A_{\Delta_u}\|_2 \le \theta_{|\T|,|\Delta_u|}$ is small. If the noise $w$ is also small, clearly $\beta_\T: = I_\T{}'\beta$ will be small and hence $\beta$ will be approximately supported on $\Delta_u$. When $|\Delta_u| \ll |\N|$,  the approximate support of $\beta$ is much smaller than that of $x$. Thus, one expects LS-CS to have smaller reconstruction error than BP-noisy when fewer measurements are available. This statement is quantified for the error upper bounds in \cite{just_lscs} (see Theorem 1 and its corollary and the discussion that follows).%

However, notice that the support size of $\beta$ is $|\T| +|\Delta_u| \ge |\N|$. Since the number of measurements required for exact recovery is governed by the exact support size, LS-CS is not able to achieve exact recovery using fewer noiseless measurements than those needed by BP-noisy. 



\subsection{Modified-CS}
\label{modcs}
The search for a solution that achieves exact reconstruction using fewer measurements led to the modified-CS idea \cite{isitmodcs0,isitmodcs}. To understand the approach, suppose first that $\Delta_e$ is empty, i.e., $\N = \T \cup \Delta_u$. Then the sparse recovery problem becomes one of trying to find the vector $b$ that is sparsest outside the set $\T$ among all vectors that satisfy the data constraint. In the noise-free case, this can be written as
$$\min_{b} \|b_{\T^c}\|_0 \  \text{s.t.} \ y = A b $$
This is referred to as the {\em modified-$\ell_0$} problem \cite{isitmodcs0,isitmodcs}.
It also works if $\Delta_e$ is not empty. The following can be shown.
\begin{proposition}[{\cite{isitmodcs}}]
Modified-$\ell_0$ will exactly recover $x$ from $y:=Ax$ if every set of $(k+2u)$ columns of $A$ is linearly independent. Recall $k = |\T|$, $u = |\Delta_u|$.
\end{proposition}

Notice that $k+2u = s+u+e = |\N|+|\Delta_e|+|\Delta_u|$.
In comparison, the original $\ell_0$ program, (\ref{ell0min}), requires every set of $2s$ columns of $A$ to be linearly independent \cite{decodinglp}. This is a stronger requirement when $u \ll s$ and $e \ll s$.

Like simple $\ell_0$, the modified-$\ell_0$ program also has exponential complexity, and hence we can again replace it by the $\ell_1$ program to get
\bea
\min_{b} \|b_{\T^c}\|_1 \  \text{s.t.} \ y = A b  
\label{l1seqcs}
\eea
The above program was called {\em modified-CS} in \cite{isitmodcs0,isitmodcs} where it was introduced. As shown in the result below, this again works even when $\Delta_e$ is not empty.
The following was shown by Vaswani and Lu \cite{isitmodcs0,isitmodcs}.
\begin{theorem}[{RIP-based modified-CS  exact recovery\cite{isitmodcs}}]
Consider recovering $x$ with support $\N$ from $y:=Ax$ by solving modified-CS, i.e., (\ref{l1seqcs}).
\ben
\item $x$ is the unique minimizer of (\ref{l1seqcs}) if
\ben
\item  $\delta_{\st+\sd} < 1$ and $\delta_{2\sd} + \delta_{\st} + \theta_{\st,2\sd}^2 < 1$ and 
\label{cond1}

\item $a_{\st}(2\sd,\sd) + a_{\st}(\sd,\sd) < 1$ where $a_{\st}(i,\check{i}) := \frac{\theta_{\check{i},i} + \frac{\theta_{\check{i},\st} \ \theta_{i,\st}}{1 - \delta_{\st}}}{ 1-\delta_i - \frac{\theta_{i,\st}^2}{1 - \delta_{\st}} }$
\een
\item \label{weaker}  A weaker (but simpler) sufficient condition for exact recovery is
\bea
2\delta_{2\sd}+\delta_{3\sd}+\delta_{k}+\delta_{k+u}^2+2\delta_{k+2u}^2 < 1.
\label{modcs_exact_rip_simple}
\eea

\item \label{weakest}
An even weaker (but even simpler) sufficient condition for exact recovery is
$$\delta_{\st+2\sd} \le 0.2.$$
\een
Recall that $s:=|\N|, \ k:=|\T|, \ u:=|\Delta_u|, \ e:=|\Delta_e|$. The above conditions can also be rewritten in terms of $\sn, \sde, \sd$ by substituting $\st = \sn + \sde - \sd$ so that $k+2u = s+e+u$.
\label{modcs_exact_rip}
\end{theorem}

Compare this result with that for BP which requires  \cite{candes_rip,foucart_lai,dantzig}
$\delta_{2\sn} < \sqrt{2}-1  \text{~~or~~}  \delta_{2\sn}+\delta_{3\sn}<1.$
To compare the conditions numerically, we can use $\sd = \sde = 0.02 \sn$ which is typical for time series applications (see Fig. \ref{suppchange}). Using $\delta_{cr} \le c \delta_{2r}$ \cite[Corollary 3.4]{cosamp}, it can be show that Modified-CS allows $\delta_{2\sd} <  0.008$. On the other hand, BP requires $\delta_{2\sd} <  0.004$ which is clearly stronger. 

An idea similar to modified-CS was independently introduced by von Borries et al. \cite{camsap07}.
For noisy measurements, one can relax the data constraint in modified-CS either using the BP-noisy approach or the BPDN approach from Sec. \ref{bgnd}.

\subsection{Weighted-$\ell_1$}
The weighted-$\ell_1$ program studied in the work of Khajehnejad et al. \cite{hassibi_isit,hassibi} and the later work of Friedlander et al. \cite{friedlander} can be interpreted as a generalization of modified-CS. The idea is to partition the index set $\{1,2, \dots m\}$ into sets $\T_1, \T_2, \dots \T_q$ and to assume that the percentage of nonzero entries in each set is known. This knowledge is used to weight the $\ell_1$ norm along each of these sets differently. Performance guarantees are obtained for the two set partition case $q=2$. Using the notation from above, the two set partition can be labeled $\T, \T^c$. In this case, weighted-$\ell_1$ solves
\bea
\min_{b} \|b_{\T^c}\|_1 + \tau \|b_\T\|_1  \  \text{s.t.} \ y = A b
\label{well1}
\eea
Clearly, modified-CS is a special case of the above with $\tau=0$.
In general, the set $\T$ contains both extra entries $\Delta_e$ and missing (unknown) entries $\Delta_u$. As long as the number of extras, $|\Delta_e|$, is small, modified-CS cannot be improved much further by weighted-$\ell_1$. However if $|\Delta_e|$ is larger or if the measurements are noisy, the weighted-$\ell_1$ generalization has smaller recovery error. This has been demonstrated experimentally in \cite{friedlander,regmodbpdn}.
For noisy measurements, one can relax the data constraint in \eqref{well1} either using the BP-noisy approach or the BPDN approach from Sec. \ref{bgnd}.

Khajehnejad et al. \cite{hassibi_isit,hassibi} obtained ``weak thresholds" on the number of measurements required to ensure exact recovery with high probability. We state and discuss this result in Sec. \ref{modcs_guar_weakthresh} below. We first give here the RIP based exact recovery condition from Friedlander et al. \cite{friedlander} since this can be easily compared with the modified-CS result from Theorem \ref{modcs_exact_rip}.
\begin{theorem}[RIP-based weighted-$\ell_1$ exact recovery \cite{friedlander}]
Consider recovering $x$ with support $\N$ from $y:=Ax$ by solving weighted-$\ell_1$, i.e., (\ref{well1}).
Let $\alpha =\frac{|\T \cap \N|}{|\T|} = \frac{(\sn - \sd)}{(\sn+\sde-\sd)}$ and $\rho = \frac{|\T|}{|\N|} = \frac{(\sn+\sde-\sd)}{\sn}$. Let $\mathbb{Z}$ denote the set of integers and let $\frac{1}{s} \mathbb{Z}$ denote the set $\{\dots, \frac{-2}{s}, \frac{-1}{s}, 0, \frac{1}{s}, \frac{2}{s}, \dots \}$.
\ben
\item  Pick an $a \in \frac{1}{s} \mathbb{Z}$ that is such that $a > \max(1,(1-\alpha)\rho)$. Weighted-$\ell_1$ achieves exact recovery if
$$\delta_{as} + \frac{a}{\gamma^2} \delta_{(a+1)s} < \frac{a}{\gamma^2} - 1$$
for $\gamma = \tau + (1-\tau)\sqrt{1+\rho-2\alpha \rho}.$

\item A weaker (but simpler) sufficient condition for exact recovery is
$$\delta_{2s} \le \frac{1}{\sqrt{2}(\tau+ (1-\tau)\sqrt{1+\rho -2\alpha \rho}) + 1}.$$
\een
\label{well1_exact_rip}
\end{theorem}

\begin{corollary}
By setting $\tau=0$ in the above result, we get an exact recovery result for modified-CS.
By setting $\tau=1$, we obtain the original results for exact recovery using BP.
\label{well1_exact_rip_cor}
\end{corollary}

Let us compare Corollary \ref{well1_exact_rip_cor} with the result from Theorem \ref{modcs_exact_rip}. To reduce the number of variables, suppose that $u=e$ so that $k=|\T|=s=|\N|$ and $\rho=1$. By Corollary \ref{well1_exact_rip_cor}, modified-CS achieves exact recovery if
\bea
\delta_{2s} \le \frac{1}{1 + 2 \sqrt{\frac{u}{s}}}.
\label{well1_exact_rip_simple}
\eea
As demonstrated in \cite{friedlander}, the above condition from Corollary \ref{well1_exact_rip_cor} is weaker than the simplified condition \eqref{modcs_exact_rip_simple} from Theorem \ref{modcs_exact_rip}. We summarize their discussion here. When $k=s$, it is clear that \eqref{modcs_exact_rip_simple} will hold {\em only if}
\bea
\delta_s + 3 \delta_s^2 < 1.
\label{eq:VaswaniRIC}
\eea
This, in turn, will hold {\em only if} $\delta_s < 0.4343$ \cite{friedlander}. Using the upper bounds for the RIC of a random Gaussian matrix from \cite{bah_tanner}, one can bound $\delta_{2s}$ and find the set of $u$'s for which (\ref{well1_exact_rip_simple}) will hold. The upper bounds on $u/s$ for various values of $n/m$ and $s/n$ for which (\ref{well1_exact_rip_simple}) holds are displayed in Table \ref{tab:RIPrecoveryGuarantee} (this is Table 1 of \cite{friedlander}). For these values of $n/m$ and $s/n$, $\delta_s = 0.4343$ and hence (\ref{modcs_exact_rip_simple}) of Theorem \ref{modcs_exact_rip} does not hold. 

\begin{table}[t]
	\centering
		\begin{tabular}{|c|c|c|c|c|}
		\hline
		{$n/m$} & {$s/n$} & ${\delta_s}$ & ${\delta_{2s}}$ & {$u/s$} \\ \hline
		0.1 & 0.0029 & 0.4343 & 0.6153 & 0.0978 \\ \hline
		0.2 & 0.0031 & 0.4343 & 0.6139 & 0.0989 \\ \hline	
		0.3 & 0.003218 & 0.4343 & 0.61176 & 0.1007 \\ \hline
		0.4 & 0.003315 & 0.4343 & 0.61077 & 0.1015 \\ \hline
		0.5 & 0.003394 & 0.4343 & 0.60989 & 0.1023 \\ \hline
		\end{tabular}
	\caption{This is Table I of \cite{friedlander} translated into the notation used in this work. It shows five sets of values of $(n/m), (s/n)$ for which (\ref{modcs_exact_rip_simple}) does not hold, but (\ref{well1_exact_rip_simple}) holds as long as $u/s$ is bounded by the value given in the last column.
}
	\label{tab:RIPrecoveryGuarantee}
\end{table}


\subsubsection{Weak thresholds for high probability exact recovery for weighted-$\ell_1$ and Modified-CS} \label{modcs_guar_weakthresh}
In very interesting work, Khajehnejad et al. \cite{hassibi} obtained ``weak thresholds" on the minimum number of measurements, $n$, (as a fraction of $m$) that are sufficient for exact recovery with overwhelming probability: the probability of not getting exact recovery decays to zero as the signal length $m$ increases. The weak threshold was first defined by Donoho in \cite{donoho_large} for BP. 
%
\begin{theorem}[{weighted-$\ell_1$ weak threshold \cite[Theorem 4.3]{hassibi}}]
Consider recovering $x$ with support $\N$ from $y:=Ax$ by solving weighted-$\ell_1$, i.e., (\ref{well1}).
Let $\omega := 1/\tau$, $\gamma_1 := \frac{|\T|}{m}$ and $\gamma_2 := \frac{|\T^c|}{m} = 1 -\gamma_1$. Also let $p_1, p_2$ be the sparsity fractions on the sets $\T$ and $\T^c$, i.e., let $p_1 := \frac{|\T|-|\Delta_e|}{|\T|}$ and $p_2:=\frac{|\Delta_u|}{|\T^c|}$. Then there exists a critical threshold
$$\delta_c = \delta_c(\gamma_1, \gamma_2, p_1, p_2, \omega)$$
such that for all $\frac{n}{m} > \delta_c$, the probability that a sparse vector $x$ is not recovered decays to zero exponentially with $m$. {\cred In the above, $\delta_c(\gamma_1, \gamma_2, p_1, p_2, \omega) = \min\{\delta~|~\psi_{com}(\tau_1,\tau_2)-\psi_{int}(\tau_1,\tau_2)-\psi_{ext}(\tau_1,\tau_2)<0~ \forall ~0\leq \tau_1\leq \gamma_1(1-p_1), 0\leq \tau_2\leq \gamma_2(1-p_2), \tau_1+\tau_2 > \delta-\gamma_1p_1-\gamma_2p_2 \}$ where $\psi_{com}$, $\psi_{int}$ and $\psi_{ext}$
are obtained from the following expressions:

\noindent Define $g(x)=\frac{2}{\sqrt{\pi}}e^{-{x^2}}$,
$G(x)=\frac{2}{\sqrt{\pi}}\int_{0}^{x}e^{-y^2}dy$ and let
$\varphi(.)$ and $\Phi(.)$ be the standard Gaussian pdf and cdf
functions respectively.
\begin{enumerate}
\item(Combinatorial exponent)
\bea \psi_{com}(\tau_1,\tau_2) &=&
\left( \gamma_1(1-p_1)H(\frac{\tau_1}{\gamma_1(1-p_1)})+ \right. \nn \\
&& \left. \gamma_2(1-p_2)H(\frac{\tau_2}{\gamma_2(1-p_2)})+\tau_1+\tau_2\right)\log{2} \nn
\eea \noindent where $H(\cdot)$ is the entropy function defined by
$H(x) = -x\log{x}-(1-x)\log(1-x)$.

\item(External angle exponent)
Define $c=(\tau_1+\gamma_1p_1)+\omega ^2(\tau_2+\gamma_2p_2)$,
$\alpha_1=\gamma_1(1-p_1)-\tau_1$ and
$\alpha_2=\gamma_2(1-p_2)-\tau_2$. Let $x_0$ be the unique solution
to $x$ of the following:
\begin{equation*}
2c-\frac{g(x)\alpha_1}{xG(x)}-\frac{\omega g(\omega
x)\alpha_2}{xG(\omega x)}=0 \nn
\end{equation*}
Then
\begin{equation}
\psi_{ext}(\tau_1,\tau_2) =
cx_0^2-\alpha_1\log{G(x_0)}-\alpha_2\log{G(\omega x_0)} \nn
\end{equation}

\item(Internal angle exponent)
Let $b=\frac{\tau_1+\omega ^2\tau_2}{\tau_1+\tau_2}$,
$\Omega'=\gamma_1p_1+\omega ^2\gamma_2p_2$ and
$Q(s)=\frac{\tau_1\varphi(s)}{(\tau_1+\tau_2)\Phi(s)}+\frac{\omega
\tau_2\varphi(\omega s)}{(\tau_1+\tau_2)\Phi(\omega s)}$. Define the
function $\hat{M}(s)=-\frac{s}{Q(s)}$ and solve for $s$ in
$\hat{M}(s)=\frac{\tau_1+\tau_2}{(\tau_1+\tau_2)b+\Omega'}$. Let the
unique solution be $s^*$ and set $y=s^*(b-\frac{1}{\hat{M}(s^*)})$.
Compute the rate function $\Lambda^*(y)= sy
-\frac{\tau_1}{\tau_1+\tau_2}\Lambda_1(s)-\frac{\tau_2}{\tau_1+\tau_2}\Lambda_1(\omega
s)$ at the point $s=s^*$, where $\Lambda_1(s) = \frac{s^2}{2}
+\log(2\Phi(s))$. The internal angle exponent is then given by:
\begin{equation}
\psi_{int}(\tau_1,\tau_2) =
(\Lambda^*(y)+\frac{\tau_1+\tau_2}{2\Omega'}y^2+\log2)(\tau_1+\tau_2) \nn
\end{equation}
\end{enumerate}
}
%
\label{well1_weakthresh}
\end{theorem}

As explained in \cite{hassibi}, the above result can be used to design an intelligent heuristic for picking $\tau$ as follows. Create a discrete set of possible values of $\tau$. Use the above result to compute the weak threshold $\delta_c$ for each value of $\tau$ from this set and pick the $\tau$ that needs the smallest weak threshold. The weak threshold for the best $\tau$ also specifies the required number of measurements, $n$.



\begin{corollary}[modified-CS and BP weak threshold \cite{hassibi}]
Consider recovering $x$ from $y:=Ax$.  
Assume all notation from Theorem \ref{well1_weakthresh}.

The weak threshold for BP is given by $\delta_c(0, 1, 0, \frac{s}{m}, 1)$.

The weak threshold for modified-CS is given by $\delta_c(\gamma_1, \gamma_2, p_1, p_2, \infty)$. In the special case when $\T \subseteq \N$, $p_1=1$. In this case, the weak threshold satisfies
$$\delta_c(\gamma_1, \gamma_2, 1, p_2, \infty) = \gamma_1 + \gamma_2 \delta_c(0,1,0, p_2, 1).$$
\end{corollary}

As explained in \cite{hassibi}, when $\T \subseteq \N$, the above has a nice intuitive interpretation. In this case, the number of measurements $n$ needed by modified-CS is equal to $|\T|$ plus the number of measurements needed for recovering the remaining $|\Delta_u|$ entries from $\T^c$ using BP.

\subsection{Modified greedy algorithms and IHT-PKS}
The modified-CS  idea can also be used to modify other approaches for sparse recovery. This has been done in recent work by Stankovic et al. and Carillo et al. \cite{modomp_stankovic,modcosamp_udel} with encouraging results. They have developed and evaluated OMP with partially known support (OMP-PKS) \cite{modomp_stankovic}, Compressive Sampling Matching Pursuit (CoSaMP)-PKS and IHT-PKS \cite{modcosamp_udel}.
The greedy algorithms (OMP and CoSaMP) are modified as follows. Instead of starting with an initial empty support set, one starts with $\T$ as being the initial support set. 

IHT is modified as follows. Let $k=|\T|$ and $s=|\N|$. IHT-PKS iterates as \cite{modcosamp_udel}:
\beq
\xhat^0 = 0, \ \xhat^{i+1} = (\xhat^i)_\T + H_{s-k}( ( \xhat^i + A'(y - A \xhat^i) )_{\T^c} ).
\label{iht_pks}
\eeq
The authors also bound its error at each iteration for the special case when $\T \subseteq \N$. The bound shows geometric convergence as $i \tends \infty$ to the true solution in the noise-free case and to within noise level of the true solution in the noisy case. We summarize this result in Section \ref{noisy_pks} along with the other noisy case results. 

\subsection{An interesting interpretation of modified-CS}
The following has been shown by Bandeira et al. \cite{partial}. Assume that $A_\T$ is full rank (this is a necessary condition in any case). Let $\mathcal{P}_{\T,\perp}$ denote a projection onto the space perpendicular to $A_\T$, i.e., let
$$
\mathcal{P}_{\T,\perp}:=(I - A_\T ({A_\T}'A_\T)^{-1} A_\T').
$$
Let $\tilde{y}:= \mathcal{P}_{\T,\perp} y$, $\tilde{A}:= \mathcal{P}_{\T,\perp} A_{\T^c}$ and $\tilde{x}:=x_{\T^c}$. It is easy to see that $\tilde{y}= \tilde{A} \tilde{x}$ and $\tilde{x}$ is $|\Delta_u|$ sparse.
Thus modified-CS  can be interpreted as finding a $|\Delta_u|$-sparse vector $\tilde{x}$ of length $m-|\T|$ from $\tilde{y}:= \tilde{A} \tilde{x}$. One can then recover $x_{\T}$ as the (unique) solution of $A_\T x_{\T} = y - A_{\T^c} x_{\T^c}$.
More precisely let $\xhat_{modCS}$ denote the solution of modified-CS, i.e., (\ref{l1seqcs}). Then,
\bea
(\xhat_{modCS})_{\T^c} & \in & \arg\min_b \|b\|_1 \ s.t. \ (\mathcal{P}_{\T,\perp}y) = (\mathcal{P}_{\T,\perp} A_{\T^c})b, \nn \\
(\xhat_{modCS})_{\T} \se (A_\T)^\dag(y - A_{\T^c} (\xhat_{modCS})_{\T^c}). \nn
\eea
This interpretation can then be used to define a partial RIC.
\bd
We refer to $\delta_{u}^{k}$ as the {\em partial RIC} for a matrix $A$ if, for any set $\T$ of size $|\T|= k$, $A_\T$ is full column rank and $\delta_u^{k}$ is the smallest real number so that
\[
(1-\delta_{u}^{k})\|b\|_2^2 \le \|(\mathcal{P}_{\T,\perp} A_{\T^c}) b \|_2^2 \le (1+\delta_{u}^{k})\|b\|_2^2
\]
for all $u$-sparse vectors $b$ of length $(m-k)$ and all sets $\T$ of size $k$.
\ed

With this, any of the results for BP or BP-noisy can be directly applied to get a result for modified-CS. For example, using Theorem \ref{BPthm} we can conclude the following
\begin{theorem}
If $\delta_{2u}^k < \sqrt{2}-1$, then modified-CS achieves exact recovery.
\end{theorem}


\subsection{Related ideas: Truncated BP and Reweighted $\ell_1$}
\subsubsection{Truncated BP}
The modified-CS  program has been used in the work of Wang and Yin \cite{trunc_bp} for developing a new algorithm for simple sparse recovery. They call it {\em truncated basis pursuit (BP)} and use it iteratively to improve the recovery error for regular sparse recovery. In the zeroth iteration, they solve the BP program and compute the estimated signal's support by thresholding. This is then used to solve modified-CS  in the second iteration and the process is repeated with a specific support threshold setting scheme. 

\subsubsection{Reweighted $\ell_1$}
The reweighted $\ell_1$ approach developed by Candes et al. in much earlier work \cite{reweighted_cs} is similarly related to weighted $\ell_1$ described above. In this case again, in the zeroth iteration, one solves BP. Denote the solution by $\xhat^0$. In the next iteration one uses the entries of $\xhat^0$ to weight the $\ell_1$ norm in an intelligent fashion and solves the weighted $\ell_1$ program. This procedure is repeated until a stopping criterion is met.

\subsection{Error bounds for the noisy case} \label{noisy_pks}
It is easy to bound the error of modified-CS-noisy or weighted-$\ell_1$-noisy by modifying the corresponding RIP-based results for BP-noisy. 
Weighted-$\ell_1$-noisy solves
\bea
\min_{b} \|b_{\T^c}\|_1 + \tau \|b_\T\|_1  \  \text{s.t.} \ \|y - A b\|_2 \le \eps
\label{w_ell1_noisy}
\eea
and modified-CS-noisy solves the above with $\tau=0$. 

Let $x$ be a sparse vector with support $\N$ and let $y:=Ax+w$ with $\|w\|_2 \le \eps$.

\begin{theorem}[{modified-CS  error bound \cite[Lemma 2.7]{stab_jinchun_jp}}]
Let $\xhat$ be the solution of modified-CS-noisy, i.e., (\ref{w_ell1_noisy}) with $\tau=0$.
If $\delta_{|\T|+3|\Delta_u|} < (\sqrt{2}-1)/2$, 
then
$$\|x - \xhat\| \le C_1(|\T|+3|\Delta_u|) \eps \le 7.50 \eps, \  C_1(k) \defn \frac{4 \sqrt{1+\delta_k}}{1 - 2 \delta_k}.$$
\label{modcs_bnd}
\end{theorem}

Notice that $\delta_{|\T|+3|\Delta_u|}=\delta_{|\N|+ |\Delta_{e}|+ 2|\Delta_u|}$.
A similar result for a compressible (approximately sparse) signal and weighted-$\ell_1$-noisy was proved in \cite{friedlander}.

\begin{theorem}[{weighted-$\ell_1$ error bound \cite{friedlander}}]
Let $\xhat$ be the solution of (\ref{w_ell1_noisy}). Let $x_s$ denote the best $s$ term approximation for $x$ and let $\N = \text{support}(x_s)$. Let $\alpha$, $\rho$, and $a$ be as defined in Theorem \ref{well1_exact_rip}. Assume the exact recovery condition given in Theorem \ref{well1_exact_rip} holds for $\N$ as defined here. Then,
$$\|\xhat - x\|_2 \le C_0' \eps + C_1' s^{-1/2}( \tau \|x-x_s\|_1 + (1-\tau)\|x_{(N \cup T)^c}\|_1 ) $$
where $C_0' = C_0'(\tau, s, a, \delta_{(a+1)s}, \delta_{as})$ and $C_1' = C_1'(\tau, s, a, \delta_{(a+1)s}, \delta_{as})$ are constants specified in Remark 3.2 of \cite{friedlander}.
\end{theorem}

The following was proved for IHT-PKS. It only analyzes the special case $\T \subseteq \N$ (no extras in $\T$). 
\begin{theorem}
Recall the IHT-PKS algorithm from (\ref{iht_pks}). If $e:=|\Delta_e|=0$, i.e., if $\T \subseteq \N$, if
$\|A\|_2 < 1$ and if $\delta_{3s-2k} < 1/\sqrt{32}$, then the $i$-th IHT-PKS iterate satisfies
\beq
\|x-\xhat^{i}\|_2 \leq (\sqrt{8}\delta_{3s-2k})^i \|x\|_2 + C\epsilon,
\eeq
where $C = \sqrt{1+\delta_{2s-k}}\left(\frac{1-\alpha^t}{1-\alpha}\right)$. Recall that $s=|\N|$ and $k=|\T|$. With $e=0$, $3s-2k = s+3u$ and $2s-k = s+2u$.
\end{theorem}


\section{Exploiting slow support and slow signal value change: Sparse recovery with partial support and signal value knowledge} \label{supp_sig_knowledge}

In the previous section, we discussed approaches that only exploit slow support change. In many recursive dynamic CS applications, slow signal value change also holds. We discuss here how to design improved solutions that use this knowledge also. We begin by first stating the reformulated static problem.

\subsection{Reformulated problem: Sparse recovery with partial support and signal value knowledge} \label{probdef_ssvc}
The goal is to recover a sparse vector $x$, with support set $\N$, either from noise-free undersampled measurements, $y:= Ax$, or from noisy measurements, $y:= Ax + w$, when a signal value estimate, $\hat\mu$, is available. We let the partial support knowledge $\T$ be equal to the support of $\hat\mu$. %

The true support $\mathcal{N}$ can be written as
$$\N = \T \cup {\Delta_u} \setminus \Delta_e \ \ \text{where} \ \ \Delta_u:= \N \setminus \T, \ \Delta_e:= \T \setminus \N $$
and the true signal $x$ can be written as
\bea
(x)_{\N \cup \T} = (\hat\mu)_{\N \cup \T} + \nu, \  (x)_{\N^c} = 0.
\label{x_ssvc}
\eea
By definition, $(\hat\mu)_{\T^c} = 0$. The error $\nu$ in the prior signal estimate is assumed to be small, i.e., $\|\nu\| \ll \|x\|$.


\subsection{Regularized modified-BPDN} \label{regmodcs_sec}
Regularized modified-BPDN adds the slow signal value change constraint to modified-BPDN. In general, this can be imposed as either a bound on the max norm or on the 2-norm \cite{regmodbpdn_exact} or it can be included into the cost function as a soft constraint. {\em Regularized modified-BPDN (reg-mod-BPDN)} does the latter \cite{regmodbpdn}. It solves:
\bea
\min_{b} \gamma\|b_{\T^c}\|_1 + 0.5\|{y} - Ab\|_2^2 + 0.5 \lambda\|b_\T - \hat\mu_\T\|_2^2.
\label{regmodbpdn_eq}
\eea
Setting $\lambda=0$ in \eqref{regmodbpdn_eq} gives modified-BPDN (mod-BPDN).

\begin{remark}
Notice that the third term in \eqref{regmodbpdn_eq} acts as a regularizer on $b_\T$. It ensures that \eqref{regmodbpdn_eq} has a unique solution at least when $b$ is constrained to be supported on $\T$. In the absence of this term, one would need $A_\T$ to be full rank to ensure this. The first term is a regularizer on $b_{\T^c}$. As long as $\gamma$ is large enough, i.e., $\gamma \ge \gamma^*$, and an incoherence condition holds on the columns of the matrix $A$, one can show that \eqref{regmodbpdn_eq} will have a unique minimizer even without a constraint on $b$. We state a result of this type next. It gives a computable bound that holds always (without any sufficient conditions)\cite{regmodbpdn}. 
\end{remark}





\subsubsection{Computable error bounds that hold always}\label{erc_noisy_bnd}
In \cite{tropp}, Tropp introduced the Exact Recovery Coefficient (ERC) to quantify the incoherence between columns of the matrix $A$ and used this to obtain a computable error bound for BPDN. By modifying this approach, one can get a similar result for reg-mod-BPDN and hence also for mod-BPDN (which is a special case) \cite{regmodbpdn}. With some extra work, one can obtain a computable bound that holds always. The main idea is to maximize over all choices of the sparsified approximation of $x$ that satisfy the required sufficient conditions \cite{regmodbpdn}.
Since the bound is computable and holds without sufficient conditions and, from simulations, is also fairly tight (see \cite[Fig. 4]{regmodbpdn}), it provides a good heuristic for setting $\gamma$ for mod-BPDN and $\gamma$ and $\lambda$ for reg-mod-BPDN. We explain this in Sec. \ref{parameter_set}.

\begin{theorem}[reg-mod-BPDN computable bound that holds always]
Let $x$ be a sparse vector with support $\N$ and let $y:=Ax+w$ with $\|w\|_2 \le \eps$. Assume that the columns of $A$ have unit 2-norm. When they do not have unit 2-norm, we can use Remark \ref{A_norm}.
If $\gamma = \gamma_{\T,\lambda}^*(\tilde\Delta_u^{*}(k_{\min}))$, then reg-mod-BPDN (\ref{regmodbpdn_eq}) has a unique minimizer, $\xhat$, that is supported on $\T \cup \tilde\Delta_u^{*}(k_{\min})$, and that satisfies
$$\|x-\xhat\|_2 \le g_\lambda(\tilde\Delta_u^{*}(k_{\min})).$$
The quantities $k_{\min}$, $\tilde\Delta_u^{*}(k_{\min})$, $\gamma_{\T,\lambda}^*(\tilde\Delta_u^{*}(k_{\min}))$, $ g(\tilde\Delta_u^{*}(k_{\min}))$  are defined in Appendix \ref{erc_noisy_bnd_appendix} (have very long expressions).
\label{regmod_thm}
\end{theorem}

\begin{remark}
Everything needed for the above result can be computed in polynomial time if $x$, $\T$, and a bound on $\|w\|_\infty$ are available.
\label{set_gamma_lambda}
\end{remark}

\begin{remark}
One can get a unique minimizer of the reg-mod-BPDN program using any $\gamma \ge \gamma_{\T,\lambda}^*(\tilde\Delta_u^{*}(k_{\min}))$. By setting $\gamma = \gamma^*$, we get the smallest error bound and this is why we state the theorem as above.
\end{remark}

\begin{corollary}[computable bound for mod-BPDN, BPDN]
If $A_\T$ is full rank, and $\gamma = \gamma_{\T,0}^*(\tilde\Delta_u^{*}(k_{\min}))$, then the solution of modified-BPDN satisfies
$\|x-\xhat\|_2 \le g_0(\tilde\Delta_u^{*}(k_{\min})).$
\\
The corollary for BPDN follows by setting $\T = \emptyset$ and ${\Delta_u} = \N$ in the above result.
\end{corollary}

\subsection{Modified-BPDN-residual}
Another way to use both slow support and slow signal value knowledge is as follows. Replace $y$ by $(y - A \hat\mu)$ and solve the modified-BPDN program. Add back its solution $\hat\mu$. We refer to this as {\em modified-BPDN-residual.} It computes \cite{icipmodcs}
\bea
\xhat = \hat\mu + [\arg \min_{b} \gamma \|b_{\T^c}\|_1 + 0.5 \|y - A \hat\mu - Ab\|_2^2].
\label{modcsresidual}
\eea


\section{Recursive dynamic CS} \label{recrecon} 
To solve the original problem described in Section \ref{probdef}, it is easy to develop dynamic versions of the approaches described so far. 
To do this, two things are required. First, an accurate estimate of the sparse signal at the initial time instant is required and second an accurate support estimation technique is required. 
We discuss these two issues in the next two subsections. After this we briefly summarize the dynamic versions of the approaches from the previous two sections, followed by algorithms that were directly designed for the dynamic problem. In Sec. \ref{parameter_set}, we discuss general parameter setting strategies. Finally, in Sec. \ref{error_stab}, we summarize error stability over time results.

\subsection{Initialization} \label{initialize}
To initialize any recursive dynamic CS solution, one needs an accurate estimate of $x_0$. This can be computed using any sparse recovery technique as long as enough measurements are available to allow for exact or accurate enough reconstruction. For example, BP or BP-noisy can be used.

Alternatively, one can obtain partial support knowledge for the initial signal using other types of prior knowledge, e.g., for a wavelet-sparse image with a very small black region, most of the wavelet scaling coefficients will be nonzero. Thus the indices of these coefficients can serve as partial support knowledge at $t=0$ \cite{isitmodcs}. If this knowledge is not as accurate as the support estimate from the previous time instant, more measurements would be required at $t=0$.

In most applications where sparse recovery is used, such as MRI, it is possible to use different number of measurements at different times. In applications where this is not possible, one can use a batch sparse recovery technique for the first short batch of time instants. For example, one could solve the  multiple measurement vectors' (MMV) problem that attempts to recover a batch of sparse signals with common support from a batch of their measurements \cite{wipf2007empirical,tropp2006algorithms2,chen2006theoretical,kim2012compressivemusic,lee_bresler,eldar2009compressed}. Or one could use the approach of Zhang and Rao \cite{zhang_rao} (called temporal SBL) that solves the MMV problem with the extra assumption that the elements in each nonzero row of the solution matrix are temporally correlated. 

\subsection{Support estimation} \label{support_est}
The simplest way to estimate the support is by thresholding, i.e., by computing
$$\Nhat = \{i: |(\xhat)_i| > \alpha \}$$
where $\alpha \ge 0$ is the zeroing threshold.
If $\xhat=x$ (exact recovery), we use $\alpha=0$. Exact recovery is possible only for sparse $x$ and enough noise-free measurements.
In other situations, we need a nonzero value. In case of accurate reconstruction of a sparse signal, we can set $\alpha$ to be a little smaller than an estimate of its smallest magnitude nonzero entry \cite{isitmodcs}. For compressible signals, one should use an estimate of the smallest magnitude nonzero entry of a sparsified version of the signal, e.g., sparsified to retain $b\%$ of the total signal energy.
In general $\alpha$ should also depend on the noise level.  We discuss the setting of $\alpha$ further in Section \ref{parameter_set}.

For  all of LS-CS, modified-CS  and weighted-$\ell_1$, it can be argued that $\xhat$ is a biased estimate of $x$: it is biased towards zero along $\Delta_u$ and away from zero along $\T$ \cite{just_lscs,stab_jinchun_jp}. As a result, the threshold $\alpha$ is either too small and does not delete all extras (subset of $\T$) or is too large and does not detect all misses (subset of $\T^c$). A partial solution to this issue is provided by a two step support estimation approach that we call Add-LS-Del. The Add-LS part of this approach is motivated by the Gauss-Dantzig selector idea \cite{dantzig} that first demonstrated that support estimation with a nonzero threshold followed by computing an LS estimate on the support significantly improves the sparse recovery error because it reduces the bias in the solution estimate. 
Add-LS-Del proceeds as follows.
\bea
\T_\dett \se \T \cup \{i : |(\xhat)_i| > \alpha_{\dett} \}  \nn \\
\xhat_{\dett} \se I_{\T_\dett}{A_{\T_\dett}}^\dag y \nn \\
\Nhat \se \T_\dett \setminus \{i: |(\xhat_{\dett})_i| \le \alpha_{\del} \} \nn \\
\xhat_{final} \se I_{\Nhat} (A_\Nhat)^\dag y
\label{addLSdel}
\eea
The addition step threshold, $\alpha_{\dett}$, needs to be just large enough to ensure that the matrix used for LS estimation, $A_{\T_\dett}$ is well-conditioned. If $\alpha_{\dett}$ is chosen properly and if the number of measurements, $n$, is large enough, the LS estimate on $\T_\dett$ will have smaller error, and will be less biased, than $\xhat$. As a result, deletion will be more accurate when done using this estimate. This also means that one can use a larger deletion threshold, $\alpha_{\del}$, which will ensure deletion of more extras. 
%

\begin{algorithm}[t]
\caption{{\bf \small Dynamic modified-BPDN \cite{regmodbpdn}}}
At $t=0$: Solve BPDN with sufficient measurements, i.e., compute $\xhat_0$ as the solution of $\min_b \gamma \|b \|_1 + 0.5 \|y_0 - A_0 b \|_2^2$ and compute its support by thresholding: $\Nhat_0 = \{i: |(\xhat_{0})_i| > \alpha \}$.
For each $t>0$ do
\ben
\item Set $\T = \Nhat_{t-1}$

\item {\em Mod-BPDN} Compute $\xhat_t$ as the solution of
\[
\min_{b} \gamma\|b_{\T^c}\|_1 + 0.5\|{y}_t -  A_t b\|_2^2
\]

\item {\em Support Estimation (Simple): } $\Nhat_t =  \{i: |(\xhat_t)_i| > \alpha \}$
\een
{\em Parameter setting: } see Section \ref{parameter_set} step \ref{dyn_algs}.
\label{modbpdn_algo}
\end{algorithm}

\begin{algorithm}[t]
\caption{{\bf \small Dynamic weighted-$\ell_1$ \cite{hassibi}}}
At $t=0$: Solve BPDN with sufficient measurements, i.e., compute $\xhat_0$ as the solution of $\min_b \gamma \|b \|_1 + 0.5 \|y_0 - A_0 b \|_2^2$. 
For each $t>0$ do
\ben
\item Set $\T = \Nhat_{t-1}$

\item {\em Weighted-$\ell_1$} Compute $\xhat_t$ as the solution of
\[
\min_{b} \gamma\|b_{\T^c}\|_1 + \gamma \tau \|b_\T\|_1 + 0.5\|{y}_t -  A_t b\|_2^2
\]

\item {\em Support Estimation (Simple): } $\Nhat_t =  \{i: |(\xhat_t)_i| > \alpha \}$
\een
{\em Parameter setting: } see Section \ref{parameter_set} step \ref{dyn_algs}.
\label{w_ell1_algo}
\end{algorithm}

\begin{algorithm}[t]
\caption{{\bf \small Dynamic IHT-PKS}}
At $t=0$: Compute $\xhat_0$ using (\ref{iht}) and compute its support as $\Nhat_0:=\{i: |(\xhat_{0})_i| > 0\}$.
\\ For each $t>0$ do
\ben
\item Set $\T = \Nhat_{t-1}$

\item {\em IHT-PKS} Compute $\xhat_t$ using (\ref{iht_pks}).

\item {\em Support Estimation: } $\Nhat_t =  \{i: |(\xhat_t)_i| > 0 \}$
\een
\label{iht_pks_algo}
\end{algorithm}

\subsection{Dynamic LS-CS, modified-CS, weighted-$\ell_1$, IHT-PKS}
It is easy to develop dynamic versions of all the approaches described in Sec. \ref{supp_knowledge} by (i) recovering $x_0$ as described above, (ii) for $t>0$, using $\T = \Nhat_{t-1}$ where $\Nhat_{t-1}$ is computed from $\xhat_{t-1}$ as explained above, and (iii) by using $y = y_t$ and $A = A_t$ at time $t$.
We summarize dynamic weighted-$\ell_1$, dynamic modified-BPDN and dynamic IHT-PKS in Algorithms  \ref{modbpdn_algo}, \ref{w_ell1_algo}, \ref{iht_pks_algo} respectively. In these algorithms, we have used simple BPDN at the initial time, but this can be replaced by the other approaches described in Sec. \ref{initialize}.

\subsection{Streaming ell-1 homotopy and streaming modified weighted-$\ell_1$ (streaming mod-wl1)} 
%
In \cite{asif2013sparse}, Asif et al. developed a fast homotopy based solver for the general weighted $\ell_1$ minimization problem:
\bea
\min_b \|W b\|_1 + 0.5\|y - A b\|_2^2
\label{gen well1}
\eea
where $W$ is a diagonal weighting matrix. They also developed a homotopy for solving \eqref{gen well1} with a third term $0.5 \| b - F \mu\|_2^2$. Here $F$ can be an arbitrary square matrix. For the dynamic problem, it is usually the state transition matrix of the dynamical system \cite{asif2013sparse}. We use these solvers for some of our numerical experiments. 

Another key contribution of this work was a weighting scheme for solving the recursive dynamic CS problem that can be understood as a generalization of the modified-CS idea. At time $t$, one solves \eqref{gen well1} with $W_{i,i} = \frac{\gamma}{\beta |(\xhat_{t-1})_i| + 1}$ with $\beta \gg 1$. 
The resulting algorithm, that can be called {\em streaming modified weighted-$\ell_1$ (streaming mod-wl1)}, is summarized in Algorithm \ref{dyn_ mod_wl1}.
When $x_t$'s, and hence $\xhat_{t}$'s, are exactly sparse, if we set $\beta = \infty$, we recover the modified-CS program \eqref{l1seqcs}. To understand this, let $\T = \supp(\xhat_{t-1})$. Thus, if $i \in \T$, then $W_{i,i}=0$ while if $i \notin \T$, then $W_{i,i}$ equals a constant nonzero value. This gives the modified-BPDN cost function.

The above is a very useful generalization since it uses weights that are inversely proportional to the magnitude of the entries of $\xhat_{t-1}$. Thus it incorporates support change knowledge in a soft fashion while also using signal value knowledge. It also naturally handles compressible signals for which no entry of $\xhat_{t-1}$ is zero  but many entries are small.
The weighting scheme is similar to that introduced in the reweighted $\ell_1$ approach of \cite{reweighted_cs} to solve the simple CS problem.%

\subsection{Dynamic regularized-modified-BPDN}
It is easy to develop a dynamic version of reg-mod-BPDN by using $y = y_t$ and $A = A_t$ and by recovering $x_0$ as described in Sec. \ref{initialize} above. We set  $\T = \Nhat_{t-1}$ where $\Nhat_{t-1}$ is computed from $\xhat_{t-1}$ as explained above. For $\hat\mu$, one can either use $\hat\mu = \xhat_{t-1}$, or, in certain applications, e.g., functional MRI reconstruction \cite{fmrimodcs}, where the signal values do not change much w.r.t. the first frame, using $\hat\mu = \xhat_0$ is a better idea. This is because $\xhat_0$ is estimated using many more measurements and hence is much more accurate. We summarize the complete algorithm in Algorithm \ref{regmod_algo}.%

\begin{algorithm}
\caption{\bf  Streaming modified weighted-$\ell_1$ (streaming mod-wl1) \cite{asif2013sparse}} \label{dynhom} \label{dyn_ mod_wl1}
\begin{algorithmic}[1]
\STATE At $t=0$: Solve BPDN with sufficient measurements, i.e., compute $\xhat_0$ as the solution of $\min_b \gamma \|b \|_1 + 0.5 \|y_0 - A_0 b \|_2^2$ and compute its support $\Nhat_0 = \{i: |(\xhat_{0})_i| > \alpha \}$.

\STATE For $t\geq2$, set  $$(W_t)_{ii} = \frac{\gamma}{\beta|(\hat{x}_{t-1})_i|+1},$$ where $\beta = n\frac{\|\hat{x}_{t-1}\|_2^2}{\|\hat{x}_{t-1}\|_1^2}$,
and compute $\xhat$ as the solution of $$\min_x \|W_{t}x\|_1 + \frac{1}{2}\|y_t-A_t x\|_2^2$$
\STATE $t\leftarrow t+1$, go to step 2.
\end{algorithmic}
\end{algorithm}

\begin{algorithm}[t]
\caption{{\bf \small Dynamic Regularized Modified-BPDN \cite{regmodbpdn}}}
At $t=0$: Solve BPDN with sufficient measurements, i.e., compute $\xhat_0$ as the solution of $\min_b \gamma \|b \|_1 + 0.5 \|y_0 - A_0 b \|_2^2$ and compute its support $\Nhat_0 = \{i: |(\xhat_{0})_i| > \alpha \}$.

For each $t>0$ do
\ben
\item Set $\T = \Nhat_{t-1}$

\item {\em Reg-Mod-BPDN} Compute $\xhat_t$ as the solution of
\[
\min_{b} \gamma\|b_{\T^c}\|_1 + 0.5\|{y}_t - A_t b\|_2^2 + 0.5 \lambda\|b_\T - \hat\mu_\T\|_2^2
\]

\item {\em Support Estimation (Simple): } $\Nhat_t =  \{i: |(\xhat_t)_i| > \alpha \}$
\een
{\em Parameter setting: } see Section \ref{parameter_set} step \ref{dyn_algs}.
\label{regmod_algo}
\end{algorithm}

\subsection{Kalman filtered Modified-CS (KF-ModCS) and KF-CS} 


Kalman Filtered CS-residual (KF-CS) was introduced for solving the recursive dynamic CS problem in \cite{kfcsicip} and in fact this was the first solution to this problem. With the modified-CS approach and results now known, a much better idea than KF-CS is {\em Kalman Filtered Modified-CS-residual (KF-ModCS).} This can be understood as modified-BPDN-residual but with $\hat\mu$ obtained as a Kalman filtered estimate on the previous support estimate $\T = \Nhat_{t-1}$.
For the KF step, one needs to assume a model on signal value change. In the absence of specific information, a Gaussian random walk model with equal change variance in all directions can be used \cite{kfcsicip}:
\bea
(x_{0})_{\N_0} & \sim & \n(0, \sigma_{sys,0}^2 I), \nn \\
(x_{t})_{\N_t} \se (x_{t-1})_{\N_t} + \nu_t, \ \nu_t \sim \n(0, \sigma_{sys}^2 I) \nn \\
(x_{t})_{\N_t^c} \se 0 
\label{xt_mod}
\eea
Here $\n(a,\Sigma)$ denotes a Gaussian distribution with mean $a$ and covariance matrix $\Sigma$. 
Assume for a moment that the support does not change with time, i.e., $\N_t = \N_0$, and $\N_0$ is known or can be perfectly estimated using BP-noisy followed by support estimation. Then $y_t := A_t x_t+w_t$ can be rewritten as $y_t = (A_t)_{\N_0} (x_t)_{\N_0} + w_t$. With this, if the observation noise $w_t$ being Gaussian, the KF provides a causal minimum mean squared error (MMSE) solution, i.e., it returns $\xhat_{t|t}$ which solves 
$$
\arg\min_{{\xhat}_{t|t}: ({\xhat}_{t|t})_{\N_0^c}=0} \E_{x_t|y_1,y_2, \dots y_t}[\|x_t - \xhat_{t|t}(y_1,y_2, \dots y_t)\|_2^2] \ \ \
$$
Here $\E_{x|y}[q(x)]$ denotes expectation of $q(x)$ given $y$.

Our problem is significantly more difficult because the support set $\N_t$ changes with time and is unknown. To solve it, KF-ModCS is a practical heuristic that combines the modified-BPDN-residual idea for tracking the support with an adaptation of the regular KF algorithm to the case where the set of entries of $x_t$ that form the state vector for the KF change with time: the KF state vector at time $t$ is $(x_t)_{\N_t}$ at time $t$.
Unlike the regular KF for a fixed dimensional linear Gaussian state space model, KF-CS or KF-ModCS do not enjoy any optimality properties. One expects KF-ModCS to outperform modified-BPDN when accurate prior knowledge of the signal values is available. 
We summarize it in Algorithm \ref{kfcsalgo}.%

An open question is how to analyze KF-CS or KF-ModCS and get a meaningful performance guarantee? This is a hard problem because the KF state vector is $(x_t)_{\N_t}$ at time $t$ and $\N_t$ changes with time. In fact, even if we assume that the sets $\N_t$ are given, there are no known results for the resulting ``genie-aided KF".

\begin{algorithm}[t]
\caption{{\bf \small KF-ModCS: use of mod-BPDN-residual to replace BPDN in the KF-CS algorithm of \cite{kfcsicip}}} 
{\em Parameters: } $\sigma_{\text{sys}}^2$, $\sigma_{obs}^2$, $\alpha$, $\gamma$ \\
At $t=0$: Solve BPDN with sufficient measurements, i.e., compute $\xhat_0$ as the solution of $\min_b \gamma \|b \|_1 + 0.5 \|y_0 - A_0 b \|_2^2$. Denote the solution by $\xhat_0$.
Estimate its support, $\T = \Nhat_0= \{i: |(\xhat_{0})_i| > \alpha \}$.
\\
Initialize $P_0  = \sigma_{sys}^2 I_{\T} {I_\T}'$.
\\
For each $t>0$ do
\ben
\item Set $\T = \Nhat_{t-1}$
\item {\em Mod-BPDN-residual:}
\[
\xhat_{t,mod}=  \xhat_{t-1} + [\arg\min_b \gamma \|b_{\T^c} \|_1 + \|y_t -  A_t \xhat_{t-1} -  A_t b \|^2]
\]


\item {\em Support Estimation - Simple Thresholding: }
\bea
\Nhat_t =  \{i: |(\xhat_{t,mod})_i| > \alpha \}
\eea

\item {\em Kalman Filter: }
\bea
\hat{Q}_t \se \sigma_{sys}^2 I_{\Nhat_t} {I_{\Nhat_t}}' \nn \\
K_t \se (P_{t-1} + \hat{Q}_t) A_t' \left(  A_t (P_{t-1} + \hat{Q}_t)  A_t' + \sigma_{obs}^2 I \right)^{-1} \nn \\
P_t \se (I - K_t  A_t)(P_{t-1} + \hat{Q}_t) \nn \\
\xhat_t \se (I - K_t  A_t) \xhat_{t-1} + K_t y_t
\eea

\een
{\em Parameter setting: } see Section \ref{parameter_set} step \ref{dyn_algs}.
\label{kfcsalgo}
\end{algorithm}

\subsection{Dynamic CS via approximate message passing(DCS-AMP)}
Another approximate Bayesian approach was developed in very interesting recent work by Ziniel and Schniter \cite{schniter_track,schniter_track_jp}. They introduced the dynamic CS via approximate message passing (DCS-AMP) algorithm by developing the recently introduced AMP approach of Donoho et al. \cite{amp} for the dynamic CS problem.
The authors model the dynamics of the sparse vectors over time using a stationary Bernoulli Gaussian prior as follows: for all $i=1,2, \dots m$,
$$
(x_t)_i = (s_t)_i (\theta_t)_i
$$
where $(s_t)_i$ is a binary random variable that forms a stationary Markov chain over time and $(\theta_t)_i$ follows a stationary first order autoregressive (AR) model with nonzero mean. Independence is assumed across the various indices $i$. Let $\lambda:=\Pr((s_t)_i=1)$ and $p_{10}:=\Pr((s_t)_i=1|(s_{t-1})_i=0).$ Using stationarity, this tells us that $p_{01}:=\Pr((s_t)_i=0|(s_{t-1})_i=1) = \frac{\lambda p_{10}}{1-\lambda}.$
The following stationary AR model with nonzero mean is imposed on $\theta_t$.
$$
(\theta_t)_i = \zeta_i + (1-\alpha) ((\theta_{t-1})_i - \zeta_i) + \alpha (v_t)_i
$$
with $0 < \alpha \le 1.$ In the above model, $v_t$ is the i.i.d. white Gaussian perturbation, $\zeta$ is the mean, and $(1-\alpha)$ is the AR coefficient.
The choice of $\alpha$ controls how slowly or quickly the signal values change over time. The choice of $p_{10}$ controls the likelihood of new support addition(s). This model results in a Bernoulli-Gaussian or ``spike-and-slab'' distribution of $(x_t)_i$, which is known to be an effective sparsity promoting prior \cite{schniter_track,schniter_track_jp}.%

Exact computation of the minimum mean squared error (MMSE) estimate of $x_t$ cannot be done under the above model. On one end, one can try to use sequential Monte Carlo techniques (particle filtering) to approximate the MMSE estimate. Some attempts to do this are described in \cite{pafimocs_icassp,pafimocs_asilomar}. But these can get computationally expensive for high dimensional problems and it is never clear what number of particles is sufficient to get an accurate enough estimate. The AMP approach developed in \cite{schniter_track_jp} is also approximate but is extremely fast and hence is useful.
The complete DCS-AMP algorithm taken from \cite[Table II]{schniter_track_jp} is summarized in Table \ref{tab:algorithm_equations} (with permission from the authors). Its code is available at \url{http://www2.ece.ohio-state.edu/~schniter/DCS/index.html}.

\subsection{Parameter Setting} \label{parameter_set}
In this section we discuss parameter setting for the algorithms described above.
\ben
\item For DCS-AMP, an expectation maximization (EM) algorithm is proposed for parameter learning \cite{schniter_track_jp}. We summarize this in Table \ref{tab:em_updates}.


\item \label{dyn_algs} For the rest of the algorithms, the following procedure is proposed. This assumes that, for the first two time instants, enough measurements are available so that a simple CS technique such as BPDN can be used to recover $x_1$ and $x_2$ accurately. 
\ben

\item Let $\xhat_1, \xhat_2$ denote the estimates of $x_1,x_2$ obtained using BPDN (and enough measurements). 

\item Consider the simple single-threshold based support estimation scheme. Let $\alpha_0$ be the smallest real number so that $\sum_{i: |(\xhat_1)_i| > \alpha_0} (\xhat_1)_i^2 < 0.999 \sum_i (\xhat_1)_i^2$. We set the threshold $\alpha = c \alpha_0$ with $c=1/12$. In words, $\alpha$ is a fraction of the smallest magnitude nonzero entry of $\xhat_1$ sparsified to $99.9\%$ energy. We used $99.9\%$ energy for sparsification and $c=1/12$ in our experiments, but these numbers can be changed.
\label{support_step}

\item Consider dynamic reg-mod-BPDN (Algorithm \ref{regmod_algo}). It has three parameters $\alpha, \gamma, \lambda$. We set $\alpha$ as in step \ref{support_step}.
We set $\gamma$ and $\lambda$ using a heuristic motivated by Theorem \ref{regmod_thm}. This theorem can be used because, for a given choice of $\lambda$, it gives us an error bound that is computable in polynomial time and that holds always (without any sufficient conditions). To use this heuristic, we need $\T, \N, \hat\mu,x, \Delta_u, \Delta_e$ and an estimate of $\|w\|_\infty$. We compute $\Nhat_1 = \{i: |(\xhat_1)_i|>\alpha\}$ and $\Nhat_2 = \{i: |(\xhat_2)_i|>\alpha\}$. We set $\T = \Nhat_1$, $\N = \Nhat_2$, $\hat\mu = \xhat_1$, $x = \xhat_2$, $\Delta_u= \N \setminus \T$, and $\Delta_e = \T \setminus \N$. We use $\|y_2 - A\xhat_2\|_\infty$ as an estimate of $\|w\|_\infty$. With these, for a given value of $\lambda$, one can compute $g_\lambda(\tilde\Delta_u^*(k_{\min}))$ defined in Theorem \ref{regmod_thm}. We pick $\lambda$ by selecting the one that minimizes $g_\lambda(\tilde\Delta_u^*(k_{\min}))$ out of a discrete set of possible values. In our experiments we selected $\lambda$ out of the set $\{0.5, 0.2, 0.1, 0.05, 0.01, 0.005, 0.001, 0.0001\}$. We then use this value of $\lambda$ to compute $\gamma_{\T,\lambda}^*(\tilde\Delta_u^*(k_{\min}))$ given in Theorem \ref{regmod_thm}. This gives us the value of $\gamma$.
\label{gamma_lambda_step}

\item Consider modified-BPDN. It has two parameters $\alpha, \gamma$. We set $\alpha$ as in step \ref{support_step}.
Modified-BPDN is reg-mod-BPDN with $\lambda=0$. Thus one would expect to be able to just compute $\gamma = \gamma_{\T,0}^*$, i.e., use the above approach with $\lambda=0$. However with $\lambda=0$, certain matrices used in the computation of $\gamma_{\T,\lambda}^*$ become ill-conditioned when $A_\T$ is ill-conditioned (either it is not full rank or is full rank but is ill-conditioned). Hence we instead compute $\gamma = \gamma_{\T,0.0001}^*$, i.e., we use $\lambda = 0.0001$ and step \ref{gamma_lambda_step}.
\label{gamma_step}

\item Consider weighted-$\ell_1$. It has three parameters $\alpha, \gamma, \tau$. We set $\alpha$ as in step \ref{support_step}.
Even though Theorem \ref{regmod_thm} does not apply for it directly, we still just use the heuristic of step \ref{gamma_step} for setting $\gamma$.
As suggested in \cite{friedlander}, we set $\tau$ equal to the ratio of the estimate of the number of extras to the size of the known support, i.e., $\tau = \frac{|\Delta_e|}{|\N|}$.

\item Consider KF-ModCS. It has parameters $\alpha, \gamma, \sigma_{sys}^2, \sigma_{sys,0}^2, \sigma_{obs}^2$. We set $\alpha$ as in step \ref{support_step}. We set $\gamma$ as in step \ref{gamma_step}. We assume $\sigma_{sys,0}^2 = \sigma_{sys}^2$ and we set $\sigma_{sys}^2$ by maximum likelihood estimation, i.e., we compute it as $\frac{1}{|\Nhat_2|}\sum_{i \in \Nhat_2 } (\hat{x}_{2}-\hat{x}_{1})_i^2$.  We use $\|y_2 - A \xhat_2\|_2^2/m$ as an estimate for $\sigma_{obs}^2$.
\label{kfmod_step}

\item For initial sparse recovery, we use BPDN with $\gamma$ set as suggested in \cite{asif2013sparse}: $\gamma = \max\{10^{-2}\|A_1^T[y_1\ y_2]\|_{\infty}, \sigma_{\text{obs}}\sqrt{m}\}$. 
\een

\een

Lastly consider the add-LS-del support estimation approach. This is most useful for exactly sparse signal sequences. This needs two thresholds $\alpha_\add$, $\alpha_\del$. We can set $\alpha_{\add,t}$ using the following heuristic. It is not hard to see (see \cite[Lemma 2.8]{stab_jinchun_jp}) that $(x_t - \xhat_{t,\add})_{\T_{\add,t}} = ({A_{\T_{\add,t}}}'A_{\T_{\add,t}})^{-1} [ {A_{\T_{\add,t}}}' w_t + {A_{\T_{\add,t}}}' A_{\Delta_{\add,t}}  (x_t)_{\Delta_{\add,t}}]$. To ensure that this error is not too large, we need $A_{\T_{\add,t}}$ to be well conditioned. To ensure this we pick $\alpha_{\add,t}$ as the smallest number such that $\sigma_{\min}({A_{\T_{\add,t}}})\geq 0.4$.
If one could set $\alpha_\del$ equal to the lower bound on $x_{\min,t}-\|(x_t - \xhat_{t,\add})_{\T_{\add,t}}\|_{\infty}$, there will be zero misses. Here $x_{\min,t}= \min_{i \in \N_t}|(x_t)_i|$ is the minimum magnitude nonzero entry. Using this idea, we  let $\alpha_{\del,t}$ be an estimate of the lower bound of this quantity. As explained in  \cite[Section VII-A]{stab_jinchun_jp}, this leads to $\alpha_{\del,t}=0.7\hat{x}_{\min,t}-\|A_{\T_{\add,t}}^\dag (y_t-A\xhat_{t,\text{modcs}})\|_{\infty}$ where $\hat{x}_{\min,t}=\min_{i \in \Nhat_{t-1}} |(\xhat_{t-1})_i|$.
An alternative approach useful for videos is explained in \cite[Algorithm 1]{rrpcp_tsp}.

 \newcommand{\limit}[2]{\ensuremath{\lim_{#1 \rightarrow #2}}}
 \newcommand{\uvec}[1]{\ensuremath{\underline{\boldsymbol{#1}}}}
 \newcommand{\tvec}[1]{\ensuremath{\Tilde{\boldsymbol{#1}}}}
 \newcommand{\ovec}[1]{\ensuremath{\Bar{\boldsymbol{#1}}}}
 \newcommand{\hvec}[1]{\ensuremath{\Hat{\boldsymbol{#1}}}}
 \newcommand{\bvec}[1]{\ensuremath{\Breve{\boldsymbol{#1}}}}
 \renewcommand{\vec}[1]{\ensuremath{\boldsymbol{#1}}}
 \newcommand{\mat}[1]{\ensuremath{\begin{bmatrix}#1\end{bmatrix}}}
 \newcommand{\smallmat}[1]{\ensuremath{
        \left[\begin{smallmatrix}#1\end{smallmatrix}\right]}}
 \newcommand{\ip}[1]{\ensuremath{\left\langle #1 \right\rangle}}
 \newcommand{\norm}[1]{\ensuremath{\| #1 \|}}
 \newcommand{\mc}[1]{\ensuremath{\mathcal{#1}}}
\renewcommand{\st}{{~\text{s.t.}~}}
 \newcommand{\barst}[1]{\ensuremath{\text{\raisebox{-0.5mm}{$\bigl|_{#1}$}}}}
 \newcommand{\Barst}[1]{\ensuremath{\text{\raisebox{-0.5mm}{$\biggl|_{#1}$}}}}
 \newcommand{\zci}{{\textstyle\frac{1}{z^*}}}
 \newcommand{\Real}{{\mathbb{R}}}
 \newcommand{\Complex}{{\mathbb{C}}}
 \newcommand{\Int}{{\mathbb{Z}}}
 \newcommand{\Nat}{{\mathbb{N}}}
 \newcommand{\Rat}{{\mathbb{Q}}}
 \newcommand{\Field}{\mathbb{F}}
 \newcommand{\Ell}{\mathcal{L}}
\renewcommand{\kron}{\otimes}
 \newcommand{\conv}{\ast}
 \newcommand{\cconv}{\circledast}
 \newcommand{\modulo}[1]{\left\langle #1 \right\rangle}
 \newcommand{\floor}[1]{\left\lfloor #1 \right\rfloor}
 \renewcommand{\ceil}[1]{\left\lceil #1 \right\rceil}
 \newcommand{\of}[1]{^{(#1)}}
 \newcommand{\oft}[1]{^{{(#1)}T}}
 \newcommand{\ofc}[1]{^{{(#1)}*}}
 \newcommand{\ofH}[1]{^{{(#1)}H}}

 %

 \renewcommand{\eqref}[1]{(\ref{eq:#1})}
 \newcommand{\Eqref}[1]{Equation~(\ref{eq:#1})}
 \newcommand{\Figref}[1]{Figure~\ref{fig:#1}}
 \newcommand{\figref}[1]{Fig.~\ref{fig:#1}}
 \newcommand{\tabref}[1]{Table~\ref{tab:#1}}
 \newcommand{\secref}[1]{Section~\ref{sec:#1}}
 \newcommand{\Secref}[1]{Section~\ref{sec:#1}}
 \newcommand{\appref}[1]{Appendix~\ref{app:#1}}
 \newcommand{\lemref}[1]{Lemma~\ref{lem:#1}}
 \newcommand{\thmref}[1]{Theorem~\ref{thm:#1}}
 \newcommand{\corref}[1]{Corollary~\ref{cor:#1}}
 \newcommand{\conref}[1]{Conjecture~\ref{con:#1}}
 \newcommand{\exaref}[1]{Example~\ref{exa:#1}}
 \newcommand{\etal}{et al.\ }
 \newcommand{\ie}{i.e., }
 \newcommand{\eg}{e.g., }

 \newcommand{\textr}[1]{\textcolor{black}{#1}}
 \newcommand{\textb}[1]{\textcolor{blue}{#1}}

 \newcounter{comment}[section]
 \newcommand{\comment}[1]{\vspace{2mm}
   \noindent\refstepcounter{comment}\textsl{\underline{Comment
   \thesection.\thecomment\mbox{}} #1:} }
 \newcounter{texthead}[section]
 \newcommand{\texthead}[1]{\medskip
   \noindent\refstepcounter{texthead}\textsl{\underline{%
   \thesection.\thetexthead) #1}:} }
 \renewenvironment{itemize}
   {\begin{list}{\labelitemi}{\topsep 0.05in \itemsep 0in}}{\end{list}}

 \newcommand{\mus}{\vec{\mu}(\vec{s})}
 \newcommand{\musn}{\vec{\mu}(\vec{s}')}
 \newcommand{\Rs}{\vec{R}(\vec{s})}
 \newcommand{\Rsn}{\vec{R}(\vec{s}')}
 \newcommand{\Phis}{\vec{\Phi}(\vec{s})}
 \newcommand{\Phisn}{\vec{\Phi}(\vec{s}')}
 \newcommand{\ls}{\vec{l}(\vec{s})}
 \newcommand{\mmse}{_{\text{\sf mmse}}}
 \newcommand{\ammse}{_{\text{\sf ammse}}}
 \newcommand{\map}{_{\text{\sf map}}}
 \newcommand{\amap}{_{\text{\sf amap}}}
 \newcommand{\SNR}{{\text{\sf SNR}}}

 \newcommand{\pre}{^{\text{\sf pre}}}
 \newcommand{\preH}{^{\text{\sf pre}H}}
 \newcommand{\rt}{^{\text{\sf root}}}
 \newcommand{\rtH}{^{\text{\sf root}H}}

 \newcommand{\fwd}[1]{\accentset{\rightharpoonup}{#1}}
 \newcommand{\bwd}[1]{\accentset{\leftharpoonup}{#1}}
 \newcommand{\ubar}[1]{\underaccent{\bar}{#1}}
 \newcommand{\poz}{p_{{}_{10}}}
 \newcommand{\pzo}{p_{{}_{01}}}
 \newcommand{\intd}{\,\mathrm{d}}
 \newcommand{\mf}[1]{\mathfrak{#1}}
 \renewcommand{\vect}[1]{\vec{#1}^{(t)}}
 \newcommand{\nt}[1]{{#1}_n^{(t)}}
 \newcommand{\msg}[2]{\nu_{{#1} \to {#2}}}

\begin{table}[t]
\centering
\scriptsize
\setlength{\tabcolsep}{2pt}
\setlength{\belowcaptionskip}{3ex}
\begin{tabular}{|llrclr|}
	\hline
	\multicolumn{5}{|l}{$\textsf{\% Define soft-thresholding functions: }$}&\\
	& \multicolumn{4}{l}{$\textit{F}_{nt}(\phi; c) \triangleq (1 + \gamma_{nt}(\phi; c))^{-1}
		\Big( \frac{\bwd{\psi}_n^{(t)} \phi + \bwd{\xi}_n^{(t)} c}{\bwd{\psi}_n^{(t)} + c} \Big)$}	& \threshcnt{eq:thresh_start}\\[2ex]
	& \multicolumn{4}{l}{$\textit{G}_{nt}(\phi; c) \triangleq (1 + \gamma_{nt}(\phi; c))^{-1}
		\Big( \frac{\bwd{\psi}_n^{(t)} c}{\bwd{\psi}_n^{(t)} + c} \Big) + \gamma_{nt}(\phi; c) |\textit{F}_{nt}(\phi; c)|^2$}	& \threshcnt{}\\[1ex]
	& \multicolumn{4}{l}{$\textit{F}_{nt}'(\phi; c) \triangleq \tfrac{\partial}{\partial \phi} \textit{F}_{nt}(\phi,c) = \tfrac{1}{c} \textit{G}_{nt}(\phi; c)$}	& \threshcnt{}\\[1ex]
	& \multicolumn{4}{l}{$\gamma_{nt}(\phi; c) \triangleq \Big( \frac{1 - \bwd{\pi}_n^{(t)}}{\bwd{\pi}_n^{(t)}} \Big) \Big( \frac{\bwd{\psi}_n^{(t)} + c}{c} \Big) $} &\\
	&&& \multicolumn{2}{l}{$\quad \times \exp\Big( - \Big[ \frac{\bwd{\psi}_n^{(t)} |\phi|^2 + \bwd{\xi}_n^{(t)\,*} c \phi +
		\bwd{\xi}_n^{(t)} c \phi^* - c |\bwd{\xi}_n^{(t)}|^2}{c(\bwd{\psi}_n^{(t)} + c)} \Big] \Big)$}	& \threshcnt{eq:thresh_end}\\[1ex]
	\hline
	\multicolumn{5}{|l}{$\textsf{\% Begin passing messages} \ldots$} &\\[-1ex]
  	\multicolumn{5}{|l}{$\textsf{for } t=1,\ldots,T:$}&\\[-1ex]
	&\multicolumn{4}{l}{$\quad \textsf{\% Execute the } \textbf{(into)} \textsf{ phase} \ldots$} &\\
	&\multicolumn{4}{l}{$\quad \nt{\bwd{\pi}} = \frac {\nt{\fwd{\lambda}} \cdot \nt{\bwd{\lambda}}}
		{(1 - \nt{\fwd{\lambda}}) \cdot (1 - \nt{\bwd{\lambda}}) + \nt{\fwd{\lambda}} \cdot \nt{\bwd{\lambda}}} \quad \forall n$}	& \algcnt{} \\[2mm]
	&\multicolumn{4}{l}{$\quad \bwd{\psi}_{n}^{(t)} = \frac{\fwd{\kappa}_{n}^{(t)} \cdot \bwd{\kappa}_{n}^{(t)}}
		{\fwd{\kappa}_{n}^{(t)} + \bwd{\kappa}_{n}^{(t)}} \quad \forall n$}	& \algcnt{}\\[2mm]
	&\multicolumn{4}{l}{$\quad \bwd{\xi}_{n}^{(t)} = \bwd{\psi}_{n}^{(t)} \cdot \Big(\frac{\fwd{\eta}_{n}^{(t)}}
		{\fwd{\kappa}_{n}^{(t)}} + \frac{\bwd{\eta}_{n}^{(t)}}{\bwd{\kappa}_{n}^{(t)}}\Big) \quad \forall n$}	& \algcnt{}\\

	&\multicolumn{4}{l}{$\quad \textsf{\% Initialize AMP-related variables} \ldots$} &\\
	&\multicolumn{4}{l}{$\quad \forall m: z_{mt}^1 = y_{m}^{(t)}, \forall n: \mu_{nt}^1 = 0, \textsf{ and } c_t^1 = 100 \cdot \textstyle \sum_{n=1}^N \psi_n^{(t)}$}&\\

	&\multicolumn{5}{l|}{$\quad \textsf{\% Execute the } \textbf{(within)} \textsf{ phase using AMP} \ldots$} \\
  	&\multicolumn{4}{l}{$\quad \textsf{for $i=1,\ldots,I$, }:$}&\\
  	&&$\qquad \phi_{nt}^i$ &$=$& $\sum_{m=1}^M A_{mn}^{(t)\,*} z_{mt}^i + \mu_{nt}^i \quad \forall n$ & \algcnt{eq:amp_start}\\
	&&$\qquad \mu_{nt}^{i+1}$ &$=$& $\textit{F}_{nt}(\phi_{nt}^i; c_t^i) \quad \forall n$	&\algcnt{eq:amp_mu_defn}\\
	&&$\qquad v_{nt}^{i+1}$ &$=$& $\textit{G}_{nt}(\phi_{nt}^i; c_t^i) \quad \forall n$	& \algcnt{eq:amp_v_defn}\\
	&&$\qquad c_t^{i+1}$ &$=$& $\sigma_e^2 + \tfrac{1}{M} \sum_{n=1}^N v_{nt}^{i+1}$	& \algcnt{}\\
	&&$\qquad z_{mt}^{i+1}$ &$=$& $y_m^{(t)} - \vec{a}_{m}^{(t)\,\textsf{T}}\! \vec{\mu}_t^{i+1} + \tfrac{z_{mt}^i}{M} \sum_{n=1}^N \textit{F}_{nt}'(\phi_{nt}^i; c_t^i) \quad \forall m$	& \algcnt{eq:amp_end}\\
  	&\multicolumn{2}{l}{$\quad \textsf{end}$}&&&\\
	&\multicolumn{4}{l}{$\quad \hat{x}_n^{(t)} = \mu_{nt}^{I+1} \quad \forall n \qquad \textsf{\% Store current estimate of } x_n^{(t)}$}	& \algcnt{eq:amp_x_hat}\\

	&\multicolumn{5}{l|}{$\quad \textsf{\% Execute the } \textbf{(out)} \textsf{ phase} \ldots$} \\
	&\multicolumn{4}{l}{$\quad \fwd{\pi}_n^{(t)} = \Big(1 + \Big( \frac{\bwd{\pi}_n^{(t)}}{1 - \bwd{\pi}_n^{(t)}} \Big) \gamma_{nt}(\phi_{nt}^{I}; c_{t}^{I+1}) \Big)^{-1} \quad \forall n$}	& \algcnt{}\\[1.5ex]
	&\multicolumn{4}{l}{$\,\,\,\,(\fwd{\xi}_n^{(t)}, \fwd{\psi}_n^{(t)}) = \left\{
		\begin{array}{ll}
		(\phi_n^I/\eps, c_t^{I+1}/\eps^2), & \bwd{\pi}_n^{(t)} \le \tau \\
		(\phi_n^I, c_t^{I+1}), & \text{o.w.}
		\end{array} \right.  \quad \forall n \quad (\eps \ll 1)$}	& \algcnt{} \\[0ex]
	&\multicolumn{5}{l|}{$\quad \textsf{\% Execute the } \textbf{(across)} \textsf{ phase forward in time} \ldots$} \\
	&\multicolumn{4}{l}{$\quad \fwd{\lambda}_n^{(t+1)} = \frac{\poz (1-\fwd{\lambda}_n^{(t)})(1 - \fwd{\pi}_n^{(t)}) + (1-\pzo) \fwd{\lambda}_n^{(t)} \fwd{\pi}_n^{(t)}} {(1-\fwd{\lambda}_n^{(t)})(1 - \fwd{\pi}_n^{(t)}) + \fwd{\lambda}_n^{(t)} \fwd{\pi}_n^{(t)}} \quad \forall n$}	& \algcnt{}\\[2ex]
	&\multicolumn{4}{l}{$\quad \fwd{\eta}_n^{(t+1)} = (1-\alpha) \Big(\frac{\fwd{\kappa}_{n}^{(t)} \fwd{\psi}_{n}^{(t)}}{\fwd{\kappa}_{n}^{(t)} + \fwd{\psi}_{n}^{(t)}}\Big) \Big(\frac{\fwd{\eta}_{n}^{(t)}}{\fwd{\kappa}_{n}^{(t)}} + \frac{\fwd{\xi}_{n}^{(t)}}{\fwd{\psi}_{n}^{(t)}}\Big) + \alpha \zeta \quad \forall n$}	& \algcnt{}\\[2ex]
	&\multicolumn{4}{l}{$\quad \fwd{\kappa}_{n}^{(t+1)} = (1-\alpha)^2 \Big(\frac{\fwd{\kappa}_{n}^{(t)} \fwd{\psi}_{n}^{(t)}}{\fwd{\kappa}_{n}^{(t)} + \fwd{\psi}_{n}^{(t)}}\Big) + \alpha^2 \rho \quad \forall n$}	& \algcnt{}\\
	\multicolumn{5}{|l}{$\textsf{end}$}&\\
	\hline

\end{tabular}

\caption{The DCS-AMP algorithm (filtering mode) from \cite{schniter_track_jp}. Using notation from \cite{schniter_track_jp}, $n=1,2,\dots, N$ denotes the indices of $x_t$ and $m=1,2, \dots, M$ denotes the indices of $y_t$. As explained in \cite{schniter_track_jp}, $\epsilon = 10^{-7}$, $\tau=0.99$, $I=25$. The model parameters are estimated using the EM algorithm given in Table \ref{tab:em_updates}. The algorithm output at time $t$ is $\hat{x}^{(t)}$.}
\label{tab:algorithm_equations}
\label{dcs_amp}
\end{table}

\begin{table}
\centering
\scriptsize
\setlength{\tabcolsep}{2pt}
\setlength{\belowcaptionskip}{2ex}
\begin{tabular}{|llr|}
	\hline
	\multicolumn{3}{| l |}{$\textsf{\% Define key quantities obtained from AMP-MMV at iteration } k \textsf{:}$}\\
	$\text{E}\big[s_n^{(t)} \big| \ovec{y}\big] = \frac{\big(\fwd{\lambda}_n^{(t)} \fwd{\pi}_n^{(t)} \bwd{\lambda}_n^{(t)}\big)}{\big(\fwd{\lambda}_n^{(t)} \fwd{\pi}_n^{(t)} \bwd{\lambda}_n^{(t)} + (1\!-\!\fwd{\lambda}_n^{(t)})(\!1-\!\fwd{\pi}_n^{(t)})(\!1-\!\bwd{\lambda}_n^{(t)})\big)}$ & & (Q1)\\[2ex]
	$\text{E}\big[s_n^{(t)} s_n^{(t-1)} \big| \ovec{y}\big] = p\big(s_n^{(t)}=1, s_n^{(t-1)}=1 \big| \ovec{y}\big)$ & & (Q2)\\[2ex]
	$\tilde{v}_n^{(t)} \triangleq \text{var}\{\theta_n^{(t)} | \ovec{y}\} = \left(\frac{1}{\fwd{\kappa}_n^{(t)}} + \frac{1}{\fwd{\psi}_n^{(t)}} + \frac{1}{\bwd{\kappa}_n^{(t)}} \right)^{-1}$ & & (Q3)\\[2ex]
	$\tilde{\mu}_n^{(t)} \triangleq \text{E}[\theta_n^{(t)} | \ovec{y}] = \tilde{v}_n^{(t)} \cdot  \left(\frac{\fwd{\eta}_n^{(t)}}{\fwd{\kappa}_n^{(t)}} + \frac{\fwd{\xi}_n^{(t)}}{\fwd{\psi}_n^{(t)}} + \frac{\bwd{\eta}_n^{(t)}}{\bwd{\kappa}_n^{(t)}} \right)$ & & (Q4)\\[2ex]
 	$v_n^{(t)} \triangleq \text{var}\big\{x_n^{(t)} \big| \ovec{y}\big\} \qquad \text{\% See \eqref{amp_v_defn} of \tabref{algorithm_equations}}$ & & \\
	$\mu_n^{(t)} \triangleq \text{E}\big[x_n^{(t)} \big| \ovec{y}\big] \qquad \quad \text{\% See \eqref{amp_mu_defn} of \tabref{algorithm_equations}}$ & & \\[1ex]
 	\hline
	\multicolumn{3}{| l |}{$\textsf{\% EM update equations:}$}\\
	$\lambda^{k+1} = \tfrac{1}{N} \sum_{n=1}^{N} \text{E}\big[s_n^{(1)} \big| \ovec{y}\big]$ & & \emcnt{} \\[1ex]

	$\pzo^{k+1} = \frac{\sum_{t=2}^{T} \sum_{n=1}^{N} \text{E}\big[s_n^{(t\!-\!1)} \big| \ovec{y}\big] - \text{E}\big[s_n^{(t)} s_n^{(t\!-\!1)} \big| \ovec{y}\big] } {\sum_{t=2}^{T} \sum_{n=1}^{N} \text{E}\big[s_n^{(t\!-\!1)} \big| \ovec{y}\big] }$ & & \emcnt{} \\[2ex]
	
	$\zeta^{k+1} = \left( \tfrac{N(T-1)}{\rho^k} + \tfrac{N}{(\sigma^2)^k} \right)^{-1} \Big( \tfrac{1}{(\sigma^2)^k} \sum_{n=1}^N \tilde{\mu}_n^{(1)}$ & & \\[1ex]
	$\qquad \quad + \sum_{t=2}^T \sum_{n=1}^N \tfrac{1}{\alpha^k \rho^k} \big(\nt{\tilde{\mu}} - (1 - \alpha^k) \tilde{\mu}_n^{(t-1)}\big) \Big)$ & & \emcnt{} \\[1ex]
	
	$\alpha^{k+1} = \tfrac{1}{4N(T-1)} \Big(\mf{b} - \sqrt{\mf{b}^2 + 8N(T-1) \mf{c}}\Big)$ & & \emcnt{} \\[1ex]
	\quad where: & & \\[0ex]
	$\quad \mf{b} \triangleq \tfrac{2}{\rho^k} \sum_{t=2}^T \sum_{n=1}^N \mf{Re}\big\{ E[{\nt{\theta}}^{*} \theta_n^{(t-1)} | \ovec{y}] \big\}$  & & \\[0ex]
	$\qquad \qquad - \mf{Re}\{(\nt{\tilde{\mu}} - \tilde{\mu}_n^{(t-1)})^{*} \zeta^k\} - \tilde{v}_n^{(t-1)} - |\tilde{\mu}_n^{(t-1)}|^2$ & & \\[0ex]
	$\quad \mf{c} \triangleq \tfrac{2}{\rho^k} \sum_{t=2}^T \sum_{n=1}^N \nt{\tilde{v}} + |\nt{\tilde{\mu}}|^2 + \tilde{v}_n^{(t-1)} + |\tilde{\mu}_n^{(t-1)}|^2$ & & \\[0ex]
	$\qquad \qquad - 2 \mf{Re}\big\{ E[{\nt{\theta}}^{*} \theta_n^{(t-1)} | \ovec{y}] \big\}$ & & \\[0ex]
	$\rho^{k+1} = \tfrac{1}{(\alpha^k)^2 N (T-1)} \sum_{t=2}^T \sum_{n=1}^N \nt{\tilde{v}} + |\nt{\tilde{\mu}}|^2$ & & \\[0ex]
	$\qquad \qquad  + (\alpha^k)^2 |\zeta^k|^2 - 2 (1 - \alpha^k) \mf{Re}\big\{ E[{\nt{\theta}}^{*} \theta_n^{(t-1)} | \ovec{y}] \big\}$ & & \\[0ex]
	$\qquad \qquad - 2 \alpha^k \mf{Re}\big\{ \tilde{\mu}_n^{(t)*} \zeta^k \big\} + 2 \alpha^k (1-\alpha^k) \mf{Re}\big\{ \tilde{\mu}_n^{(t-1)*} \zeta^k \big\}$ & & \\[0ex]
	$\qquad \qquad + (1 - \alpha^k) (\tilde{v}_n^{(t-1)} + |\tilde{\mu}_n^{(t-1)}|^2)$ & & \emcnt{} \\[0ex]
	$(\sigma_e^2)^{k+1} = \tfrac{1}{TM} \left( \sum_{t=1}^T \|\vect{y} - \vec{A} \vect{\mu}\|^2 + \vec{1}_N^T \vect{v} \right)$ & & \emcnt{} \\[0ex]
	\hline
\end{tabular}
\caption{EM update equations \cite[Table III]{schniter_track_jp} for the signal model parameters used for the DCS-AMP algorithm from \cite{schniter_track_jp}. As explained in \cite{schniter_track_jp}, at iteration $k$, the DCS-AMP algorithm from Table \ref{tab:algorithm_equations} is first run and its outputs are used for the $k$-th EM iteration steps given above. As explained in \cite[Section VI]{schniter_track_jp}, for the first EM iteration, one initializes $\alpha$ using equation (12) of \cite{schniter_track_jp} while the other parameters are initialized using the approach of \cite[Section V]{schniter_EM_AMP}.}
\label{tab:em_updates}
\end{table}

\begin{algorithm}[t!]
\caption{{\bf \small Dynamic Modified-CS -noisy}}
%
At $t=0$: Solve BP-noisy with sufficient measurements, i.e., compute $\xhat_0$ as the solution of $\min_{b} \|b\|_1 \  \text{s.t.} \ \|y - A_0 b\|_2 \le \eps$ and compute its support: $\Nhat_0 = \{i: |(\xhat_{0})_i| > \alpha \}$.

For $t > 0$ do
\ben
\item Set $\T = \Nhat_{t-1}$
\item {\em Modified-CS -noisy. }  Compute $\xhat_t$ as the solution of
\[
\min_{b} \|b_{\T^c}\|_1 \  \text{s.t.} \ \|y -  A_t b\|_2 \le \eps
\]

\item {\em Support Estimation - Simple Thresholding. }
\bea
\Nhat_t =  \{i: |(\xhat_t)_i| > \alpha \}
\eea

\een
Dynamic modified-CS-Add-LS-del: replace the support estimation step by the Add-LS-Del procedure of (\ref{addLSdel}).
\label{modcs_algorithm}
\end{algorithm}

\subsection{Error stability over time}\label{error_stab}
It is easy to apply the results from the previous two sections to obtain exact recovery conditions or the noisy case error bounds at a given time $t$. In the noise-free case, these directly also imply error stability at all times. For example, consider dynamic Modified-CS given in Algorithm \ref{modcs_algorithm}. A direct corollary of Theorem \ref{modcs_exact_rip} is the following.
\begin{corollary}[dynamic Modified-CS  exact recovery]
Consider recovering $x_t$ with support $\N_t$ from $y_t := A_t x_t$ using Algorithm \ref{modcs_algorithm} with $\eps=0$ and $\alpha=0$. Let $u_t:=|\N_t \setminus \N_{t-1}|$,  $e_t = |\N_{t-1} \setminus \N_t|$ and $s_t =  |\N_t|$. If $\delta_{2s_0}(A_0) \le 0.2$ and if, for all $t > 0$, $\delta_{s_t+u_t+e_t}(A_t) \le 0.2$, then $\xhat_t = x_t$ (exact recovery is achieved) at all times $t$. 
\end{corollary}

Similar corollaries can be obtained for Theorem \ref{well1_exact_rip} and  Theorem \ref{well1_weakthresh} for weighted-$\ell_1$. For the noisy case, the error bounds are functions of $|\T_t|,|\Delta_{u,t}|$ (or equivalently of $|\N_t|, |\Delta_{u,t}|, |\Delta_{e,t}|$). The bounds are small as long as $|\Delta_{e,t}|$ and $|\Delta_{u,t}|$ are small for a given $|\N_t|$. But, unless we obtain conditions to ensure time-invariant bounds on $|\Delta_{e,t}|$ and $|\Delta_{u,t}|$, the size of these sets may keep growing over time resulting in instability. We obtain these conditions in the error stability over time results described next.
%
So far, such results exist only for LS-CS \cite[Theorem 2]{just_lscs} and for dynamic modified-CS-noisy summarized in Algorithm \ref{modcs_algorithm}. \cite{stability_allerton,stab_jinchun,stab_jinchun_jp}. We give here two sample results for the latter from Zhan and Vaswani \cite{stab_jinchun_jp}.
The first result below does not assume anything about how signal values change while the second assumes a realistic signal change model. 
%

\begin{theorem}[{modified-CS  error stability: no signal model \cite{stab_jinchun_jp}}]
\label{theorem_modcs_simplified}
Consider recovering $x_t$'s from $y_t$ satisfying (\ref{obsmod}) using Algorithm \ref{modcs_algorithm}.
Assume that the support size of $x_t$ is bounded by $s$ and that there are at most $s_a$ additions and at most $s_a$ removals at all times. 
If
\ben
\item {\em (support estimation threshold) }  $\alpha =  7.50 \eps$, 

\item {\em (number of measurements)} 
$\delta_{s + 6s_a}(A_t)  \le 0.207$,
\label{measmodel_simplified}

\item {\em (number of small magnitude entries)}
$| \{i\in \N_t:  |(x_t)_i| \leq \alpha+ 7.50 \eps\} | \le s_a$
\label{minmag_simplified}
\item {\em (initial time)} at $t=0$, $n_0$ is large enough to ensure that $|\N_0 \setminus \Nhat_0|=0$, $|\Nhat_0 \setminus \N_0|=0$,
\label{initass_modcs_simplified}
\een
then for all $t$,
\bi
\item $|\Delta_{u,t}|\leq 2s_a$, $|\Delta_{e,t}| \leq s_a$, $|\T_t|\leq s$,
\item $\|x_t-\hat{x}_t\|\leq 7.50\epsilon$,
\item $|\N_t \setminus \Nhat_t|\leq s_a$, $|\Nhat_t \setminus \N_t|=0$, $|\tT_t| \leq s$
\ei
\end{theorem}
The above result is the most general, but it does not give us practical models on signal change that would ensure the required upper bound on the number of small magnitude entries.
Next we give one realistic model on signal change followed by a stability result for it.
Briefly, this model says the following. At any time $t$, $x_t$ is a sparse vector with support set $\N_t$ of size $s$ or less. At most $s_a$ elements get added to the support at each time $t$ and at most $s_a$ elements get removed from it. At time $t$, a new element $j$ gets added at an initial magnitude $a_{j}$. 
Its magnitude increases for the next $d_j$ time units with $d_j \ge d_{\min} > 0$. Its magnitude increase at time $\tau$ is $r_{j,\tau}$. Also, at each time $t$, at most $s_a$ elements out of the ``large elements" set (defined in the signal model) leave the set and begin to decrease. These elements keep decreasing and get removed from the support in at most $b$ time units. In the model as stated above, we are implicitly allowing an element $j$ to get added to the support at most once. In general, $j$ can get added, then removed and then added again. To allow for this, we let $\tjset$ be the {\em set} of time instants at which $j$ gets added; we replace $a_j$ by $a_{j,t}$ and we replace $d_{j}$ by $d_{j,t}$ (both of which are nonzero only for $t \in \tjset$).
\begin{sigmodel}[Model on signal change over time (parameter $\ell$)]
Assume the following model on signal change
\ben
\item At $t=0$, $|\N_0| = s_0$. 
\item At time $t$, $s_{a,t}$ elements are added to the support set. Denote this set by $\mathcal{A}_t$. A new element $j$ gets added to the support at an initial magnitude $a_{j,t}$ and its magnitude increases for at least the next $d_{\min}$ time instants. At time $\tau$ (for $t < \tau \le t+ d_{\min}$), the magnitude of element $j$ increases by $r_{j,\tau}$.
\\
Note: $a_{j,t}$ is nonzero only if element $j$ got added at time $t$, for all other times, we set it to zero.

\item For a given scalar $\ell$, define the ``large set" as
\[
\mathcal{L}_t(\ell):=\{j \notin \cup_{\tau=t-d_{\min}+1}^t \mathcal{A}_\tau: |(x_t)_j| \ge \ell \}.
\]
    Elements in $\mathcal{L}_{t-1}(\ell)$ either remain in $\mathcal{L}_{t} (\ell)$ (while increasing or decreasing or remaining constant) or decrease enough to leave $\mathcal{L}_{t} (\ell)$.
   At time $t$, we assume that $s_{d,t}$ elements out of $\mathcal{L}_{t-1}(\ell)$ decrease enough to leave it. All these elements continue to keep decreasing and become zero (removed from support) within at most $b$ time units. 

\item At all times $t$, $0 \le s_{a,t} \le s_a$, $0 \le s_{d,t} \le \min\{s_a,|\mathcal{L}_{t-1}(\ell)|\}$, and the support size, $s_t:=|\N_t| \le s$ for constants $s$ and $s_a$ such that $s + s_a \le m$.
\een
\label{sigmodgen}
\end{sigmodel}
Notice that an element $j$ could get added, then removed and added again later. Let
$$\tjset:=\{t: a_{j,t}\neq 0 \}$$
denote the set of time instants at which the index $j$ got added to the support. Clearly, $\tjset=\emptyset$ if $j$ never got added.
Let
$$a_{\min}:= \min_{j:\tjset\neq \emptyset} \min_{t \in \tjset, t > 0} a_{j,t}$$
denote the minimum of $a_{j,t}$ over all elements $j$ that got added at $t>0$. We are excluding coefficients that never got added and those that got added at $t=0$.
Let
$$r_{\min}(d):= \mathop{\min}_{j: \tjset\neq \emptyset} \min_{t \in \tjset, t > 0}\min_{\tau \in[t+1, t+d]} r_{j,\tau}$$
denote the minimum, over all elements $j$ that got added at $t>0$, of the minimum of $r_{j,\tau}$ over the first $d$ time instants after $j$ got added.

\begin{theorem}[dynamic Modified-CS  error stability] \cite{stab_jinchun_jp}
Consider recovering $x_t$'s from $y_t$ satisfies (\ref{obsmod}) using Algorithm \ref{modcs_algorithm}. Assume that Model \ref{sigmodgen} on $x_t$ holds with $$\ell = a_{\min}+d_{\min}r_{\min}(d_{\min})$$
where $a_{\min}$, $r_{\min}$ and the set $\tjset$ are defined above.
If there exists a $d_0 \le d_{\min}$ such that the following hold:
\ben
\item (algorithm parameters)  $\alpha =  7.50 \eps$,
\item (number of measurements)  $\delta_{s+ 3(b + d_0 + 1)s_a}(A_t) \le 0.207$,

\item (initial magnitude and magnitude increase rate)
\bea
&& \min\{ \ell,\min_{j: \tjset \neq \emptyset} \ \min_{t \in \tjset} (a_{j,t}+\sum_{\tau=t+1}^{t+d_0}r_{j,\tau}) \} \nn \\
&& > \alpha + 7.50 \eps, \nn
%
\eea

\item at $t=0, n_0$ is large enough to ensure that $|\N_0 \setminus \Nhat_0|=0$, $|\Nhat_0 \setminus \N_0|=0$, 
\een
then, for all $t$,
\bi
\item $|\Delta_{u,t}| \le b s_a + d_0 s_a + s_a$, $|\Delta_{e,t}|\le s_a$, $|\T_t| \le s$,
\item $\|x_t - \xhat_t\| \le 7.50 \eps$,
\item $|\N_t \setminus \Nhat_t| \le b s_a + d_0 s_a$, $|\Nhat_t \setminus \N_t|= 0$, $|\tT_t| \le s$
\ei
\label{modcsthm}
\end{theorem}
Results similar to the above two results also exist for dynamic modified-CS-Add-LS-Del \cite{stab_jinchun_jp}. Their main advantage is that they require a weaker condition 3).

\subsubsection{Discussion}
Notice that $s$ is a bound on the support size at any time. As long as the number of new additions or removals, $s_a \ll s$, i.e., slow support change holds, the above result shows that the worst case number of misses or extras is also small compared to the support size. This makes it a meaningful result. The reconstruction error bound is also small compared to the signal energy as long as the signal-to-noise ratio is high enough ($\eps^2$ is small compared to $\|x_t\|^2$). 
Observe that both the above results need a bound on the RIC of $A_t$ of order $s+\mathcal{O}(s_a)$. On the other hand, BP-noisy needs the same bound on the RIC of $A_t$ of order $2s$ (see Theorem \ref{BPthm}). This is stronger when $s_a \ll s$ (slow support change).

As explained in \cite{stab_jinchun_jp}, Model \ref{sigmodgen} allows for both slow and fast signal magnitude increase or decrease. Slow magnitude increase and decrease would happen, for example, in an imaging problem when one object slowly morphs into another with gradual intensity changes. Or, in case of brain regions becoming ``active" in response to stimuli, the activity level gradually increases from zero to a certain maximum value within a few milliseconds (10-12 frames of fMRI data), and similarly the ``activity" level decays to zero within a few milliseconds. In both of the above examples, a new coefficient will get added to the support at time $t$ at a small magnitude $a_{j,t}$ and increase by $r_{j,t}$ per unit time for sometime after that. A similar thing will happen for the decay to zero of the brain's activity level. 
On the other hand, the model also allows support changes resulting from motion of objects, e.g., translation. In this case, the signal magnitude changes will typically not be slow. As the object moves, a set of new pixels enter the support and another set leave. The entering pixels may have large enough pixel intensity and their intensity may never change. For our model, this means that the pixel enters the support at a large enough initial magnitude $a_{j,t}$ but its magnitude never changes i.e., $r_{j,t} = 0$ for all $t$. If all pixels exit the support without their magnitude first decreasing, then $b = 1$.

The only thing that the above result requires is that (i) for any element $j$ that is added, either $a_{j,t}$ is large enough or $r_{j,t}$ is large enough for the initial few ($d_0$) time instants so that condition 3) holds; and (ii) a decaying coefficient decays to zero within a short delay, $b$. Condition (i) ensures that every newly added support element gets detected either immediately or within a finite delay; while (ii) ensures removal within finite delay of a decreasing element. For the moving object case, this translates to requiring that $a_{j,t}$ be large enough. For the morphing object example or the fMRI activation example, this translates to requiring that $r_{j,t}$ be large enough for the first $d_0$ frames after index $j$ gets added and $b$ be small enough.%


\newcommand{\hhat}{{\hat{\mathbf{h}}}}
\newcommand{\vecx}{{\mathbf{x}}}
\newcommand{\vecd}{{\mathbf{d}}}
\newcommand{\matx}{{\mathbf{X}}}
\newcommand{\h}{{\mathbf{h}}}
\newcommand{\ai}{{\mathbf{a}_i}}
\newcommand{\aaa}{{\mathbf{a}}}

\section{Tracking-based and adaptive-filtering-based solutions for recursive dynamic CS} \label{track_adapt_filt}
As explained in Section \ref{motiv}, it is possible to use ideas from the tracking literature or the adaptive filtering literature along with a sparsity constraint to get solutions for the recursive dynamic CS problem. These only use the slow signal value change assumption and sparsity without explicitly using slow support change. We describe the tracking-based solutions next followed by the adaptive filtering based solutions.

\subsection{Tracking-based recursive dynamic CS solutions} 
We discuss here two tracking-based solutions.

\subsubsection{BP-residual}\label{csres_sec}
To explain the idea, we use BP-noisy. The exact same approach can also be applied to get BPDN-residual, IHT-residual or OMP-residual.
BP-residual replaces $y_t$ by the measurement residual $y_t - A \xhat_{t-1}$ in the BP or the BP-noisy program, i.e., it computes
\bea
\xhat_t \se \xhat_{t-1}  + \arg \min_\beta [ \|\beta\|_1 \ \text{s.t.} \ \|y_t - A \xhat_{t-1} - A \beta\|_2 \le \eps ] \ \ \ \ \ \ 
\label{CSres}
\eea
with setting $\eps=0$ in the noise-free case. This is related to BP on observation differences idea used in \cite{reddy}.
%
Observe that this is using the assumption that the difference $(x_t - x_{t-1})$ is small. 
However, if $x_t$ and $x_{t-1}$ are $k$-sparse, then $(x_t - x_{t-1})$ will also be at least $k$-sparse unless the difference is exactly zero along one or more coordinates. There are very few practical situations where one can hope to get perfect prediction along a few dimensions. One possible example is the case of coarsely quantized $x_t$'s. In the results known so far, the number of measurements required for exact sparse recovery depend only on the support size of the sparse vector, e.g., see \cite{decodinglp,candes_rip,iht_blum,iht_blum2}. Thus, when using BP-residual, this number will be as much or more than what BP needs. This is also observed in Monte Carlo based computations of the probability of exact recovery in Fig. \ref{phasetrans}. As can be seen, both BP and BP-residual need the same $n$ for exact recovery with Monte Carlo probability equal to one. This number is much larger than what modified-CS and weighted-$\ell_1$, that exploit slow support change, need.
On the other hand, as we will see in Fig. \ref{nmse_larynx}, for compressible signal sequences, its performance is not much worse than that of approaches that do explicitly use slow support change.

\subsubsection{Pseudo-measurement based CS-KF (PM-CS-KF)}
In \cite{ibm}, Carmi et al. introduced the pseudo-measurement based CS-KF (PM-CS-KF) algorithm. It uses an indirect method called the pseudo-measurement (PM) technique \cite{pm_tech} to include the sparsity constraint while trying to minimize the estimation error in the KF update step. To be precise, it uses PM to approximately solve the following
$$\min_{\xhat_{k|k}} \E_{x_k|y_1,y_2, \dots y_k}[\|x_k - \xhat_{k|k}\|_2^2] \ \text{s.t.} \ \|\xhat_{k|k}\|_1 \le \eps$$
The idea of PM is to replace the constraint $\|\xhat_{k|k}\|_1 \le \eps$ by a linear equation of the form
$$\tilde{H} x_k - \eps = 0$$
where $\tilde{H}= \text{diag}\left(\text{sgn}((x_k)_1), \text{sgn}((x_k)_2), \dots \text{sgn}((x_k)_m)\right)$
and $\eps$ serves as measurement noise. It then uses an extended KF approach iterated multiple times to enforce the sparsity constraint. As explained in  \cite{ibm}, the covariance of the pseudo noise $\eps$, $R_{\eps}$, is a tuning parameter that can be chosen in the same manner as the process noise covariance is determined for an extended KF.
The complete algorithm \cite[Algorithm 1]{ibm} is summarized in Algorithm \ref{pmcskf_algo} (with permission from the authors). The code is not posted on the authors' webpages but is available from this article's webpage \url{http://www.ece.iastate.edu/~namrata/RecReconReview.html}.

\begin{algorithm}[t]
\caption{\bf \small PM-CS-KF (Algorithm 1 of \cite{ibm}} \label{pmcskf_algo}
\begin{algorithmic}[1]
\STATE Prediction
\begin{equation}
\begin{array}{ll}
&\hat{z}_{k+1|k} = A\hat{z}_{k|k}\\
&P_{k+1|k} = AP_{k|k}A^T + Q_k
\end{array}
\end{equation}
\STATE Measurement Update
\begin{equation}
\begin{array}{ll}
&K_k = P_{k+1|k}H'^T\left(H'P_{k+1|k}H'^T + R_k\right)^{-1}\\
&\hat{z}_{k+1|k+1} = \hat{z}_{k+1|k} + K_k(y_k - H'\hat{z}_{k+1|k})\\
&P_{k+1|k+1} = (I-K_kH')P_{k+1|k}
\end{array}
\end{equation}
\STATE CS Pseudo Measurement: Let $P^1 = P_{k+1|k+1}$ and $\hat{z}^1 = \hat{z}_{k+1|k+1}$.
\STATE for $\tau = 1, 2, \cdots, N_\tau-1$ iterations do
\STATE \begin{equation}
\begin{array}{ll}
&\bar{H}_\tau = [sign(\hat{z}^\tau(1)), \cdots, sign(\hat{z}^{\tau}(n))]\\
&K^\tau = P^\tau\bar{H}_\tau^T\left(\bar{H}_\tau P^\tau\bar{H}_\tau^T + R_\epsilon\right)^{-1}\\
&\hat{z}^{\tau+1} = (I-K^\tau\bar{H}_\tau)\hat{z}^\tau\\
& P^{\tau+1} = (I-K^\tau\bar{H}_\tau)P^\tau.
\end{array}
\end{equation}
\STATE end for
\STATE Set $P_{k+1|k+1} = P^{N_\tau}$ and $\hat{z}_{k+1|k+1} = \hat{z}^{N_\tau}$.
\end{algorithmic}
\end{algorithm}


\subsection{Sparsity-aware adaptive filtering and its application to recursive dynamic CS} \label{sparse_adaptive}
A problem related to the recursive dynamic CS problem is that of sparsity-aware adaptive filtering. This is, in fact, a special case of a more general problem of recovering a single sparse signal from sequentially arriving measurements. 
The general problem is as follows. An unknown sparse vector $\h$ needs to be recovered from sequentially arriving measurements
\bea \label{genprob}
d_i: = \ai'\h + v_i, i = 1,2, \dots, n
\eea
To connect this with our earlier notation, $y:=[d_1, d_2, \dots d_n]'$, $A:= [\aaa_1, \aaa_2, \dots \aaa_n]'$, $x:=\h$ and $w:=[v_1, v_2, \dots v_n]'$.
We use different notation in this section so as to keep it consistent with what is used in the adaptive filtering literature.  
Let $\hhat^i$ denote the estimate of $\h$ based on the first $i$ measurements. The goal is to obtain $\hhat^i$ from $\hhat^{i-1}$ without re-solving the sparse recovery problem. Starting with the works of Malioutov et al. \cite{seqcs_wilsky} and Garrigues et al. \cite{ghaoui}, various homotopy based solutions have been introduced in recent literature to do this \cite{romberg,giannakis}. 
Some of these works also provide a stopping criterion that tells the user when to stop taking new measurements. 



A special case of the above problem is sparsity-aware adaptive filtering for system identification (e.g., estimating an unknown communication channel). In this case, a known discrete time signal $x[i]$ is sent through an unknown linear time-invariant system (e.g., a communication channel) with impulse response denoted by $\h$. The impulse response vector $\h$ is assumed to be sparse.
One can see the output signal $d[i]$. The goal is to keep updating the estimates of $\h$ on-the-fly so that the output after passing $x[i]$ through the estimated system/channel is close to the observed output $d[i]$.
Let $x[i]$ denote a given input sequence and define the vector
$$\vecx[i]:= [x[i], x[i-1], x[i-2], \dots x[i-m+1]]'$$
Then sparsity-aware adaptive filtering involves recovering $\h$ from sequentially arriving $d[i]$ satisfying (\ref{genprob}) with $d_i = d[i]$ and $\ai = \vecx[i]$.
%
This problem has been studied in a sequence of recent works. 
One of the first solutions to this problem, called zero-attracting least mean squares (ZA-LMS) \cite{chen2009sparse,jin2010stochastic}, modifies the standard LMS algorithm by including an $\ell_1$-norm constraint in the cost function. Let $\hhat^i$ denote the estimate of $\h$ at time $i$ (based on the past $i$ measurements) and let
$$e[i]:=d[i] -  \ai'\hhat^i.$$ 
with $\ai = \vecx[i]$.
At time $i$, regular LMS takes one gradient descent step towards minimizing $L(i) = 0.5 e[i]^2$. ZA-LMS does almost the same thing for $L(i) = 0.5 e[i]^2 + \gamma  \|\hhat^i\|_1$. 
Let $\mu$ denote the step size. Then, ZA-LMS computes $\hhat^{i+1}$ from $\hhat^{i}$ using
\[
\hhat^{i+1} = \hhat^{i} + \mu e[i]\ai - \mu \gamma \sgn(\hhat^i)
\]
where $\sgn(z)$ is a component-wise sign function: $(\sgn(z))_i = z_i/|z_i|$ if $z_i \neq 0$ and $(\sgn(z))_i = 0$ otherwise. The last term in the above update attracts the coefficients of the estimate of $\h$ to zero. We should point out here that the term $\|\hhat^i\|_1$ is not differentiable and hence it does not have gradient. ZA-LMS is actually replacing its gradient by one possible vector from its sub-gradient set. 
The work of Jin et al. \cite{jin2010stochastic} significantly improves the basic ZA-LMS algorithm by including a differentiable approximation to the $\ell_0$ norm to replace the $\ell_1$ norm used in the cost function for ZA-LMS. In particular they use $L(i)= 0.5 e[i]^2  + \gamma \sum_{k=1}^m(1 - \exp(-\alpha |\h_k|))$ and then derived a similar algorithm. Their algorithms were analyzed in \cite{su2012performance}.


In later work by Babadi et al. \cite{babadi2010sparls}, the SPARLS algorithm was developed for solving the sparse recursive least squares (RLS) problem. This was done for RLS with an exponential forgetting factor. At time $i$, RLS with an exponential forgetting factor computes $\hhat^i$ as the minimizer of $L(i)= \sum_{j=1}^i \lambda^{i-j} e[j]^2$ with $e[j]:=d[j] - \mathbf{a}_j'\hhat^i$ and $\mathbf{a}_j = \vecx[j]$. By defining the vector $\vecd[i]:= [d_1, d_2, \dots d_i]'$, the matrix $\matx[i]:=[\mathbf{a}_1, \mathbf{a}_2, \dots \mathbf{a}_i]'$, and a diagonal matrix $\mathbf{D}[i]:= diag(\lambda^{i-1}, \lambda^{i-2}, \dots 1)$, $L(i)$ can be rewritten as $L(i)= \|\mathbf{D}[i]^{1/2} \vecd[i] - \mathbf{D}[i]^{1/2} \matx[i] \hhat^i\|_2^2$. In the above definitions, we use $\ai = \vecx[i]$.
This is now a regular least squares (LS) problem. When solved in a recursive fashion, we get the RLS algorithm for regular adaptive filtering. Sparse RLS adds an $\ell_1$ term to the cost function. It solves
\bea
\min_{\hhat^i}  \frac{1}{2\sigma^2} \|\mathbf{D}[i]^{1/2} \vecd[i] - \mathbf{D}[i]^{1/2} \matx[i] \hhat^i\|_2^2 + \gamma \|\hhat^i\|_1
\label{sparse_rls_cost}
\eea
%
Instead of solving the above using conventional convex programming techniques, the authors of \cite{babadi2010sparls} developed a low-complexity Expectation Minimization (EM) algorithm motivated by an earlier work of Figueirado and Nowak \cite{wavelet_EM}.
Another parallel work by Angelosante et al. \cite{angelosante2010online} also starts with the RLS cost function and adds an $\ell_1$ cost to it to get the sparse RLS cost function. Their solution approach involves developing an online version of the cyclic coordinate descent algorithm. They, in fact, developed sparse RLS algorithms with various types of forgetting factors.

While all of the above algorithms are designed for a recovering a fixed sparse vector $\h$, they also often work well for situations where $\h$ is slow time-varying \cite{jin2010stochastic}. In fact, this is how they relate to the problem studied in this article. 
However, all the theoretical results for sparse adaptive filtering are for the case where $\h$ is fixed. For example, the result from \cite{babadi2010sparls} for SPARLS states the following.%
\begin{theorem}[{\cite[Theorem 1]{babadi2010sparls}}]
Assume that $\h$ is fixed and that the input sequence $x[i]$ and the output sequence $d[i]$ are realizations of a jointly stationary random process. Then the estimates $\hhat^i$ generated by the SPARLS algorithm converge almost surely to the unique minimizer of (\ref{sparse_rls_cost}).
\end{theorem}

Another class of approaches for solving the above problem involves the use of the set theoretic estimation technique \cite{murakami2010sparse, kopsinis2011online, slavakis2013generalized}. Instead of recursively trying to minimize a cost function, the goal, in this case, is to find a set of solutions that are in agreement with the available measurements and the sparsity constraints. For example, the work of \cite{kopsinis2011online} develops an algorithm that finds all vectors $\h$ that belong to the intersection of the sets
$S_j(\epsilon):= \{\h: |\vecx[j]'\h - d[j]| \le \epsilon\}$ for all $j \le i$ and the weighted-$\ell_1$ ball $B(\delta):= \{\h: \sum_{k=1}^m \omega_k |\h_k| \le \delta\}$. Thus at time $i$, it finds the set of all $\h$'s that belong to $\cap_{j=1}^i S_j(\epsilon) \cap B(\delta)$. 

Other work on the topic includes \cite{mileounis2010adaptive} (Kalman filter based sparse adaptive filter), \cite{paleologu2010efficient} (``proportionate-type algorithms" for online sparsity-aware system identification problems), \cite{lima2014sparsity} (combine sparsity-promoting schemes with data-selection mechanisms), \cite{dumitrescu2012greedy} (greedy sparse RLS), \cite{chen2012recursive} (recursive $\ell _{1,\infty}$ group LASSO), \cite{themelis2014variational} (variational Bayes framework for sparse adaptive filtering) and \cite{abdolee2014estimation, chouvardas2012sparsity} (distributed adaptive filtering).

\subsubsection{Application to recursive dynamic CS}
While the approaches described above are designed for a single sparse vector $\h$, as mentioned above, they can also be used to track a time sequence of sparse vectors that are slow time-varying. Instead of using a time-sequence of vector measurements (as in Sec. \ref{probdef}), in this case, one uses a single sequence of scalar measurements indexed by $i$. To use these algorithms for recursive dynamic CS, for time $t$, the index $i$ will vary from $nt+1$ to $nt+n$. For a given $i$, define $t(i)=\lfloor \frac{i}{n} \rfloor$ and $k(i)=i - n t$. Then one can use the above algorithms with the mapping $d_i = (y_{t(i)})_{k(i)}$ and with $\ai$ being the $k(i)$-th row of the matrix $A_t$. Also, one assumes $\h_i = x_{t(i)}$, i.e., $\h_i$ is is equal to $x_t$ for all $i=nt+1, nt+2, \dots nt+n$. The best estimate of $x_t$ is then given by $\hhat^{nt+n}$, i.e., we use $\xhat_t = \hhat^{nt+n}$. We use this formulation to develop ZA-LMS to solve our problem and show experiments with it in the next section.

We should point out that ZA-LMS or SPARLS are among the most basic sparse-adaptive-filtering (SAF) solutions in literature. In the later work cited above, their performance has been greatly improved. We have chosen to describe these because they are simple and hence easily explain the overall idea of using SAF for recursive dynamic CS.

\subsection{Pros and cons of sparse-adaptive-filtering (SAF) methods}

The advantage of the above sparse-adaptive-filtering (SAF) approach is that it is very quick and needs very little memory. Its computational complexity is linear in the signal length. Moreover, since SAF processes one scalar observation entry at a time, its performance can be significantly improved by iteratively processing the observation vector entries more than once, in randomized order as is done in the work of Jin et al. \cite{jin2010stochastic}. Finally, a disadvantage of most of the methods discussed in this article is that they need more measurements at the initial time instant ($n_0 > n_t$) so that an accurate estimate of the first signal is obtained. Since SAF methods operate on one scalar measurement at a time, as noted by a reviewer, this requirement is less stringent for SAF methods.

The disadvantage of SAF methods is that they do not explicitly leverage the slow support change assumption and hence may not work well in situations where there is some support change at every time or very frequently. However, as noted by an anonymous reviewer, this issue may not occur if the SAF approach converges fast enough. A second important disadvantage of SAF methods is that all the theoretical guarantees are for the case where $\h$ is {\em fixed}.

In general SAF methods are much faster than the other approaches discussed here. The exception is problems in which the transformation of the signal to both the sparsity basis and to the space in which measurements are taken can be computed using fast transforms. A common example is MRI with the image being assumed to be wavelet sparse. Since very fast algorithms exist for computing both the discrete Fourier and the discrete wavelet transform, for large-sized problems, the methods discussed in this work (which process a measurements' vector at a time) would be faster than SAF which process the measurements one scalar at a time.

\setlength{\tabcolsep}{0.2em}

\section{Numerical Experiments} \label{sims}
We report three sets of experiments. The first studies the noise-free case and compares the exact recovery performance of algorithms that only exploit slow support change. This problem can be reformulated as one of sparse recovery with partial support knowledge. In case of exact recovery performance, the previous signal's nonzero entries' values do not play a role and hence we only simulate the static problem with partial support knowledge. 
%
%
In the second experiment, we study the noisy case and the dynamic problem for a simulated exactly sparse signal sequence and random Gaussian measurements. In the third experiment, we study the same problem but for a real image sequence (the larynx MRI sequence shown in Fig. \ref{examples}) and simulated MRI measurements. This image sequence is only approximately sparse in the wavelet domain.
In the second and third experiment, we compare all types of algorithms - those that exploit none, one or both of slow support change and slow signal value change.

\begin{figure*}
\centerline{
\subfigure[$e=0$]{\label{phase_d0}
\includegraphics[width=6cm]{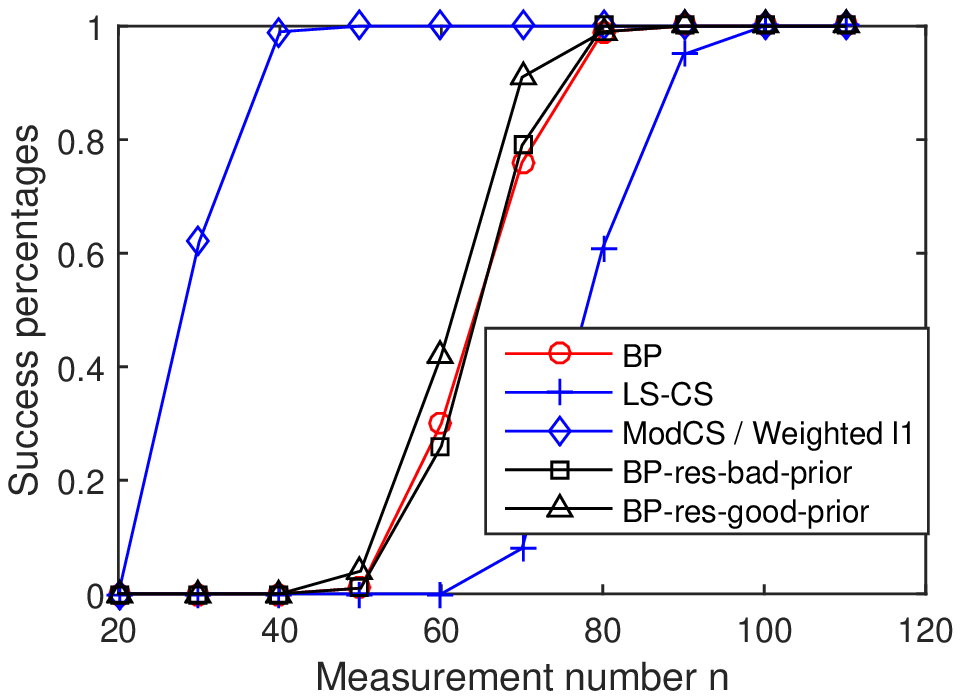} 
}
\subfigure[$e=u$]{\label{phase_d1}
\includegraphics[width=6cm]{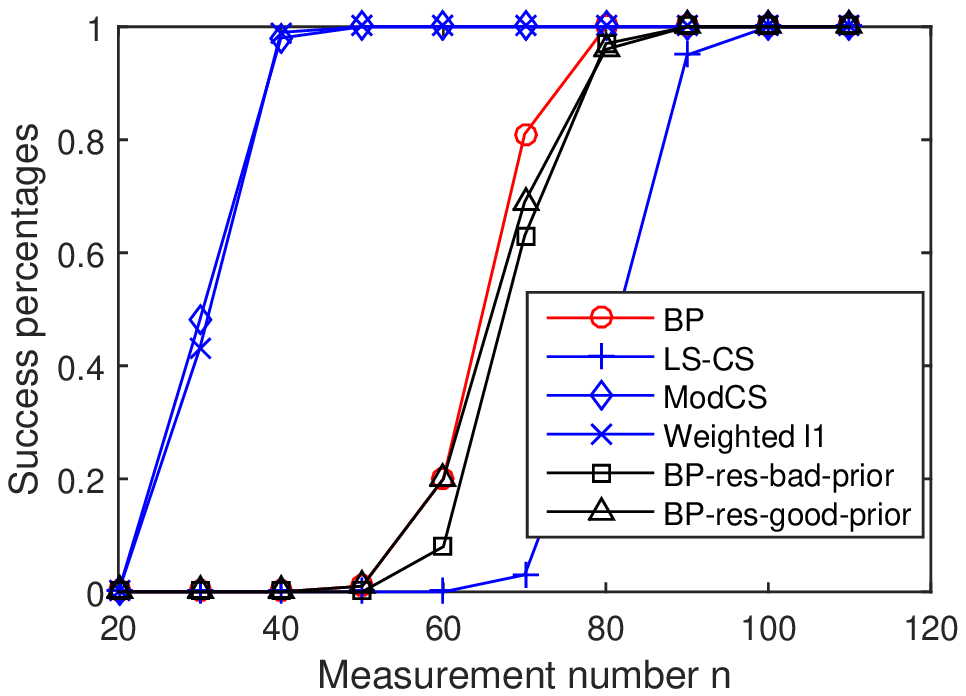} 
}
\subfigure[$e=4u$]{\label{phase_d4}
\includegraphics[width=6cm]{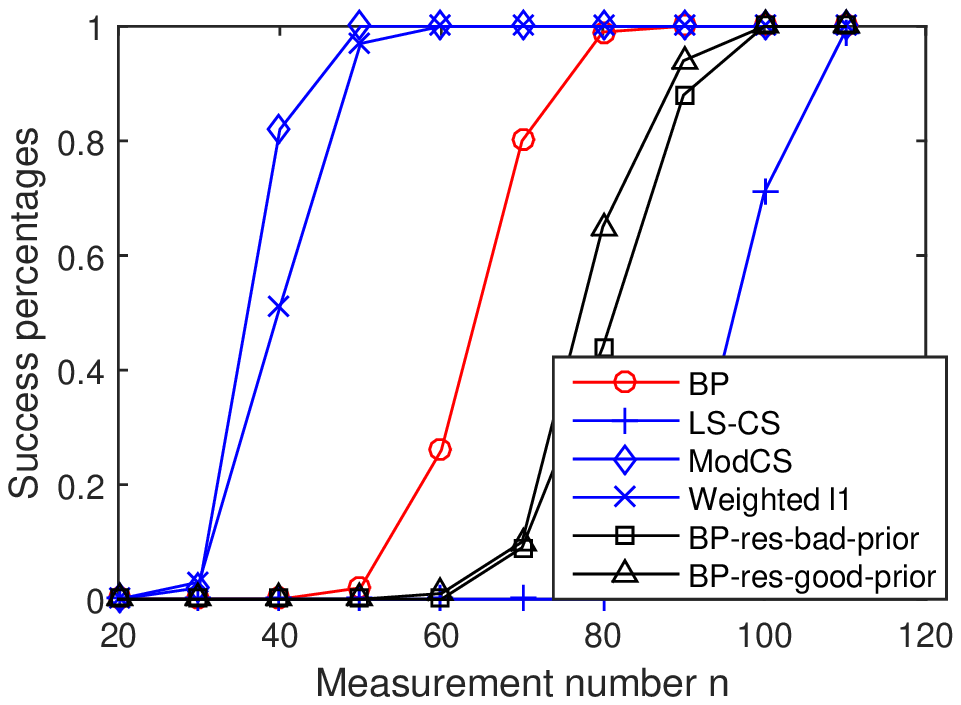}
}
}
%
\centerline{
\subfigure[BP]{\label{phase_cs}
\includegraphics[width=5cm]{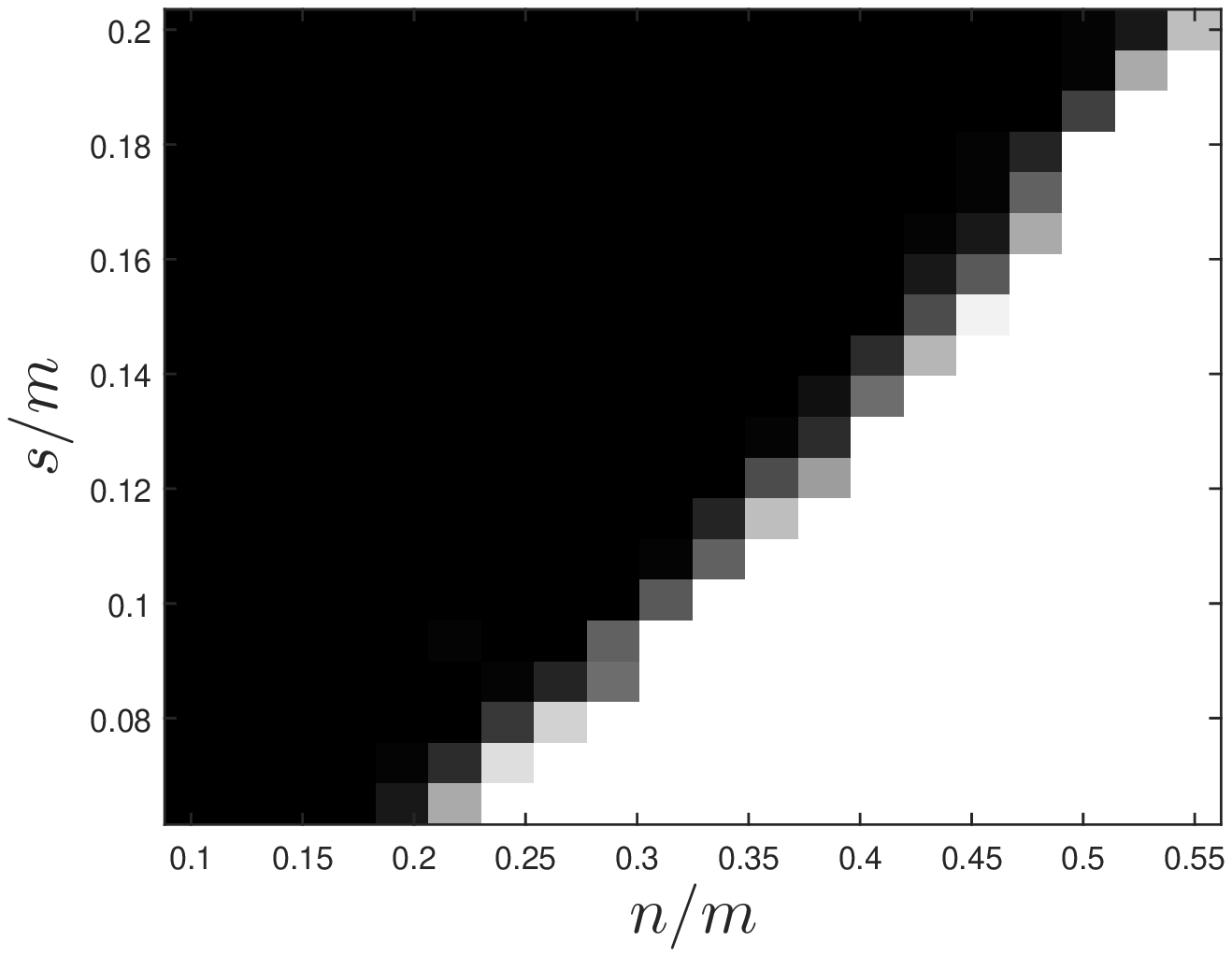} 
}
\subfigure[modified-CS, $e=4u$]{\label{phase_modcs_e4}
\includegraphics[width=5cm]{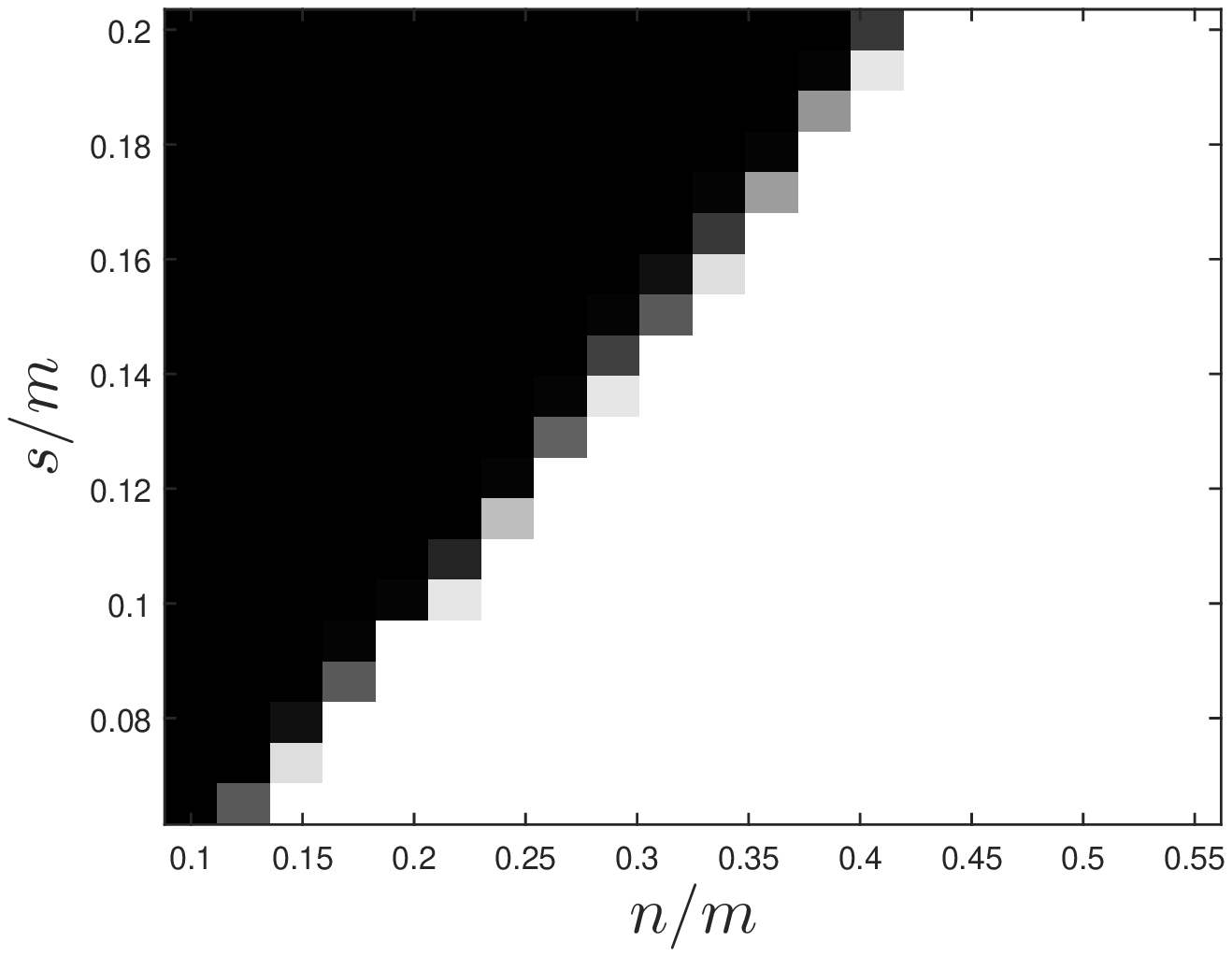} 
}
\subfigure[weighted-$\ell_1$, $e=4u$]{\label{phase_weightedl1_e4}
\includegraphics[width=5cm]{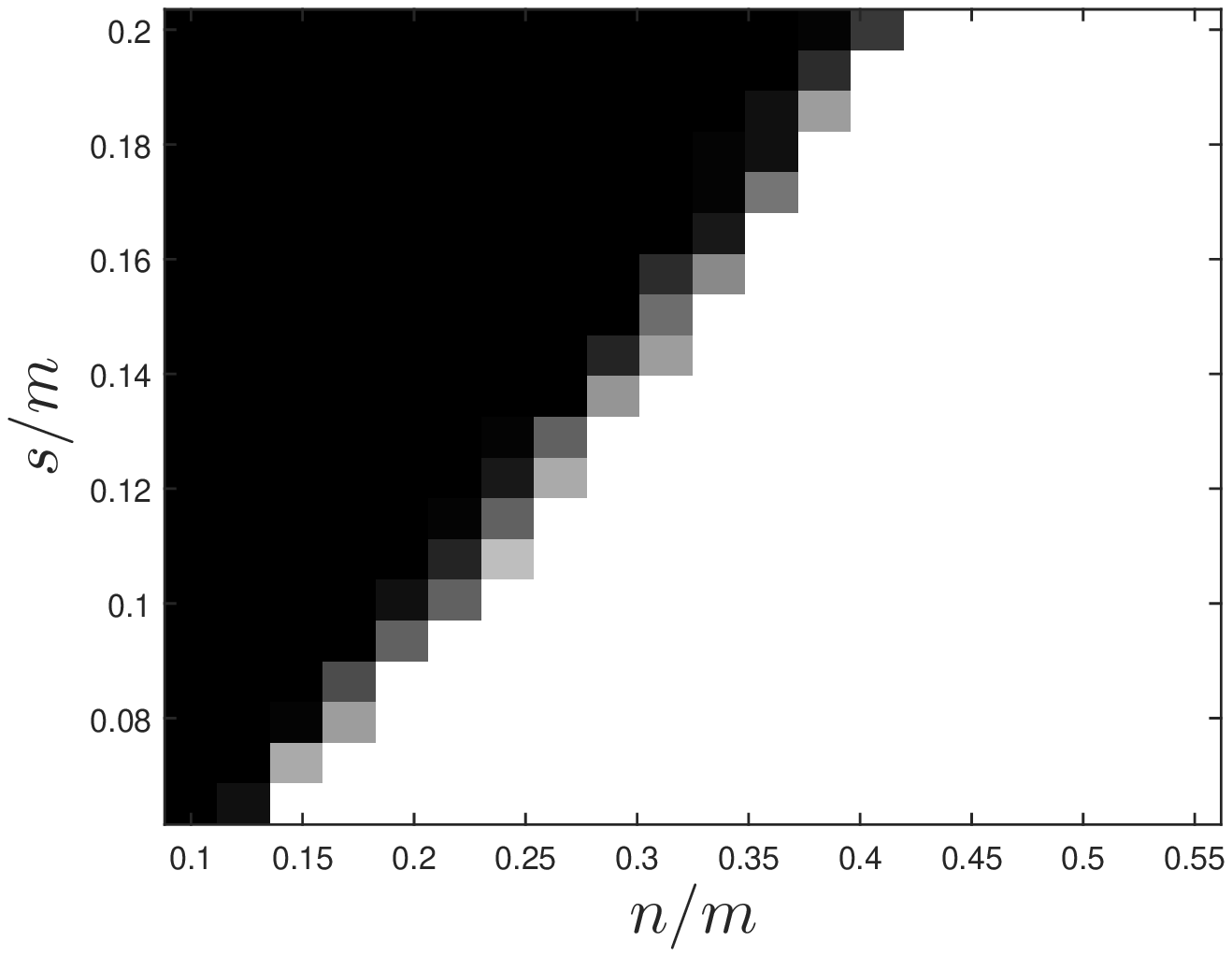} 
}
}
\caption{\small{
Phase transition plots. In all figures we used signal length $m=200$. In the top row figures, we plot the Monte Carlo estimate of the probability of exact recovery against $n$ for
$s=0.1m$, $u=0.1s$ and three values for $e$. Since the plots are 1D plots, we can compare all figures in a single figure. We compare BP, LS-CS, modified-CS, weighted-$\ell_1$ with $\tau=e/s$ and BP-residual with two types of prior signal value knowledge - good and bad.
In the ``images" shown in the second row, we display the Monte Carlo estimate of the probability of exact recovery for various values of $n$ (varied along x-axis) and $s$ (varied along y-axis). The grey scale intensity for a given $(n/m),(s/m)$ point is proportional to the computed probability of exact recovery for that $n,s$.
}}
\label{phasetrans}
\end{figure*}

\subsection{Sparse recovery with partial support knowledge: phase transition plots for comparing $n$ needed for exact recovery} \label{sims_phase}
We compare BP and BP-residual with all the approaches that only use partial support knowledge --  Modified-CS  and weighted-$\ell_1$ -- using phase transition plots shown in Fig. \ref{phasetrans}. BP-residual is an approach that is motivated by tracking literature and it uses signal value knowledge but not support knowledge. We include this in our comparison to demonstrate that it cannot use a smaller $n$ than BP for exact recovery with probability one.
For a given support size, $s$, number of misses in the support knowledge, $u$, number of extras, $e$, and number of measurements, $n$, we use Monte Carlo to estimate the probability of exact recovery of the various approaches. We generated the true support $\N$ of size $s$ uniformly at random from $\{1,2,\dots,m\}$. The nonzero entries of $x$ were generated from a Gaussian $\n(0,\sigma_x^2)$ distribution. The support knowledge $\T$ was generated as $\T = \N \cup \Delta_e \setminus \Delta_u$ where $\Delta_e$ is generated as a set of size $e$ uniformly at random from $\N^c$ and $\Delta_u$ was generated as a set of size of $u$ uniformly at random from $\N$. We generated the observation vector $y = Ax$ where $A$ is an $n \times m$ random Gaussian matrix.
%
Since BP-residual uses signal value knowledge, for it, we generate a ``signal value knowledge" $\hat\mu$ as follows. Generate $\hat\mu_{\T}  = x_{\T} + \nu_{\T}$ with $\nu \sim \n(0, \sigma_\nu^2  I)$, and set $\hat\mu_{\T^c} = 0$.  We generated two types of prior signal knowledge, the good prior case with  $\sigma_\nu^2 = 0.0001 \sigma_x^2$ and the bad prior case with $\sigma_\nu^2 = \sigma_x^2$.

Modified-CS solved (\ref{l1seqcs}) which has no parameters. Weighted-$\ell_1$  solved (\ref{well1}) with $\tau = e/s$ \cite{friedlander}. BP-residual computed $\xhat = \hat\mu + [ \arg\min_{b} \|b\|_1  \text{ s.t. }  y - A \hat\mu = A b ]$, where $\hat\mu$ was generated as above. For LS-CS, we did the same thing but $\hat\mu$ was now  $ I_{\T} A_\T^\dag y$ (LS estimate on $\T$).
All convex programs were solved using CVX.


For the first row figures of Fig. \ref{phasetrans}, we used $m = 200$, $s = 0.1m$, $u = 0.1s$, $\sigma_x^2 = 5$ and three values of $e$: $e=0$ (Fig. \ref{phase_d0}) and $e=u$ (Fig. \ref{phase_d1}) and $e = 4u$ (Fig. \ref{phase_d4}). The plot varies $n$ and plots the Monte Carlo estimate of the probability of exact recovery.
The probability of exact recovery was computed by generating 100 realizations of the data and counting the number of times $x$ was exactly recovered by a given approach. We say that $x$ is exactly recovered if $\frac{\|x-\hat{x}\|}{\|x\|} < 10^{-6}$ (precision used by CVX).
We show three cases, $e=0$, $e = u$ and $e = 4u$ in Fig. \ref{phasetrans}. In all cases we observe the following.  (1) LS-CS needs more measurements for exact recovery than either of BP or BP-residual. This is true even in the $e=0$ case (and this is something we are unable to explain).
(2) BP-residual needs as many or more measurements as BP to achieve exact recovery with (Monte Carlo) probability one. This is true even when very good prior knowledge is provided to BP-residual.
(3) Weighted-$\ell_1$ and modified-CS significantly outperform all other approaches -- they need a significantly smaller $n$ for exact recovery with (Monte Carlo) probability one.
(4) From this set of simulations, it is hard to differentiate modified-CS and weighted-$\ell_1$ for the noise-free case. It has been observed in other works \cite{friedlander,regmodbpdn} though, that weighted-$\ell_1$ has a smaller recovery error than modified-CS when $e$ is larger and the measurements are noisy. 

In the second row, we show a 2D phase transition plot that varies $s$ and $n$ and displays a grey-scale intensity proportional to the Monte Carlo estimate of the exact recovery probability. This is done to compare BP, modified-CS and weighted-$\ell_1$ in detail. Its conclusions are similar to the ones above. Notice that the white area (the region where an algorithm works with probability nearly one) is smallest for BP.%



\subsection{Recursive recovery of a simulated sparse signal sequence from noisy random-Gaussian measurements} \label{sim_seq}
We generated a sparse signal sequence, $x_t$, that satisfied the assumptions of Model \ref{sigmodgen} with $m=256$, $s=0.1m=25$, $s_a=1$, $b=4$, $d_{\min}=2$, $a_{\min} = 2$, $r_{\min} = 1$. The specific generative model to generate the $x_t$'s was very similar to the one specified in \cite[Appendix I]{stab_jinchun_jp} with one change: the magnitude of {\em all} entries in the support (not just the newly added ones) changed over time. We briefly summarize this here.
The support set $\N_t$ consisted of three subsets. The first was the ``large set", $\{i: |(x_t)_i|>a_{\min}+d_{\min}r_{\min}\}$. The magnitude of an element of this set increased by $r_{j,t}$ (for element $j$ at time $t$) until it exceeded $a_{\min} + 6d_{\min}r_{\min}$.
The second set was the ``decreasing set", $\{i: 0 < |(x_t)_i| < |(x_{t-1})_i| \text{ and } |(x_t)_i|<a_{\min}+d_{\min}r_{\min}\}$. At each time, there was a 50\% probability that some entry from the large set entered the decreasing set (the entry itself was chosen uniformly at random). All entries of the decreasing set decreased to zero within $b=4$ time units. The third set was the ``increasing set", $\{i: |(x_t)_i| > |(x_{t-1})_i| \text{ and } |(x_t)_i|<a+d_{\min}r_{\min}\}$. At time $t$, if $|\N_{t-1}|<s$, then an element out of $\N_{t-1}^c$ was selected uniformly at random to enter the increasing set. Element $j$ entered the increasing set at initial magnitude $a_{j,t}$ at time $t$. The magnitude of entries in this set increased for at least $d_{\min}$ time units with rate $r_{j,t}$ (for element $j$ at time $t$). For all $j,t$, $r_{j,t}$ was i.i.d. uniformly distributed in the interval $[r_{\min}, 2r_{\min}]$ and the initial magnitude $a_{j,t}$ was i.i.d. uniformly distributed in the interval $[a_{\min}, 2a_{\min}]$.
With the above model, $s_{d,t}$ was zero roughly half the time and so was $s_{a,t}$ except for the initial few time instants. For example, for a sample realization, $s_{a,t}$ was equal to 1 for 43 out of 100 time instants while being zero for the other 57.
From the $x_t$'s we generated measurements, $y_t = Ax_t+w_t$ where $w_t$ was Gaussian noise with zero mean and variance $\sigma_{obs}^2= 0.0004$ and $A$ was a random Gaussian matrix of size $n_1 \times m$ for $t=1,2$ and of size $n_3 \times m$ for $t \ge 3$. We used $n_1=n_2=180$, and $n_t=n_3 = 0.234m=60$ for $t \ge 3$. More measurements were used for the first two time instants because we used simple BPDN to estimate $x_1$ and $x_2$ for these time instants. These were used for parameter estimation as explained in Section \ref{parameter_set}. 

We compare BPDN, BPDN-residual, PM-CS-KF (Algorithm \ref{pmcskf_algo}) \cite{ibm}, modified-BPDN (Algorithm \ref{modbpdn_algo}), weighted-$\ell_1$ (Algorithm \ref{w_ell1_algo}), streaming modified weighted-$\ell_1$ (streaming mod-wl1) \cite{asif2013sparse}, reg-mod-BPDN (Algorithm \ref{regmod_algo}) KF-ModCS (Algorithm \ref{kfcsalgo}), DCS-AMP \cite{schniter_track,schniter_track_jp} (algorithm in Table \ref{dcs_amp}), CS-MUSIC \cite{kim2012compressivemusic} and  Temporal SBL \cite{zhang_rao}. BPDN solved (\ref{ubpdn}) with  $y = y_t$, $A = A_t$ at time $t$ and $\gamma =\max\{10^{-2}\|A'[y_1 \ y_2]\|_{\infty}, \sigma_{\text{obs}}\sqrt{\log m}\}$ \cite{asif2013sparse}. BPDN-residual solved (\ref{ubpdn}) with $y = y_t - A_t \xhat_{t-1}$, $A = A_t$ at time $t$. 
Among these, BPDN uses no prior knowledge; BPDN-residual and PM-CS-KF are tracking-based methods that only use slow signal value change and sparsity; modified-BPDN and weighted-$\ell_1$ use only slow support change; reg-mod-BPDN, KF-ModCS, DCS-AMP use both slow support and slow signal value change; and streaming mod-wl1 enforces a ``soft" version of slow support change by also using the previous signal values' magnitudes.
CS-MUSIC \cite{kim2012compressivemusic} and temporal SBL \cite{zhang_rao} are two batch algorithms that solve the MMV problem (assumes the support of $x_t$ does not change with time).  Since $A$ is fixed, we are able to apply these as well. Temporal SBL \cite{zhang_rao} additionally also uses temporal correlation among the nonzero entries while solving the MMV problem. The MMV problem is discussed in detail in Sec. \ref{mmv_dynmmv} (part of related work and future directions).
%
We used the authors' code for DCS-AMP, PM-CS-KF, Temporal-SBL and CS-MUSIC. 
For the others, we wrote our own MATLAB code and used the $\ell_1$-homotopy solver of Asif and Romberg \cite{asif2013sparse} to solve the convex programs. 
With permission from the authors and with links to the authors' own webpages, the code for all algorithms used to generate the figures in the current article is posted at \url{http://www.ece.iastate.edu/~namrata/RecReconReview.html}.

\begin{figure}
\centering
\includegraphics[width=9cm, height=6cm]{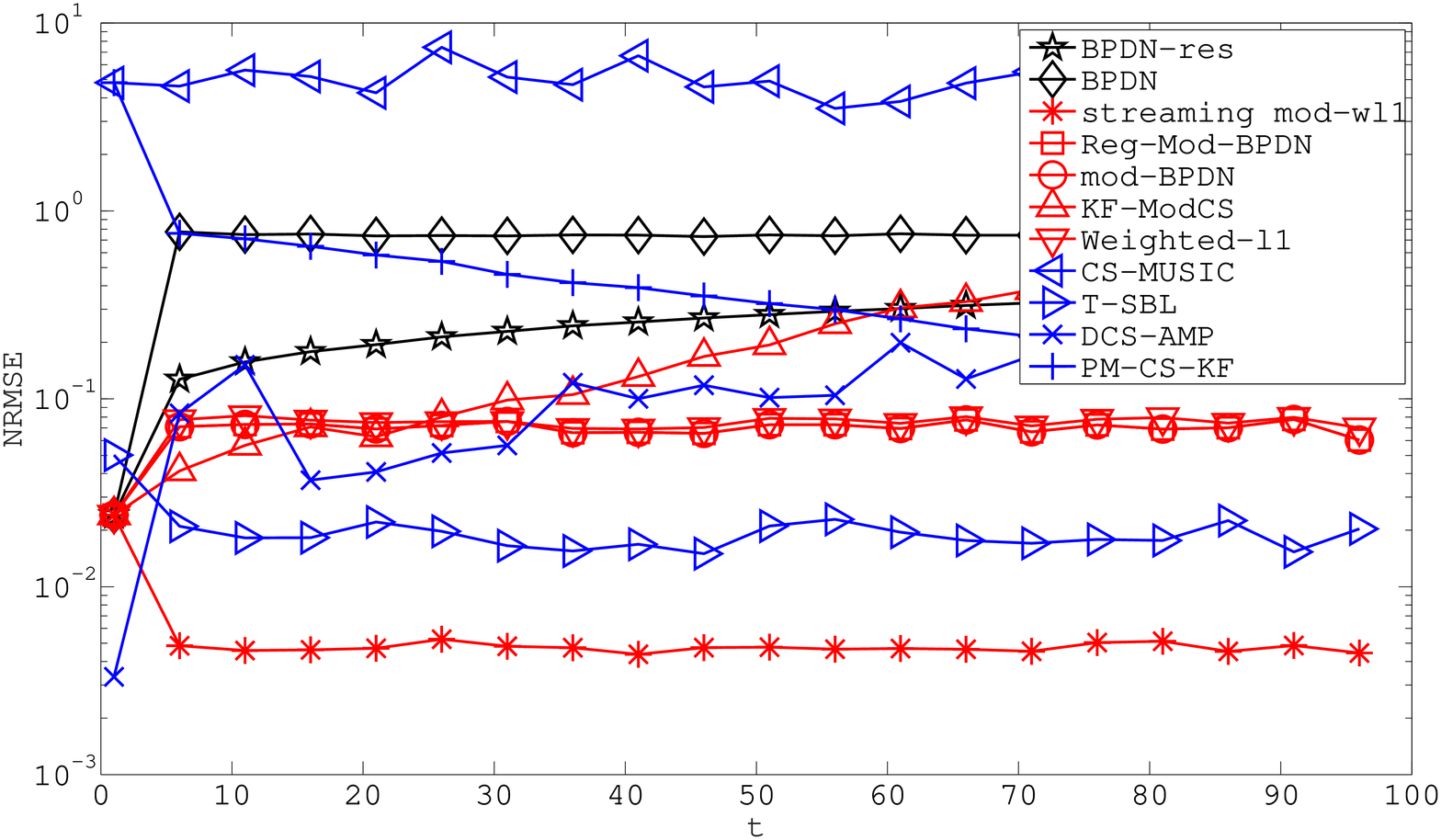}
\caption{\small{
NMSE plot for a simulated sparse signal sequence generated as explained in Section \ref{sim_seq}, solved by $\ell_1$-Homotopy \cite{asif2013sparse}.
}}
\label{nmse_simulated_l1homotopy}
\end{figure}

We plot the normalized root mean squared error (NRMSE), $\sqrt{\E[\|x_t -\xhat_t\|_2^2]}/\sqrt{\E[\|x_t\|_2^2]}$, in Fig. \ref{nmse_simulated_l1homotopy}. Here $\E[.]$ denotes the Monte Carlo estimate of the expectation. We averaged over 100 realizations. The average time taken by each algorithm is shown in Table \ref{t_simulated_l1homotopy}.
As can be seen from Fig. \ref{nmse_simulated_l1homotopy}, streaming mod-wl1 has the smallest error followed by temporal SBL and then reg-mod-BPDN, mod-BPDN and weighted-$\ell_1$. The errors of all these are also stable (do not increase with time) and are below 10\% at all times. 
Since enough measurements are used ($n=60$ for $s_t \le s=25$), in this figure, one cannot see the advantage of reg-mod-BPDN over mod-BPDN.
In terms of speed,DCS-AMP is the fastest but it has unstable or large errors. Streaming mod-wl1, mod-BPDN, reg-mod-BPDN and weighted-$\ell_1$ take similar amounts of time and all are 5-6 times slower than DCS-AMP. However, this difference in speed disappears for large sized problems (see Table \ref{t_larynx}, bottom). 


\begin{table*}[t!]
\centering
\begin{tabular}{ | c | c  | c | c | c | c | c | c | c |  c | c | c | c | c | c | c | c | c | c|} \hline
   reg-mod-BPDN & mod-BPDN &  BPDN-res & BPDN & KF-ModCS   & DCS-AMP & weighted-ell1 & PM-CS-KF & str-mod-wl1 &T-SBL  \\ \hline
   1.2340 &  1.0788 &  2.0748 &  4.4624 &  1.9184 &     0.1816 & 0.9018 &  1.6463 &  0.9933 & 28.8252  \\ \hline
\end{tabular}
\vspace{0.1in}
\caption{Averaged time taken (in seconds) by different algorithms to recover the simulated sequence, averaged over 100 simulations, solved by $\ell_1$-Homotopy.}
\label{t_simulated_l1homotopy}
\end{table*}

\begin{figure}
\centering
\subfigure[32x32 piece sequence]{
\label{nmse_larynx_l1homotopy}
\includegraphics[width=9cm, height=6cm]{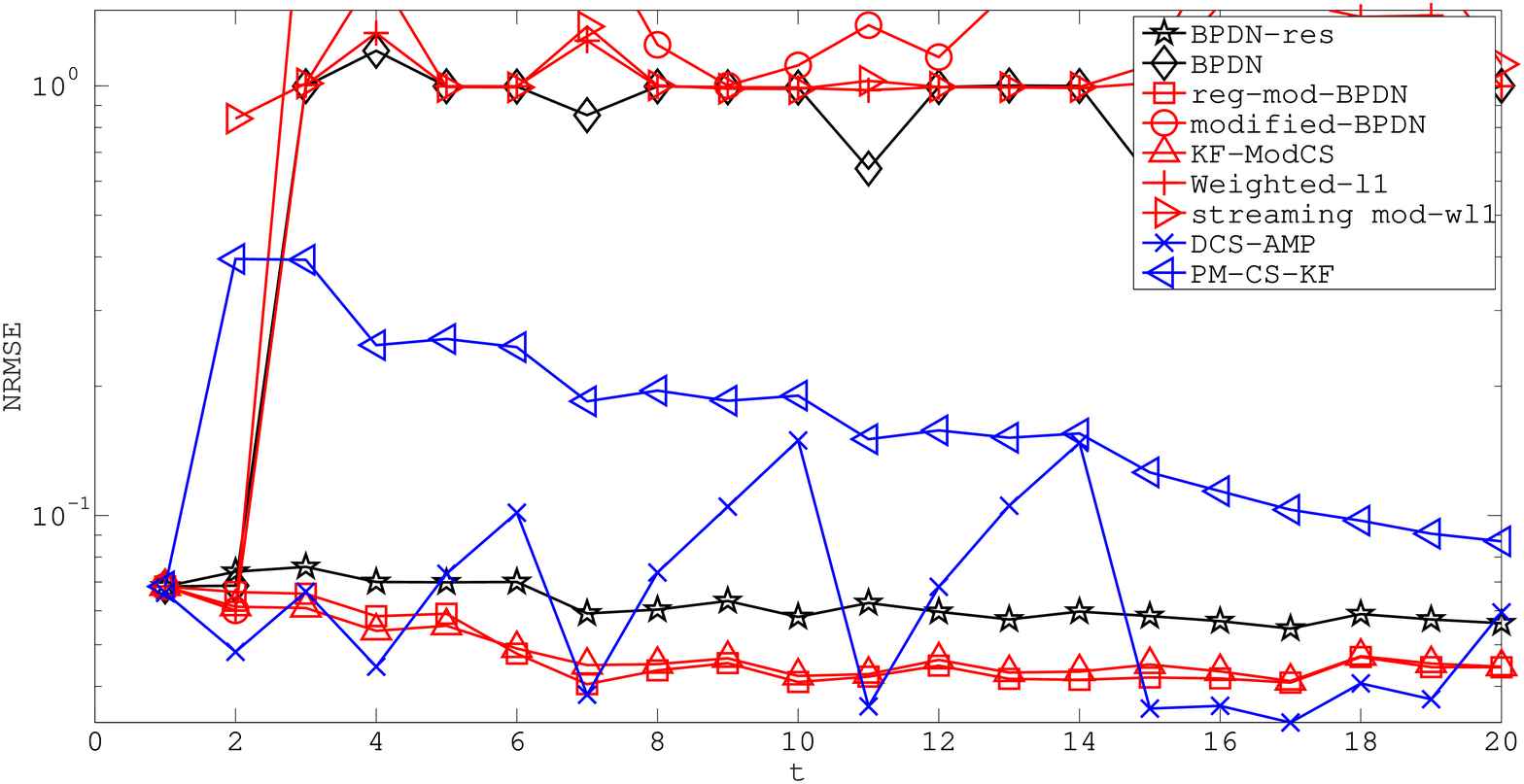}
}
%
%
\subfigure[full image (256x256) sequence]{
\label{nmse_larynx_full}
\includegraphics[width=9cm, height=6cm]{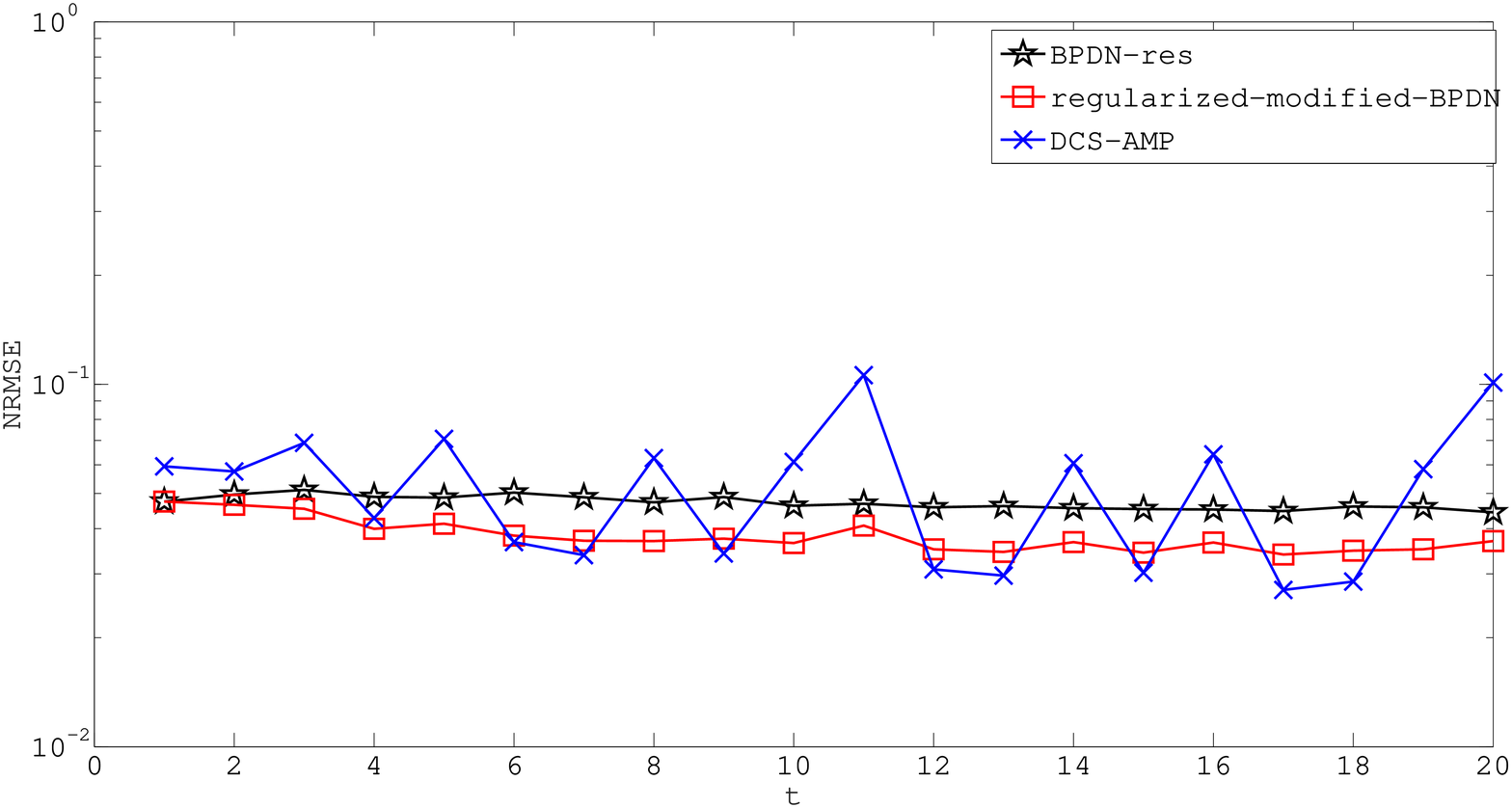}
}
\caption{\small{
NRMSE plot for recovering the larynx image sequence from simulated partial Fourier measurements corrupted by Gaussian noise. We used $n_1= n_2 = 0.18m$ and $n_t = 0.06m$ for $t>2$
(a): recovery error plot for a 32x32 sub-image sequence, used the $\ell_1$-homotopy solver. (b): recovery error plot for the full-sized image sequence, used the Yall1 solver.
}}
\label{nmse_larynx}
\end{figure}

\subsection{Recursive recovery of a real vocal tract dynamic MRI sequence (approximately sparse) from simulated partial Fourier measurements} \label{sims_mri}
We show two experiments on the larynx vocal tract dynamic MRI sequence (shown in Fig. \ref{examples}).
In the first experiment we select a $32 \times 32$ block of it that contains most of the significant motion. In the second one, we use the entire $256 \times 256$ sized image sequences. In the first experiment, the indices of the selected block were (60 : 91, 60 : 91). Thus $m=1024$. This was done to allow us to compare all algorithms including those that cannot work for large-scale problems. Let $z_t$ denote an $m_1 \times m_2$ image at time $t$ arranged as a 1D vector of length $m=m_1m_2$. We simulated MRI measurements as $y_t = H_t z_t + w_t$ where $H_t = I_{\mathcal{O}_t}{}' F_{2D,m} $ where $F_{2D,m}:=F_{m_1} \kron F_{m_2}$ corresponds to an $m_1 \times m_2$ 2D-DFT operation done using a matrix vector multiply. The observed set $\mathcal{O}_t$ consisted of $n_t$ rows selected using a modification of the low-frequency random undersampling scheme of Lustig et al. \cite{sparseMRI} and $w_t$ was zero mean i.i.d. Gaussian noise with variance $\sigma_{obs}^2 = 10$. We used $n_1= n_2 = 0.18m=184$ and $n_t = 0.06m=62$ for $t>2$. {To get 6\% measurements, we generated three mask matrices with 50\%, 40\% and 30\% measurements each using the low-frequency random undersampling scheme, multiplied them, and selected 62 rows out of the resulting matrix uniformly at random}. More measurements were used for the first two time instants because we used simple BPDN to estimate $x_1$ and $x_2$ for these time instants. These were used for parameter estimation as explained in Section \ref{parameter_set}.
For all algorithms, the sparsity basis that we used was a two-level Daubechies-4 wavelet. Thus $\Phi$ was the inverse wavelet transform corresponding to this wavelet written as a matrix. We compare all algorithms from the previous subsection except CS-MUSIC and T-SBL which require the matrix $A_t = H_t \Phi$ to be constant with time. The NRMSE is plotted in Fig. \ref{nmse_larynx_l1homotopy} and the time comparisons are shown in Table \ref{t_larynx}, top.

In the second experiment, we used the full 256x256 larynx image sequence (so $m=65536$) and generated simulated MRI measurements as above. In actual implementation, this was done by computing the 2D FFT of the image at time $t$ followed by retaining coefficients with indices in the set $\mathcal{O}_t$. We again used $n_1 = n_2 =  0.18m = 11797$ and $n_t = 0.06m=3933$ for $t>2$ and  $\sigma_{obs}^2 = 10$. We select the best algorithms from the ones compared in the first experiment and compare their NRMSE in Fig. \ref{nmse_larynx_full}. By ``best" we mean algorithms that had small error in the previous experiment and that can be implemented for the large-scale problem. We compare reg-mod-BPDN, DCS-AMP and BPDN-residual. For reg-mod-BPDN, we used $\gamma,\lambda$ from the previous experiment since these cannot be computed for this large problem. For solving the convex programs of reg-mod-BPDN and BPDN-residual, we used the YALL-1 solver \cite{yall1} which allows the user to work with partial Fourier measurements and with a DWT sparsity basis, without having to ever explicitly store the measurement matrix $H_t$ or the sparsity basis matrix $\Phi$ (both are too large to fit in memory for this problem). So, for example, it computes $H_t z$ by computing the FFT of $z$ followed by only keeping the entries with indices in $\mathcal{O}_t$. Similarly, for a vector $y$, it computes $H_t'y$ by computing the inverse FFT of $y$ and retaining the entries with indices in $\mathcal{O}_t$ to get the row vector $(H_t'y)$. 
The time comparison for this experiment is shown in Table \ref{t_larynx}, bottom.

As can be seen from Fig. \ref{nmse_larynx} and Table \ref{t_larynx}, reg-mod-BPDN has the smallest error although it is not the fastest. However it is only a little slower than DCS-AMP for the large scale problem (see Table \ref{t_larynx}, bottom). This is, in part, because the YALL-1 solver is being used for it and that is much faster. DCS-AMP error is almost as small for many time instants but not all. Its performance seems to be very dependent on the specific $A_t$ used. Another thing to notice is that algorithms such as BPDN-residual also do not have error that is too large since this example consists of compressible signal sequences.

\begin{table}[b!]
\centering
\begin{tabular}{| c | c  | c |  c | c | c | c | c | c | c | c | c | c | c | c | c | c | c | c } \hline
  reg-mod-BPDN & mod-BPDN &  BPDN-res & BPDN & KF-ModCS  \\ \hline
 16.4040  &  6.5424 & 2.0383 & 2.3205 &  10.8591  \\ \hline
 DCS-AMP & Weighted $\ell_1$ & PM-CS-KF   & & \\ \hline
  0.1515 & 2.4815 & 9.9332 & &  \\ \hline
 \end{tabular}
\vspace{0.1in}
\\
%
%
\begin{tabular}{  | c | c  |  c | c | c | c | c | c | c | c | c | c | c | c | c | c | c | c | c } \hline \label{t_larynx_full}
  reg-mod-BPDN & BPDN-res  & DCS-AMP \\ \hline
 6.6700  & 17.4458  &  4.9160  \\ \hline
 \end{tabular}
\vspace{0.1in}
\caption{\small{Averaged time taken (in seconds) by different algorithms to recover  a 32x32 piece of the larynx sequence (top) and to recover the full 256x256 larynx sequence (bottom), averaged over 100 simulations.}}
\label{t_larynx}
\end{table}

\section{Related Work and Future Directions} \label{future}
We split the discussion in this section into four parts - more general signal models, more general measurement models, open questions for the algorithms described here, and how to do sequential detection and tracking using compressive measurements.

\subsection{Signal Models}

\subsubsection{Structured Sparsity} 
There has been a lot of recent work on structured sparsity for a single signal. Dynamic extensions of these ideas should prove to be useful in applications where the structure is present and changes slowly. Two common examples of structured sparsity are block sparsity \cite{stojnic_hassibi,stojnic2} and tree structured sparsity (for wavelet coefficients) \cite{modelcs}. A length $m$ vector is block sparse if it can be partitioned into length $k$ blocks such that a lot of the blocks are entirely zero. One way to recover block sparse signals is by solving the $\ell_2$-$\ell_1$ minimization problem \cite{stojnic_hassibi,stojnic2}.
Block sparsity is valid for many applications, e.g., for the foreground image sequence of a video consisting of one or a few moving objects, or for the activation regions in brain fMRI. In both of these cases, it is also true that the blocks do not change arbitrarily and hence the block support from the previous time instant should be a good estimate of the current block support. In this case one, can again use a modified-CS type idea applied to the blocks. This was developed by Stojnic \cite{modblockcs_stojnic}. It solves
$$\min_b \sum_{j=1, j \notin \T}^{m/k} \sqrt{ \sum_{i=1}^{k-1} b_{jk+i}^2 } \ \text{ s.t. } \ y = Ab $$
where $\T$ is the set of known nonzero blocks.
Similarly, it should be possible to develop dynamic extensions of the tree-structured IHT algorithm of Cevher et al. \cite{modelcs} or of the approximate model-based IHT algorithm of Hegde et al. \cite{modelcs_approx}. 
Another related work \cite{emd_modelcs} assumes that the earth-mover's distance between the support sets of consecutive signals is small and uses this to design a recursive dynamic CS algorithm. 

\subsubsection{MMV and dynamic MMV} \label{mmv_dynmmv}
In the MMV problem \cite{wipf2007empirical,tropp2006algorithms2,chen2006theoretical,eldar2009compressed,kim2012compressivemusic,lee_bresler,davies_eldar}, the goal is to recover a set of sparse signals with a common support but different nonzero signal values from a set of their measurements (all obtained using the same measurement matrix). 
This problem can be interpreted as that of block-sparse signal recovery and hence one commonly used solution is the $\ell_2$-$\ell_1$ program. Another more recent set of solutions is inspired by the MUSIC algorithm and are called CS-MUSIC \cite{kim2012compressivemusic} or iterative-MUSIC \cite{lee_bresler}.
For signals with time-varying support (the case studied in this article), one can still use the MMV approaches as long as the size of their joint support (union of their supports), $\N:= \cup_t \N_t$, is small enough. This may not be true for the entire sequence, but will usually be true for short durations. One could design a modified-MMV algorithm that utilizes the joint support of the previous duration to solve the current MMV problem better.
A related idea was explored in a recent work \cite{dyn_supp_track}. 

Another related work is that of Zhang and Rao \cite{zhang_rao}. In it, the authors develop what they call the temporal SBL (T-SBL) algorithm. This is a batch (but fast) algorithm that solves the MMV problem with temporal correlations. It assumes that the support of $x_t$ {\em does not} change over time and the signal values are correlated over time. The various indices of $x_t$ are assumed to be independent. For a given index, $i$, it is assumed that $[(x_1)_i, (x_2)_i, \dots (x_{t_{\max}})_i]' \sim \n(0, \gamma_i B)$. All the $x_t$'s are strung together to get a long $t_{\max} m$ length vector $x$ that is Gaussian with block diagonal covariance matrix $\text{diag}(\gamma_1 B, \gamma_2 B, \dots \gamma_m B)$. T-SBL develops the SBL approach to estimate the hyper-parameters $\{\sigma^2,\gamma_1, \gamma_2, \dots \gamma_m, B\}$ and then compute an MAP estimate of the sparse vector sequence.

Another related work \cite{giannakis_2} develops and studies a causal but batch algorithm for CS for time-varying signals.


\subsubsection{Sparse Transform Learning and Dictionary Learning}
In certain applications involving natural images, the wavelet transform provides a good enough sparsifying basis. However, for many other applications, while the wavelet transform is one possible sparsifying basis, it can be significantly improved. There has been a large amount of recent work both on dictionary learning, e.g., \cite{ksvd}, and more recently on sparsifying transform learning from a given dataset of sparse signals \cite{sparse_tfm_learn}. The advantage of the latter work is that it is much faster than existing dictionary learning approaches. For dynamic sparse signals, an open question of interest is how to learn a sparsifying transform that is optimized for signal sequences with slow support change?

\subsection{Measurement Models} 

\subsubsection{Recursive Dynamic CS in Large but Structured Noise} \label{reclarge}
The work discussed in this article solves the recursive dynamic CS problem either in the noise-free case or in the small noise case. Only in these cases, one can show that the reconstruction error is small compared to the signal energy. In fact, this is true for almost all work on sparse recovery; one can get reasonable error bounds only for the small noise case.

However, in some applications, the noise energy can be much larger than that of the sparse signal. If the noise is large but has no structure then nothing can be done. But if it does have structure, that can be exploited. This was done for outliers (modeled as sparse vectors) in the work of Wright and Ma \cite{dense_error_correct}. Their work, and then many later works, showed exact sparse recovery from large but sparse noise (outliers) as long as the sparsity bases for the signal and the noise/outlier are ``different" or ``incoherent" enough. In fact, as noted by an anonymous reviewer, more generally, the earlier works on sparsity in unions of bases, e.g.,  \cite{donoho_huo,elad_bruckstein,gribonval_neilson}, can also be interpreted in this fashion.
Recent work on robust PCA by Candes et al. \cite{rpca} and Chandrasekharan et al. \cite{rpca2} posed robust PCA as a problem of separating a low-rank matrix and a sparse matrix from their sum and proposed a convex optimization solution to it.
%
In more recent work \cite{rrpcp_allerton,rrpcp_perf,rrpcp_tsp,rrpcp_aistats,grass_undersampled,xu_nips2013_1}, the recursive or online robust PCA problem was solved. This can be interpreted as a problem of recursive dynamic CS in large but structured noise (noise that is dense and lies in a fixed or ``slowly changing" low-dimensional subspace of the full space). An open question is how can the work described in this article be used in this context and what we say about performance guarantees for the resulting algorithm? For example, in \cite{rrpcp_allerton11,rrpcp_tsp,rrpcp_isit}, the authors have attempted to use weighted-$\ell_1$ to replace BP-noisy with encouraging results.

\subsubsection{Recursive Recovery from Nonlinear Measurements and Dynamic Phase Retrieval}
The work described in this article focuses on sparse recovery from linear measurements. However, in many applications such as computer vision, the measurement (e.g., image) is a nonlinear function of the sparse signal of interest (e.g., object's boundary which is often modeled as being Fourier sparse). Some recent work that has studied the static version of this ``nonlinear CS" problem includes \cite{nonlinear_cs1,nonlinear_cs2,nonlinear_cs3,nlcs_blumensath,nlcs_eldar}. These approaches use an iterative linearization procedure along with adapting standard iterative sparse recovery techniques such as IHT. An extension of these techniques for the dynamic case can potentially be developed using an extended Kalman filter and the modified-IHT (IHT-PKS) idea from \cite{modcosamp_udel}.
There has been other work on solving the dynamic CS problem from nonlinear measurements by using particle filtering based approaches \cite{pafimocs_asilomar,nlcs_pf_sastry,sparse_dynamic_sys,pafimocs_icassp}.

An important special case of the nonlinear CS problem is sparse phase retrieval,  i.e., recover a sparse $x$ from $y:= |Ax|$. Here $|.|$ takes the element-wise magnitude of the vector $(Ax)$. The special case of this problem where $A$ is the Fourier matrix occurs in applications such as astronomical imaging, optical imaging and X-ray crystallography where one can only measure the magnitude of the Fourier coefficients of the unknown quantity. This problem has been studied in a series of recent works \cite{candes_phaselift,phase_sastry,phase_baraniuk, jaganathan2012recovery,jaganathan2013sparse,candes2013phase,jaganathan2012recovery,wirtinger_sparse}. An open question is how to design and analyze an approach for phase retrieval for a time sequence of sparse signals and how much will the use of past information help? For example, the work of Jaganathan et al.  \cite{jaganathan2012recovery} provides a provably correct approach for sparse Fourier phase retrieval. It involves first using a combinatorial algorithm to estimate the signal's support, followed by using a lifting technique to get a convex optimization program for positive semi-definite matrices.
As explained in \cite{jaganathan2012recovery}, the lifting based convex program cannot be used directly because of the location ambiguity introduced by the Fourier transform. Consider the dynamic sparse phase retrieval problem. An open question is whether the support and the signal value estimates from the previous time instant can help regularize this problem enough to ensure a unique solution when directly solving the resulting lifting-based convex program? If the combinatorial support recovery algorithm can be eliminated, it would make the solution approach a lot faster. 

\subsection{Algorithms}

The work done on this topic so far consists of good algorithms that improve significantly over simple-CS solutions, and some of them come with provable guarantees. However, it is not clear if any of these are ``optimal" in any sense. An open question is, can we develop an ``optimal" algorithm or can we show that an algorithm is close enough to an ``optimal" one? For sparse recovery techniques, it is not even clear what a tractable measure of ``optimality" is?

The original KF-CS algorithm \cite{kfcsicip} was developed with this question in mind; however so far there has not been any reasonable performance bound for it or for its improved version, KF-ModCS. An open question is, can we show that KF-ModCS comes within a bounded distance of the genie-aided causal MMSE solution for this problem (the causal MMSE solution assuming that the support sets at each time are known)?

In fact, none of the approaches that utilize both slow support and signal value change have stability results so far. This is an important question for future work. In particular, it would be interesting to analyze streaming mod-wl1 (Algorithm \ref{dyn_ mod_wl1}) \cite{asif2013sparse} since it has excellent performance in simulations.%

There is other very recent work on necessary and sufficient conditions for weighted-$\ell_1$ \cite{w_ell1_sharp}.

\subsection{Compressive Sequential Signal Detection and Tracking}
In many signal processing applications, the final goal is to use the recovered signal for detection, classification, estimation or to filter out a certain component of the signal (e.g. a band of frequencies or some other subspace). The question is can we do this directly with compressive measurements without having to first recover the sparse signal? This has been studied in some recent works such as the work of Davenport et al. \cite{compressive_det}. 
For a time sequence of sparse signals, a question of interest is how to do the same thing for compressive sequential detection/classification or sequential estimation (tracking)? The question to answer would be how to use the previously reconstructed/detected signal to improve compressive detection at the current time and how to analyze the performance of the resulting approach?


%

\section{Conclusions} \label{conclude}
This article reviewed the literature on recursive recovery of sparse signal sequences or what can be called ``recursive dynamic CS". Most of the literature on this topic exploits one or both of the practically valid assumptions of slow support change and slow signal value change. While slow signal value change is commonly used in a lot of previous tracking and adaptive filtering literature, the slow support change is a new assumption introduced for solving this problem. As shown in Fig. \ref{suppchange}, this is indeed valid in applications such as dynamic MRI.
We summarized both theoretical and experimental results that demonstrate the advantage of using these assumptions. In the section above, we also discussed related problems and open questions for future work.

A key limitation of almost all the work reviewed here is that the algorithms assume more measurements are available at the first time instant. This is needed in order to get an accurate initial signal recovery using simple-CS solutions. In applications such as dynamic MRI or functional MRI, it is possible to use more measurements at the initial time (the scanner can be configured to allow this). Other solutions to this issue have been described in Section \ref{initialize}.

\appendices
\section{Computable error bound for reg-mod-BPDN, mod-BPDN, BPDN} \label{erc_noisy_bnd_appendix}
We define here the terms used in Theorem \ref{regmod_thm}.
Let $I_{\T,\T}$ denote the identity matrix on the row, column indices $\T,\T$ and let $\mathbf{0}_{\T,S}$ be a zero matrix on the row, column indices $\T,S$.
Define
\bea
\text{maxcor}(\tDelta) \sdefn \max_{i\notin (T\cup \tDelta)^c}\|{A_i}'A_{\T\cup \tDelta}\|_2, \nn \\
Q_{\T,\lambda}(S) \sdefn {A_{\T\cup S}}'A_{\T\cup S}+\lambda  \left[
                                                 \begin{array}{cc}
                                                   I_{\T,\T}\ & \mathbf{0}_{\T,S} \\
                                                   \mathbf{0}_{S,\T}\ & \mathbf{0}_{S,S} \\
                                                 \end{array}
                                               \right]  \nn \\
ERC_{\T,\lambda}(S) \sdefn 1-\max_{\omega \notin T\cup S}\|P_{\T,\lambda}(S){A_{S}}'M_{\T,\lambda}A_{\omega}\|_1, \nn \\
P_{\T,\lambda}(S)\sdefn ({A_{S}}'M_{\T,\lambda}A_{S})^{-1}  \nn \\
M_{\T,\lambda} \sdefn I-A_\T({A_\T}'A_\T+\lambda I_{\T,\T})^{-1}{A_\T}'  \nn
\eea
and
\bea
f_1(\tilde\Delta_u) \sdefn   \sqrt{\|({A_\T}'A_\T + \lambda I_{\T})^{-1}   {A_\T}'A_{\tilde\Delta_u}P_{\T,\lambda}(\tilde\Delta_u)\|_{2}^2 +  \|P_{\T,\lambda}(\tilde\Delta_u)\|_{2}^2}, \nn \\
f_2(\tilde\Delta_u) \sdefn    \|Q_{\T,\lambda}(\tilde\Delta_u)^{-1}\|_2 \nn \\
f_3(\tilde\Delta_u) \sdefn \|Q_{\T,\lambda}(\tilde\Delta_u)^{-1}{A_{\T\cup \tilde\Delta_u}}'\|_2, \nn \\
f_4(\tDelta) \sdefn \sqrt{\|Q_{\T,\lambda}({\tilde\Delta_u})^{-1}{A_{\T\cup {\tilde\Delta_u}}}'A_{\tilde\Delta_u \setminus {\tilde\Delta_u}}\|_2^2+1}. \nn \\
g_1(\tDelta) \sdefn   \lambda f_2(\tDelta) (\frac{\sqrt{|\tDelta|}f_1(\tDelta) \text{maxcor}(\tDelta)}{ERC_{\T,\lambda}(\tDelta)}+1 ), \nn \\
g_2(\tDelta) \sdefn  \frac{\sqrt{|\tDelta|}f_1(\tDelta)f_3(\tDelta) \text{maxcor}(\tDelta) }{ERC_{\T,\lambda}(\tDelta)}+f_3(\tDelta), \nn \\
g_3(\tDelta) \sdefn \frac{\sqrt{|\tDelta|}f_1(\tDelta)f_4(\tDelta) \text{maxcor}(\tDelta) }{ERC_{\T,\lambda}(\tDelta)}+f_4(\tDelta), \nn \\
g_4(\tDelta) \sdefn \frac{\sqrt{|\tDelta|}\|{A_{(T\cup \tDelta)^c}}\|_\infty \|w\|_{\infty}f_1(\tDelta)}{ERC_{\T,\lambda}(\tDelta)} \nn
\eea

For a set $\tilde\Delta_u \subseteq \Delta_u$, define
\bea
\label{gamma_expression}
\gamma^*_{\T,\lambda}(\tilde\Delta_u) \sdefn \frac{\text{maxcor}(\tDelta) }{\text{ERC}_{\T,\lambda}(\tilde\Delta_u)} \left[ \lambda f_2(\tDelta)\|x_{\T}-\hat{\mu}_\T\|_2 +  \right.  \nn \\
&& \left. f_3(\tDelta)\|w\|_2  + f_4(\tDelta)\|x_{\Delta_u \setminus \tDelta}\|_2 \right]  \nn \\
&& + \frac{\|w\|_{\infty}}{\text{ERC}_{\T,\lambda}(\tilde\Delta_u)}, \\
g_\lambda(\tilde\Delta_u) \sdefn g_1(\tDelta) \|x_\T - \hat{\mu}_\T\|_2 + g_2(\tDelta) \|w\|_2 \nn \\
            &&
+ g_3(\tDelta) \|x_{\Delta_u \setminus \tDelta}\|_2       +g_4(\tDelta)
\eea

For an integer $k$, define
\bea
\tilde\Delta_u^{*}(k) \sdefn \arg\min_{\tilde\Delta_u \subseteq \Delta_u, |\tilde\Delta_u|=k}  \|x_{\Delta_u \setminus \tilde\Delta_u}\|_2 \label{tilde_Delta_star}
\eea
This is a subset of $\Delta_u$ of size $k$ that contains the $k$ largest magnitude entries of $x$.

Let $k_{\min}$ be the integer $k$ that results in the smallest error bound $g_\lambda(\tilde\Delta_u^{*}(k))$ out of all integers $k$ for which $\text{ERC}_{\T,\lambda}(\tilde\Delta_u^{*}(k)) > 0$ and $Q_{\T,\lambda}(\tilde\Delta_u^{*}(k)) \ \text{is invertible}$. Thus,
\bea
k_{\min} \sdefn \arg \min_k B_k   \ \text{where}  \nn \\
B_k \sdefn
\left\{ \begin{array}{cc}
g(\tilde\Delta_u^{*}(k)) \ & \ \text{if} \  \text{ERC}_{\T,\lambda}(\tilde\Delta_u^{*}(k)) > 0  \ \text{and} \nn \\
                         & \ Q_{\T,\lambda}(\tilde\Delta_u^{*}(k)) \ \text{is invertible} \nn \\
\infty \ & \ \text{otherwise} \nn
\end{array}
\right. \nn
\eea

%
Notice that $\arg\min_{\tilde\Delta_u \subseteq \Delta_u, |\tilde\Delta_u|=k}  \|x_{\Delta_u \setminus \tilde\Delta_u}\|_2$ in (\ref{tilde_Delta_star}) is computable in polynomial time by just sorting the entries of $x_{\Delta_u}$ in decreasing order of magnitude and retaining the indices of its largest $k$ entries. Hence everything in the above result is computable in polynomial time if $x$, $\T$ and a bound on $\|w\|_\infty$ are available. Thus if training data is available, the above theorem can be used to compute good choices of $\gamma$ and $\lambda$.

\bibliographystyle{IEEETran}%
\bibliography{../bib/tipnewpfmt_kfcsfullpap,../bib/tipnewpfmt_cq}
\end{document}